\documentclass[usenatbib]{mn2e}
\usepackage{natbib,amssymb,amsmath,times,graphicx,url,aas_macros}
\usepackage{wasysym}	  
\usepackage{setspace}      
\usepackage{epstopdf}       
\usepackage{verbatim}       
\usepackage{bm}		  
\usepackage{fleqn}		  

\newcommand{\zetaos}{$\zeta~=~10^{-17}$~s$^{-1}$}
\newcommand{\zetaoe}{$\zeta~=~10^{-18}$~s$^{-1}$}

\newcommand{\dexoe}{$10^{-18}$~s$^{-1}$}
\newcommand{\bopos}{$\bm{B}_0\cdot\bm{\Omega}_0>0$}
\newcommand{\boneg}{$\bm{B}_0\cdot\bm{\Omega}_0<0$}
\newcommand{\bo}{$\bm{B}_0\cdot\bm{\Omega}_0$}
\newcommand{\lores}{$\sim$3$\times 10^5$}
\newcommand{\hires}{$\sim$$10^6$}
\defcitealias{pricebate07}{PB07}

\title[Can non-ideal MHD save discs?]{Can non-ideal magnetohydrodynamics solve the magnetic braking catastrophe?}

\author[Wurster, Price \& Bate]{James Wurster$^{1}$\thanks{james.wurster@monash.edu}, Daniel J. Price$^{1}$\thanks{daniel.price@monash.edu} and Matthew R. Bate$^{2,1}$ \\
$^{1}$Monash Centre for Astrophysics and School of Physics and Astronomy, Monash University, Vic 3800, Australia \\
$^{2}$School of Physics, University of Exeter, Stocker Rd, Exeter EX4 4QL, UK \\
}
\date{Submitted: Revised: Accepted: }
\pagerange{\pageref{firstpage}--\pageref{lastpage}} \pubyear{2015}

\begin{document}
\label{firstpage}
\bibliographystyle{mn2e}
\maketitle

\begin{abstract}
 We investigate whether or not the low ionisation fractions in molecular cloud cores can solve the `magnetic braking catastrophe', where magnetic fields prevent the formation of circumstellar discs around young stars. We perform three-dimensional smoothed particle non-ideal magnetohydrodynamics (MHD) simulations of the gravitational collapse of one solar mass molecular cloud cores, incorporating the effects of ambipolar diffusion, Ohmic resistivity and the Hall effect alongside a self-consistent calculation of the ionisation chemistry assuming 0.1$\micron$ grains.  When including only ambipolar diffusion or Ohmic resistivity, discs do not form in the presence of strong magnetic fields, similar to the cases using ideal MHD.  With the Hall effect included, disc formation depends on the direction of the magnetic field with respect to the rotation vector of the gas cloud.  When the vectors are aligned, strong magnetic braking occurs and no disc is formed.  When the vectors are anti-aligned, a disc with radius of 13~AU can form even in strong magnetic when all three non-ideal terms are present, and a disc of 38~AU can form when only the Hall effect is present; in both cases, a counter-rotating envelope forms around the first hydrostatic core.  For weaker, anti-aligned fields, the Hall effect produces massive discs comparable to those produced in the absence of magnetic fields, suggesting that planet formation via gravitational instability may depend on the sign of the magnetic field in the precursor molecular cloud core.
\end{abstract}


\section{Introduction}
\label{intro}
 The `magnetic braking catastrophe' refers to the failure of numerical star formation calculations to produce rotationally supported Keplerian discs when magnetic fields of strengths comparable to those observed in molecular clouds \citep[e.g.][]{HeilesCrutcher2005} are accounted for. This has been found in analytic studies \citep{als03,gallietal06}, axisymmetric numerical models \citep{mellonli08} and in 3D calculations using ideal magnetohydrodynamics (MHD)  (\citealt{pricebate07,hennebellefromang08,duffinpudritz09,hennebelleciardi09,commerconetal10,SeifriedEtAl2011}). By contrast, recent observations suggest the presence of massive, 50-100 AU discs and evidence for associated outflows in the earliest (Class 0) stages of star formation around both low and high mass stars \citep[e.g.][]{dunhametal11,LindbergEtAl2014,TobinEtAl2015}.
 
 Two primary solutions have been proposed: turbulence and non-ideal magnetohydrodynamics. \citet{SeifriedBanerjeePudritzKlessen2012} showed in calculations of the collapse of a massive 100~M$_{\odot}$ core, that 100~AU-scale disc formation in the presence of strong magnetic fields was indeed possible, with some argument over whether this is caused by turbulent reconnection \citep{SantoslimaEtAl2012,SantoslimaEtAl2013} or another mechanism \citep{SeifriedBanerjeePudritzKlessen2013,lietal14}. \citet{JoosEtAl2013} found, using simulations of collapsing 5~M$_{\odot}$ cores, that turbulence diffuses the strong magnetic field out of the inner regions of the core, and that the non-zero angular momentum of the turbulence causes a misalignment between the rotation axis and the magnetic field.  Both of these effects reduce the magnetic braking, and allow a massive disc to form. However, their turbulent discs remain smaller than the discs formed without magnetic fields.
 
The other possible solution is to include non-ideal MHD, which is the focus of our study.  The initial studies all assumed ideal MHD: The gas is fully ionised with the ions and electrons being tied to the magnetic field lines.  This has been known to be a poor approximation to the true conditions in molecular cloud cores since at least \citet{mestelspitzer56}. Detailed models for the ionisation fraction in dense cores find values as low as $n_\text{e}/ n_{\text{H}_2} = 10^{-14}$ (\citealt{nakanoumebayashi86,umebayashinakano90}, our Fig.~\ref{fig:AllBaro}). Partial ionisation leads to three main non-ideal MHD effects: \emph{Ohmic resistivity} (drift between electrons and ions/neutrals; neither ions nor electrons are tied to the magnetic field), the \emph{Hall effect} (ion-electron drift; only electrons are tied to the magnetic field), and \emph{Ambipolar diffusion} (ion-neutral drift; both ions and electrons are tied to the magnetic field). The relative importance of each of these depends, amongst other things, on the gas density and magnetic field strength (e.g. \citealp{wardleng99}; \citealp{nnu02}; \citealp{tassismouschovias07a}; \citealp{Wardle2007}; \citealp{pandeywardle08}; \citealp{keithwardle14}), with \citet{Wardle2007} suggesting a steady progression through ambipolar, Hall and Ohmic-dominated regimes as gravitational collapse proceeds.

 \citet{shuetal06} found that adding a constant Ohmic resistivity indeed re-enabled Keplerian disc formation in their self-similar analytic study, but found that an anomalously high resistivity ($\eta \sim 10^{22} {\rm~cm}^{2}{\rm s}^{-1}$) was required; this is around two orders of magnitude higher than the microscopic value (see bottom panel of Fig.~\ref{fig:AllBaro}). \citet{kls10} were able to reduce the amount of resistivity required by making different assumptions, but the required resistivity is still uncomfortably high. Correspondingly, numerical simulations with Ohmic diffusion show only small, AU-scale discs \citep{DappBasu2010,MachidaInutsukaMatsumoto2011,tomidaetal13}. Similarly, a number of authors have concluded that ambipolar diffusion alone is unable to sufficiently weaken the magnetic braking to allow large, rotationally supported discs to form under realistic conditions \citep{duffinpudritz09,MellonLi2009,LiKrasnopolskyShang2011,DappBasuKunz012,TomidaOkuzumiMachida2015,TsukamotoEtAl2015}.

The Hall effect differs from both Ohmic resistivity and ambipolar diffusion since it is not dissipative (rather it introduces a new wave, the whistler mode) and is the only effect which is sensitive to the direction of the magnetic field. The fast timescale associated with the whistler wave makes the Hall effect difficult to model numerically (see Section~\ref{ssec:num:timestepping}), so it is usually neglected in numerical simulations. Instead, \citet{BraidingWardle2012sf,BraidingWardle2012accretion} presented similarity solutions to the MHD equations for rotating, isothermal gravitational collapse. They concluded that, although the Hall effect was not the dominant term in their calculations, it was the determining factor between a solution yielding a disc with a realistic surface density or a disc with a surface density much lower than required for fragmentation and planet formation.  The only difference between the two extremes was the direction of the magnetic field with respect to the rotation vector. In a related study  using idealised calculations \citet{KrasnopolskyLiShang2011}, showed that the Hall effect could indeed enable the formation of $\sim$10~AU discs depending on the sign of the magnetic field.

Here, we evaluate the influence of all three non-ideal MHD effects, including the Hall effect, on the formation of discs, using 3D non-ideal self-gravitating smoothed particle magnetohydrodynamics simulations of collapsing, low mass cores, following the original ideal MHD study of \citet{pricebate07}  (hereafter \citetalias{pricebate07}). We present the numerical formulation in Section~\ref{sec:numerics}, including the self-consistent ionisation calculations (Section~\ref{ssec:num:ionise}). Our initial conditions are given in Section~\ref{sec:ic}.  Results are presented in Section~\ref{sec:results}, with discussion and conclusions in Section~\ref{sec:discussion}.

\section{Numerical method}
\label{sec:numerics}

\subsection{Non-ideal Magnetohydrodynamics}
\label{sec:numerics:nimhd}
We solve the equations of self-gravitating, non-ideal magnetohydrodynamics given by
\begin{eqnarray}
\frac{{\rm d}\rho}{{\rm d}t} & = & -\rho \nabla\cdot \bm{v}, \label{eq:cty} \\
\frac{{\rm d} \bm{v}}{\rm{d} t} & = & -\frac{1}{\rho}\bm{\nabla} \left[\left(P+\frac{1}{2}B^2\right)I - \bm{B}\bm{B}\right] - \nabla\Phi, \label{eq:mom} \\
\frac{{\rm d} \bm{B}}{\text{d} t} & = & \left(\bm{B}\cdot\bm{\nabla}\right)\bm{v}-\bm{B}\left(\bm{\nabla}\cdot\bm{v}\right) + \left.\frac{\text{d} \bm{B}}{\text{d} t}\right|_\text{non-ideal} \label{eq:ind}, \\
\nabla^{2}\Phi & = & 4\pi G\rho, \label{eq:grav}
\end{eqnarray}
where $\frac{\text{d}}{\text{d}t} \equiv \frac{\partial}{\partial t} + \bm{v}\cdot\bm{\nabla}$ is the Lagrangian derivative,  $\rho$ is the density, ${\bm  v}$ is the velocity, $P$ the hydrodynamic pressure, ${\bm B}$ is the magnetic field, $\Phi$ is the gravitational potential, and $I$ is the identity matrix.  The magnetic field has been normalised such that the Alfv{\'e}n velocity is defined as $v_\text{A}\equiv B/\sqrt{\rho}$ in code units.  The equation set is closed by the barotropic equation of state,
\begin{equation}
\label{eq:eos}
P = \left\{ \begin{array}{l l} c_\text{s,0}^2\rho; 		    &  \rho < \rho_\text{c}, \\
                                           c_\text{s,0}^2\rho_\text{c}\left(\rho             /\rho_\text{c}\right)^{7/5};      &  \rho_\text{c} \leq \rho < \rho_\text{d}, \\
                                           c_\text{s,0}^2\rho_\text{c}\left(\rho_\text{d}/\rho_\text{c}\right)^{7/5} \left(\rho/\rho_\text{d}\right)^{11/10};      &  \rho \geq \rho_\text{d},
\end{array}\right.
\end{equation}
where $c_\text{s,0}$ is the initial isothermal sound speed, $\rho_\text{c} = 10^{-14}$ and $\rho_\text{d}~=~10^{-10}$ g~cm$^{-3}$.  Although we do not employ full radiation magnetohydrodynamics, the barotropic equation of state is designed to mimic the evolution of the equation of state in molecular clouds \citep{Larson1969,MasunagaInutsuka2000,MachidaInutsukaMatsumoto2008}.  

These threshold densities, $\rho_\text{c}$ and $\rho_\text{d}$, are the same as used in \citet{PriceTriccoBate2012} and \citet{LewisBatePrice2015}.  The chosen value of $\rho_\text{c}$ is lower than physically motivated in order to artificially heat the disc that forms to prevent it from fragmenting.  The value of $\rho_\text{d}$ is also lower than physically motivated based on the temperature at which the second collapse should start, but it will not affect our results since our densities seldom reach values greater than $\rho_\text{d}$ due to our threshold for sink particle creation (see Section \ref{sec:ic}).  We explicitly caution that these threshold densities are not satisfactory to study the second collapse, but sink particles are inserted prior to the onset of the second collapse.  

In the given equation of state, the pressure is continuous across density thresholds.  However, since the exponent on $\rho$ changes at these densities, the local sound speed and temperature, $c_\text{s}$ and $T$, respectively, will be discontinuous.  See Appendix \ref{app:eos:dis} for further discussion.

The non-ideal MHD term in \eqref{eq:ind} is the sum of the Ohmic resistivity (OR), the Hall effect (HE), and ambipolar diffusion (AD) terms, which are given by
\begin{flalign}
\left.\frac{\text{d} \bm{B}}{\text{d} t}\right|_\text{OR} &= -\bm{\nabla} \times \left[  \eta_\text{OR}      \left(\bm{\nabla}\times\bm{B}\right)\right],                                                                     &\label{eq:ohm} \\
\left.\frac{\text{d} \bm{B}}{\text{d} t}\right|_\text{HE} &= -\bm{\nabla} \times \left[  \eta_\text{HE}       \left(\bm{\nabla}\times\bm{B}\right)\times\bm{\hat{B}}\right],                                         &\label{eq:hall} \\
\left.\frac{\text{d} \bm{B}}{\text{d} t}\right|_\text{AD} &=  \bm{\nabla} \times \left\{ \eta_\text{AD}\left[\left(\bm{\nabla}\times\bm{B}\right)\times\bm{\hat{B}}\right]\times\bm{\hat{B}}\right\}. &\label{eq:ambi}
\end{flalign}
The general form of the resistivity coefficients \citep{Wardle2007} is given by
\begin{flalign}
\eta_\text{OR} &= \frac{c^2}{4\pi\sigma_\text{O}}, &\label{eq:etaOR} \\
\eta_\text{HE} &= \frac{c^2}{4\pi\sigma_\bot}\frac{\sigma_\text{H}}{\sigma_\bot}, &\label{eq:etaHE} \\
\eta_\text{AD} &= \frac{c^2}{4\pi\sigma_\bot}\frac{\sigma_\text{P}}{\sigma_\bot} - \eta_\text{OR} = \frac{c^2}{4\pi \sigma_\text{O}}\frac{\sigma_\text{O}\sigma_\text{P} - \sigma_\bot^2}{\sigma_\bot^2}, &\label{eq:etaAD}
\end{flalign}
where $c$ is the speed of light, and $\sigma$ are the conductivities, which will be calculated in Section~\ref{ssec:num:conduct}.  The use of the magnetic unit vector, $\bm{\hat{B}}$, in \eqref{eq:hall} and \eqref{eq:ambi} is to ensure that all three coefficients have units of area per time.  As will be shown in the next two sections, the value of $\eta$ depends on the microphysics of the model, and $\eta_\text{HE}$ can be either positive or negative, whereas $\eta_\text{OR}$ and $\eta_\text{AD}$ are positive (e.g. \citealp{wardleng99}).  Moreover, since $\left(\bm{\nabla} \times \bm{B}\right) \times \bm{B}$ is perpendicular to $\bm{\nabla} \times \bm{B}$, the Hall effect is non-dissipative and breaks the degeneracy between left and right polarised Alfv{\'e}n waves (e.g. \citealp{Bai2014}).  To calculate the conductivities, the number densities and charges of all non-neutral species are required, which we calculate in the following section.

\subsection{Ionisation}
\label{ssec:num:ionise}
In ideal MHD, infinite conductivity is assumed, and the magnetic field lines are frozen into the fluid.  However, in a partially ionised plasma, diffusion of the field occurs through the relative motions of the neutral and charged particles.  We assume a partially ionised plasma containing four species: neutral gas, electrons, ions and (charged) dust grains, denoted by subscripts n, e, i and g, respectively.  The particle masses we choose are 
\begin{flalign*}
m_\text{n} =& \frac{4m_\text{p}}{2X + Y}, &\\
m_\text{i} =& 24.3 m_\text{p}, &\\
m_\text{g} =& \frac{4}{3}\pi a_\text{g}^3 \rho_\text{b}, &
\end{flalign*}
where $m_\text{p}$ is the mass of a proton, $m_\text{n} \approx 2.38m_\text{p}$ using hydrogen and helium mass fractions of $X =0.70$ and $Y =0.28$, respectively, $m_\text{i}$ is the mass of  magnesium (e.g. \citealp{AsplundEtAl2009}), and  $m_\text{g} \approx 7.51\times 10^9 m_\text{p}$ using a grain radius and grain bulk density of $a_\text{g} = 0.1 \mu$m  and $\rho_\text{b} = 3$ g~cm$^{-3}$, respectively \citep{PollackEtAl1994}.

Further, we assume the strong coupling approximation, which allows the medium to be treated using the single fluid approximation.  In this approximation, ion pressure and momentum are negligible compared to that of the neutrals, i.e. $\rho \sim \rho_\text{n}$ and $\rho_\text{i} \ll \rho$, where $\rho$, $\rho_\text{n}$ and $\rho_\text{i}$ are the total, neutral and ion mass densities, respectively.  

The electron charge is $Z_\text{e} \equiv -1$ and we assume the ion charge is $Z_\text{i} = 1$. For charge neutrality, we require
\begin{equation}
\label{eq:chargeneutrality}
n_\text{i} - n_\text{e} + Z_\text{g}n_\text{g} = 0,
\end{equation}
where we allow $Z_\text{g}$ to be a real number rather than an integer.  In general, $Z_\text{g} < 0$.

The grain number density is proportional to the total number density, $n$ \citep{keithwardle14}, according to 
\begin{equation}
\label{eq:ngrain}
n_\text{g} = \frac{m_\text{n}}{m_\text{g}}f_\text{dg} n,
\end{equation}
where $f_\text{dg} = 0.01$ is the dust-to-gas mass ratio \citep{PollackEtAl1994}.  For simplicity, we adopt a single-sized grain model.  The electron and ion number densities vary as (e.g. \citealp{UmebayashiNakano1980}; \citealp{foi11})
\begin{flalign}
\frac{{\rm d}n_\text{i}}{{\rm d}t} =& \zeta n - k_\text{ei}n_\text{i}n_\text{e} - k_\text{ig}n_\text{i}n_\text{g},  &\label{eq:dnidt}\\ 
\frac{{\rm d}n_\text{e}}{{\rm d}t} =& \zeta n - k_\text{ei}n_\text{i}n_\text{e} - k_\text{eg}n_\text{e}n_\text{g}, &\label{eq:dnedt}
\end{flalign}
where $\zeta$ is the ionisation rate and $k_{ij}$ are the charge capture rates\footnote{\citet{keithwardle14} use this form of \eqref{eq:dnidt} and \eqref{eq:dnedt} whereas \citet{foi11} use $n_\text{n}$ rather than $n$.  Given $n_\text{n} \sim n$, we will use the given form for numerical simplicity.}.  Following \citet{keithwardle14}, we assume that recombination is inefficient such that the charge capture by grains dominates (i.e. $k_\text{ei} = 0$), and that we have an approximately steady state system (i.e. $\frac{{\rm d}n_\text{i}}{{\rm d}t}  \approx \frac{{\rm d}n_\text{e}}{{\rm d}t}  \approx 0$).  This yields ion and electron number densities of
\begin{flalign}
n_\text{i} &= \frac{\zeta n}{k_\text{ig} n_\text{g}},&\\
n_\text{e} &= \frac{\zeta n}{k_\text{eg} n_\text{g}}, &
\end{flalign}
respectively.  For $Z_\text{g} < 0$, the charge capture rates for neutral grains are \citep{foi11}
\begin{flalign}
k_\text{ig}  =& \pi a_\text{g}^2 \sqrt{\frac{8k_\text{B}T}{\pi m_\text{i}}}\left(1-\frac{e^2 Z_\text{g}}{a_\text{g}k_\text{B}T}\right), &\label{eq:kig}\\
k_\text{eg} =& \pi a_\text{g}^2 \sqrt{\frac{8k_\text{B}T}{\pi m_\text{e}}}\exp\left(\frac{e^2 Z_\text{g}}{a_\text{g}k_\text{B}T}\right), &\label{eq:keg}
\end{flalign}
where $k_\text{B}$ is the Boltzmann constant, $e$ is the electron charge, and $T$ is the gas temperature.  For a given $n$ and $T$ and assuming charge neutrality (Eqn~\ref{eq:chargeneutrality}), we can construct an equation that is only dependent on $Z_\text{g}$:
\begin{flalign}
\label{eq:Zgrain}
Z_\text{g} & = \frac{\zeta n}{n_\text{g}^2}\left[\frac{1}{k_\text{ig}(Z_\text{g})} - \frac{1}{k_\text{eg}(Z_\text{g})}\right], \notag &\\
                 &= \frac{\zeta}{n}\left(\frac{m_\text{g}}{f_\text{dg}m_\text{n}}\right)^2\left[\frac{1}{k_\text{ig}(Z_\text{g})} - \frac{1}{k_\text{eg}(Z_\text{g})}\right].&
\end{flalign}
The grain charge can then be calculated by solving \eqref{eq:Zgrain} using the Newton-Raphson method.  Finally, the neutral number density is given by 
\begin{equation}
n_\text{n} = \frac{1}{m_\text{n}}\left[ \rho - \left(n_\text{i}m_\text{i} + n_\text{e}m_\text{e} \right) \right].
\end{equation}

The grain charge, ion and electron number densities are almost directly proportional to the ionisation rate, $\zeta$.  However, the values of $\zeta$ span a range of several orders of magnitude depending on the ionising source.  If the ionisation is from the decaying radionuclides from $^{26}$Al, then $\zeta = 7.6 \times 10^{-19}$ s$^{-1}$ \citep{UmebayashiNakano2009}. Canonically, if the ionisation is from cosmic rays or X-rays, then the rate is $\zeta \lesssim 10^{-17}$ and $\lesssim 10^{-18}$ s$^{-1}$, respectively, and decreases with the depth the rays penetrate into the cloud (c.f. \citealp{keithwardle14}).  However, recent studies have shown that, depending on environment, the rate can be even larger, $\sim$$10^{-16}$~s$^{-1}$ \citep{MoralesortizEtAl2014}. For our study, a fiducial value of \zetaos \ will be used, but the effect of decreasing the rate to \dexoe \ will also be studied; the former value was used in the study by \citet{wardleng99}.

\begin{figure}
\begin{center}
\includegraphics[width=0.75\columnwidth]{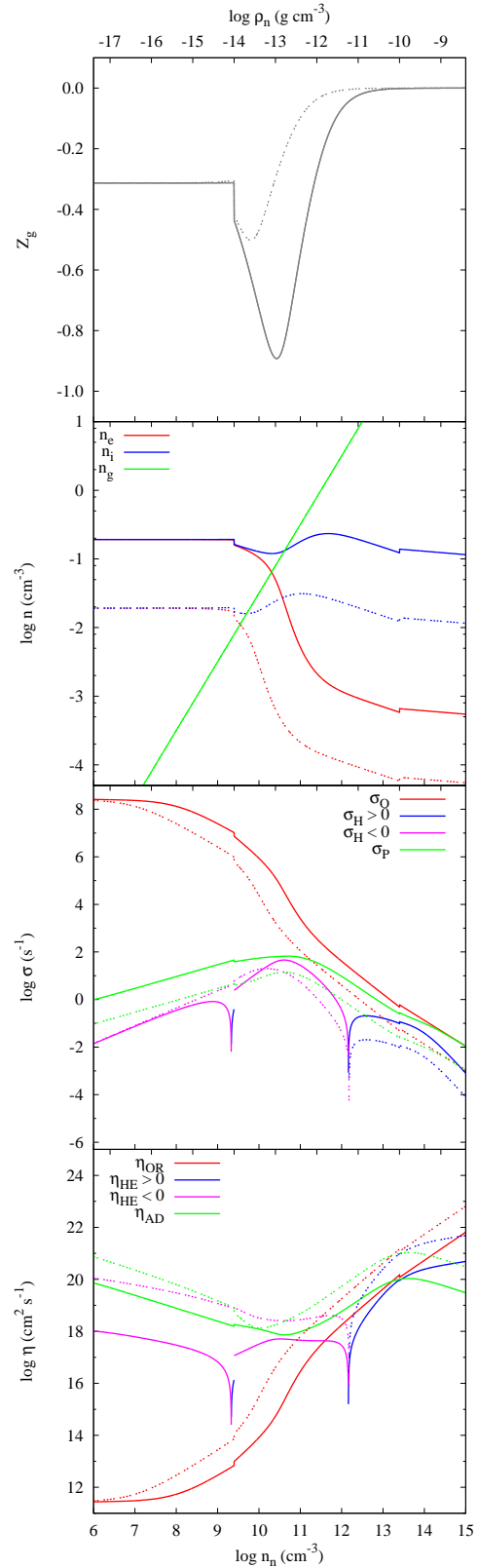}
\caption{\emph{Top to bottom}: Grain charge, charged species number densities, conductivities and resistivity coefficients, using  $\zeta = 10^{-17}$ (solid lines) and \dexoe \ (dashed lines).  The top ticks on each panel correspond to mass density (top scale), and the bottom ticks correspond to number density (bottom scale).  The discontinuities at $\rho_\text{n} = 10^{-14}$ and $10^{-10}$~g~cm$^{-3}$ correspond to the discontinuities in temperature caused by the assumed equation of state.}
\label{fig:AllBaro}
\end{center}
\end{figure}

The top two panels of Fig. \ref{fig:AllBaro} show the grain charge and species number densities, respectively.   In both panels, the solid (dashed) lines are for values using $\zeta = 10^{-17} \ (10^{-18})$~s$^{-1}$.  The discontinuities at $\rho_\text{n} = 10^{-14}$ and $10^{-10}$~g~cm$^{-3}$ correspond to the discontinuities in temperature caused by the assumed equation of state.

For all densities, $ -1 < Z_\text{g} < 0$, with $Z_\text{g} \rightarrow 0$ for increasing $n_\text{n}$.  The value of $\zeta$ is important at moderate number densities, $2\times~10^9 \lesssim~n_\text{n}/\text{cm}^{-3}~\lesssim~10^{13}$, where the absolute difference between grain charges is the highest.  At high densities, the relative difference in grain charge is the highest (i.e. being the same as the relative difference between the $\zeta$'s), but since these values are near zero, the exact value of $\zeta$ is not important. 

In our calculations, the grain number density is directly proportional to the total number density and is independent of the ionisation rate.  For $n_\text{n} \lesssim 10^{10}$~cm$^{-3}$, $n_\text{g}$ is similar to the values presented in \citet{umebayashinakano90}, but above this threshold, $n_\text{g}$ levels off for \citet{umebayashinakano90} who calculate the number density using a full treatment of the reaction rates of several different molecules.  Ion and electron number densities follow a similar trend to that presented in the top panel of fig.~1 of \citet{wardleng99}.  Our values differ quantitatively to their results due to the differences in our parameters, the choice of equation of state, and that \citet{wardleng99} used the grain densities from \citet{umebayashinakano90} rather than $n_\text{g} \propto n$.  We briefly analyse the isothermal equation of state in Appendix \ref{app:eos}, and the barotropic equation of state with $\rho_\text{c} = 10^{-13}$ g cm$^{-3}$ in Appendix~\ref{app:Beos}.  In general, the ion and electron number densities are proportional to the ionisation rate, with an exception at moderate number densities; this is the same range over which the value of $\zeta$ has the largest absolute effect on the grain charge.

\subsection{Conductivities}
\label{ssec:num:conduct}
For a charged species, i.e. $j \in \left\{\text{e},\text{i},\text{g}\right\}$, the relative magnitude between the magnetic forces and neutral drag describe the behaviour of the species.   This relation is given by the Hall parameter, whose general form is
\begin{equation}
\beta_j = \frac{|Z_j|eB}{m_j c}\frac{1}{\nu_{j\text{n}}}, \label{eq:beta}
\end{equation}
where $Z_j$ and $m_j$ are the charge\footnote{Note that some authors use a $\beta$ that includes the sign of $Z$; their conductivities, $\sigma$, are then modified accordingly.} and mass of species $j$, respectively, $B$ is the magnitude of the magnetic field and $\nu_{j\text{n}}$ is the plasma-neutral collision frequency; the Hall parameter also represents the ratio between the gyrofrequency and the neutral collision frequency (e.g. \citealp{Wardle2007}).  We have slightly modified the Hall parameter such that 
\begin{subequations}
\label{eq:betamod}
\begin{flalign}
\beta_\text{e} &= \frac{|Z_\text{e}|eB}{m_\text{e} c}\frac{1}{\nu_\text{en}+\nu_\text{ei}}, &\label{eq:betae} \\
\beta_\text{i} &= \frac{|Z_\text{i}|eB}{m_\text{i} c}\frac{1}{\nu_\text{in}+\nu_\text{ie}}. &\label{eq:betai}
\end{flalign}
\end{subequations}
With these modifications, we can recover $\eta_\text{OR}$ from \citet{pandeywardle08} and \citet{keithwardle14} under the assumption $\beta_\text{i} \ll \beta_\text{e}$.  Appendix \ref{app:beta} examines the effect of modifying the Hall parameter. 

The Hall parameter can be used to characterise different regimes where different effects are dominant.  Given the typical value of $\beta_\text{e}/\beta_\text{i} \sim 1000$, \citet{Wardle2007} defines three regimes: 
\begin{flalign*}
\beta_\text{i} \ll \beta_\text{e} \ll 1           & \text{: Ohmic resistivity}, &\\
\beta_\text{i} \ll 1                    \ll \beta_\text{e} & \text{: Hall effect}, &\\
                  1 \ll \beta_\text{i}  \ll \beta_\text{e} & \text{: Ambipolar diffusion}.  &
\end{flalign*}

The collisional frequencies, $\nu$, are empirically calculated rates.  The electron-ion rate is given by \citep{pandeywardle08}
\begin{equation}
\nu_\text{ei} = 51 \ \text{s}^{-1} \left(\frac{n_\text{e}}{\text{cm}^{-3}}\right)\left(\frac{T}{\text{K}}\right)^{-3/2}.
\end{equation}
The ion-electron rate is given by $\nu_\text{ie} = \frac{\rho_\text{e}}{\rho_\text{i}}\nu_\text{ei}$.  The plasma-neutral collisional frequency is given by
\begin{equation}
\nu_{j\text{n}} = \frac{\left< \sigma v\right>_{j\text{n}}}{m_\text{n}+m_j}\rho_\text{n},
\end{equation}
where $\left< \sigma v\right>_{j\text{n}}$ is the rate coefficient for the momentum transfer by the collision of particle of type $j$ with the neutrals.  For electron-neutral collisions, it is assumed that the neutrals are comprised of hydrogen and helium, such that the rate coefficient is 
\begin{equation}
\left< \sigma v\right>_\text{en} = X \left< \sigma v\right>_{\text{e-H}_2} + Y \left< \sigma v\right>_\text{e-He}.
\end{equation}
Following \citet{pintogalli08}, we use
\begin{flalign*}
\left< \sigma v\right>_{\text{e-H}_2} =& 3.16\times 10^{-11} \ \text{cm}^3 \ \text{s}^{-1} \left(\frac{v_\text{rms}}{\text{km} \ \text{s}^{-1}}\right)^{1.3}, &\\
\left< \sigma v\right>_\text{e-He}     =& 7.08\times 10^{-11} \ \text{cm}^3 \ \text{s}^{-1} \left(\frac{ v_\text{rms}}{\text{km} \ \text{s}^{-1}}\right), &
\end{flalign*}
with
\begin{equation}
v_\text{rms}  = \sqrt{v_\text{d}^2 + \frac{8k_\text{B}T}{\pi \mu_\text{en}} },
\end{equation}
where $v_\text{d}$ is the drift velocity between the electron and the neutral, and $\mu_\text{en}$ is the reduced mass of the electron-neutral pair; we assume $v_\text{d} = 0$ and $\mu_\text{en} \approx m_\text{e}$.
The ion-neutral rate is \citep{pintogalli08}
\begin{flalign}
\left< \sigma v\right>_\text{in} =& 2.81\times 10^{-9} \ \text{cm}^3 \ \text{s}^{-1} \ Z_\text{i}^{1/2}  &\notag \\
 \times& \left[  X\left(\frac{p_{\text{H}_2}}{\text{\AA}^3}\right)^{1/2}\left(\frac{\mu_{\text{i-H}_2}}{m_\text{p}}\right)^{-1/2}  \right. &\notag \\
+ & \left. Y \left(\frac{p_\text{He}}{\text{\AA}^3}\right)^{1/2}\left(\frac{\mu_\text{i-He}}{m_\text{p}}\right)^{-1/2} \right],&
\end{flalign}
where the values of polarisability are $p_{\text{H}_2} = 0.804 \text{\AA}^3$ and $p_\text{He} =~0.207~\text{\AA}^3$ \citep{Osterbrock1961}.

For grain-neutral collisions, the rate coefficient is given by (\citealp{wardleng99}; \citealp{pintogalli08})
\begin{equation}
\left< \sigma v\right>_\text{gn} = \pi a_\text{g}^2\delta_\text{gn}\sqrt{\frac{128k_\text{B}T}{9\pi m_\text{n}}},
\end{equation}
where $\delta_\text{gn}$ is the Epstein coefficient.  From experiments with micron-sized melamine-formaldehyde spheres, $\delta_\text{gn} \approx 1.3$ \citep{LiuEtAl2003}.
 
The Ohmic, Hall and Pedersen conductivities can now be calculated (e.g. \citealp{wardleng99}; \citealp{Wardle2007}) viz.,
\begin{flalign}
\sigma_\text{O} =& \frac{ec}{B}\sum_j n_j|Z_j|\beta_j, &\\
\sigma_\text{H} =& \frac{ec}{B}\sum_j \frac{n_j Z_j}{1+ \beta_j^2}, &\\
\sigma_\text{P} =& \frac{ec}{B}\sum_j \frac{n_j |Z_j| \beta_j}{1+ \beta_j^2}.&
\end{flalign}
We explicitly note that $\sigma_\text{O}$ and $\sigma_\text{P}$ are positive, whereas $\sigma_\text{H}$ can be positive or negative.  The total conductivity perpendicular to the magnetic field is 
\begin{equation}
\sigma_\bot = \sqrt{\sigma_\text{H}^2 + \sigma_\text{P}^2}.
\end{equation}

The third panel of Fig.~\ref{fig:AllBaro} shows the Ohmic, Hall and Pedersen conductivities for $\zeta = 10^{-17}$  (solid lines) and \dexoe \ (dashed lines).  When a magnetic field is required in the calculation, for the purpose of illustration, we use the relation used in \citet{wardleng99}, which follows the standard $n_\text{n}^{1/2}$ relation for $n_\text{n} < 10^6$~cm$^{-3}$ \citep{MyersGoodman1988} and a weaker dependence at higher densities:
\begin{equation}
\label{eq:Btest}
\left(\frac{B}{\text{mG}}\right) = \left\{ \begin{array}{l l} (n_\text{n}/10^6 \text{cm}^{-3})^{1/2}; 	    &  n_\text{n} < 10^6 \text{cm}^{-3} \\
                                                                                     (n_\text{n}/10^6 \text{cm}^{-3})^{1/4}; 	    &  \text{else} \\
\end{array}\right..
\end{equation}

The Hall conductivity is more sensitive to the value of $\zeta$ than the other two conductivities.  For both values of $\zeta$, there is a discontinuity at $n_\text{n} \approx 1.5\times 10^{12}$~cm$^{-3}$ across which the value of $\sigma_\text{H}$ switches from negative to positive.  For \zetaos, there is a small range, $2.1 \lesssim n_\text{n}/(10^9 \text{cm}^{-3}) \lesssim 2.5$, where the Hall conductivity is again positive;  at the lower end of this range, the value of $\sigma_\text{H}$ naturally switches from negative to positive, while the changeover at the upper end corresponds to the discontinuity in temperature caused by the assumed equation of state.  The behaviour of the Ohmic conductivity is similar to the behaviour of the grain charge: approximately independent of $\zeta$ at low densities and proportional to it at high densities.  The Pedersen conductivity is proportional to $\zeta$, except at medium number densities.  Our values of the conductivities differ from those in the bottom panel of their fig.~2 of  \citet{wardleng99}, however, different assumptions and values were used here; see Appendix \ref{app:eos} for further discussion.

The bottom panel of Fig.~\ref{fig:AllBaro} shows the value of the coefficients, $\eta$, for Ohmic resistivity, the Hall effect and ambipolar diffusion for both values of $\zeta$; the magnetic field is again given by \eqref{eq:Btest}.  It is clear from this figure that each non-ideal MHD term is dominant at various densities, with ambipolar diffusion being the most important at low densities and Ohmic resistivity being the most important at high densities.  The Hall effect has a small range of densities where it is dominant for \zetaoe, but it is never the dominant effect for \zetaos.  We caution that this statement is true given \eqref{eq:Btest}, but may not be true in general where $\bm{B}$ evolves differently to this (e.g. in our simulations).  Moreover, at medium densities, two or all three  effects can be simultaneously important (e.g. at $n_\text{n} \approx 2.5\times 10^{13}$~cm$^{-3}$ where all three coefficients are similar).  Given the range of densities in a typical star formation simulation at any given time, it is possible for different terms to dominate in different spatial regions.  

\subsection{Smoothed Particle Magnetohydrodynamics}
\label{ssec:num:spmhd}
To perform our simulations, we use the 3D smoothed particle magnetohydrodynamics (SPMHD) code {\sc Phantom} with the inclusion of self-gravity.  The ideal magnetohydrodynamic equations \eqref{eq:cty}--\eqref{eq:ind} are discretised into SPMHD (see review by \citealt{price12}) as 
\begin{flalign}
\rho_a  &= \sum_b m_b W_{ab}(h_a); \hspace{1em} h_{a} = h_\text{fac} \left(\frac{m_{a}}{\rho_{a}}\right)^{1/3}, \label{eq:sphcty} &\\ 
\frac{\text{d} v^i_a}{\text{d} t} &= \sum_b m_b \left[\frac{S^{ij}_a}{\Omega_a \rho_a^2} \nabla^j_a W_{ab}(h_a) + \frac{S^{ij}_b}{\Omega_b \rho_b^2}\nabla^j_a W_{ab}(h_b)\right] \notag &\\
& - f B^i_a \sum_b m_b \left[\frac{B^j_a}{\Omega_a \rho_a^2} \nabla^j_a W_{ab}(h_a) + \frac{B^j_b}{\Omega_b \rho_b^2}\nabla^j_a W_{ab}(h_b)\right] \nonumber &\\
& - \nabla \Phi_{a} + \left.\frac{\text{d} v^i_a}{\text{d} t}\right|_\text{artificial} , \label{eq:spmhdmom} &\\
\frac{\text{d} B^i_a}{\text{d} t} &= -\frac{1}{\Omega_a \rho_a} \sum_b m_b  \left[  v^i_{ab} B^j_a \nabla^j_a W_{ab}\left(h_a\right) \right.  \notag &\\
&-  \left. B^i_a v^j_{ab} \nabla^j_a W_{ab}\left(h_a\right) \right]+ \left.\frac{\text{d} B^i_a}{\text{d} t}\right|_\text{non-ideal} + \left.\frac{\text{d} B^i_a}{\text{d} t}\right|_\text{artificial},  \label{eq:spmhdind} &\\
\nabla^{2} \Phi_{a} & = 4\pi G \rho_{a}, \label{eq:poissonsph}&
\end{flalign}
where we sum over all particles $b$ within the kernel radius, $W_{ab}$ is the smoothing kernel, $\bm{v}_{ab} = \bm{v}_a - \bm{v}_b$, $\Omega_a$ is a dimensionless correction term to account for a spatially variable smoothing length $h_a$ \citep{monaghan02,springelhernquist02}, the stress tensor is given by
\begin{equation}
\label{Istress}
S^{ij} \equiv -\left(P + \frac{1}{2}B^2\right)\delta^{ij} + B^iB^j,
\end{equation}
and $\left.\frac{\text{d} v^i_a}{\text{d} t}\right|_\text{artificial}$ is the artificial viscosity, as described in \cite{pricefederrath10}.

Numerically, $\bm{\nabla}\cdot\bm{B}$ is not exactly zero.  However, this term is inherently contained in the conservative form of the momentum equation (i.e. the first line of Eqn.~\ref{eq:spmhdmom}).  When $\frac{1}{2}\bm{B}^2 > P$, the inclusion of this term can trigger the  tensile instability, which causes particles to unphysically clump together.  To correct for this, a simple approach is to subtract the source term (i.e. the second line of Eqn.~\ref{eq:spmhdmom} using $f=1$; \citealt{bot01}).  Since subtracting the term violates energy and momentum conservation (but only insofar as the divergence term in \eqref{eq:spmhdmom} is non-zero; e.g. \citealt{price12}; \citealt{triccoprice12}), \citet{bot04} introduced a variable $f$ such that $0 < f < \frac{1}{2}$.  However, \citealt{triccoprice12} showed that numerical artefacts can be produced for $ f < 1$, thus suggested $f=1$ everywhere.  Since the tensile instability is only triggered for $\frac{1}{2}\bm{B}^2 > P$, we use 
\begin{equation}
\label{eq:fdivB}
f = \left\{ \begin{array}{l l} 1 ; 	       &  \beta \le 1 , \\
                                         2 - \beta;  &  1 < \beta \le 2 \\
                                         0;             &  \beta > 2,
\end{array}\right.
\end{equation}
where $\beta = \frac{2P}{\bm{B}^2}$ is the plasma beta; $f$ is calculated for each particle, $a$, using only the properties of particle $a$.  This allows the source term to be removed where it is problematic, but maintains energy and momentum conservation elsewhere.  The function $1 < \beta \le 2$ allows a smooth decrease between the two extremes, and to avoid sharp jumps when $\beta \sim 1$.  To avoid confusion with the Hall parameters, $\beta_\text{e}$, $\beta_\text{i}$ and $\beta_\text{g}$, we will always use $\beta$ with no subscript to refer to the plasma beta.

We adopt the usual cubic spline kernel, with $h_\text{fac} = 1.2$ in \eqref{eq:sphcty} specifying the ratio of the smoothing length to the particle spacing, equivalent to $\sim$58 neighbours \citep{price12}.  Finally, the magnetic field has been normalised such that $v_\text{A}\equiv B/\sqrt{\rho}$ (see \citealt{pricemonaghan04}).  We solve (\ref{eq:poissonsph}) following \citet{pricemonaghan07} at short range, with a \emph{k}-d tree algorithm similar to that described in \citet{gaftonrosswog11} used to compute the long range gravitational interaction in an efficient manner.

To calculate the non-ideal MHD terms in \eqref{eq:spmhdind}, we follow the procedure described in \citet{WPA2014}.  First, the current density, $\bm{J} \equiv \bm{\nabla}\times\bm{B}$, is calculated using the difference operator (c.f. \citealt{price10,price12}):
\begin{equation}
\label{eq:J}
\bm{J}_a = \frac{1}{\Omega_a \rho_a} \sum_b m_b \left(\bm{B}_a - \bm{B}_b\right) \times \bm{\nabla}_a W_{ab}(h_a).
\end{equation}
The general non-ideal MHD term is then calculated using the conjugate (i.e. symmetric) operator, 
\begin{flalign}
\label{eq:D}
\left.\frac{\text{d} \bm{B}_a}{\text{d} t}\right|_\text{non-ideal} = -\rho_a \sum_b m_b  & \left[ \frac{\bm{D}_a}{\Omega_a \rho^2_a}\times \bm{\nabla}_aW_{ab}(h_a)\right. &\\ \notag
+ & \left. \frac{\bm{D}_b}{\Omega_b \rho^2_b}\times \bm{\nabla}_aW_{ab}(h_b) \right],&
\end{flalign}
where $\bm{D}_a$ is defined for each non-ideal MHD term as
\begin{flalign}
\bm{D}_a^\text{OR} &= -\eta_\text{OR} \bm{J}_a, &\\
\bm{D}_a^\text{HE} &= -\eta_\text{HE} \bm{J}_a \times \bm{\hat{B}}_a, \label{eq:etaHallSph} &\\
\bm{D}_a^\text{AD} &=  \eta_\text{AD} \left(\bm{J}_a \times \bm{\hat{B}}_a \right)\times \bm{\hat{B}}_a.&
\end{flalign}
Once $\bm{J}_a$ is calculated, $\bm{D}_a$ can be calculated without knowledge of any of particle $a$'s neighbours.   Although this algorithm is the same as in \citet{WPA2014}, here we self-consistently calculate the resistivity coefficients as described in the previous sections rather than defining them as constants for the entire simulation.  Therefore, no \emph{a priori} knowledge is required of which term is dominant.  This algorithm has been thoroughly tested for ambipolar diffusion with constant resistivity in \citet{WPA2014}; given the non-diffusive nature of the Hall effect, we present the results from two tests in Appendix \ref{app:num:hall}.

We compute the artificial resistivity term \citep{pricemonaghan04,pricemonaghan05} in \eqref{eq:spmhdind} using
\begin{flalign}
\left.\frac{\text{d} \bm{B}_a}{\text{d} t}\right|_\text{artificial}  = \frac{\rho_{a}}{2} \sum_{b} m_{b} \left(\bm{B}_a  - \bm{B}_b\right) & \left[ \frac{v^{B}_{{\rm sig}, a}}{\rho_{a}^{2}} \frac{ \hat{\bm{r}}_{ab}\cdot\bm{\nabla}_aW_{ab}(h_{a})}{\Omega_{a}}\right.  &\notag \\
& \left. + \frac{v^{B}_{{\rm sig}, b}}{\rho_{b}^{2}} \frac{ \hat{\bm{r}}_{ab}\cdot\bm{\nabla}_aW_{ab}(h_{b})}{\Omega_{b}} \right] &
\label{eq:artificialB}
\end{flalign}
where $v^\text{B}_\text{sig} = \sqrt{c_{s}^{2} + v_{\text{A}}^{2}}$ is the signal velocity, set to the fast magnetosonic speed.  Each particle has its own $\alpha_\text{B}$, set using the switch described in \citet{TriccoPrice2013}:
\begin{equation}
\alpha_\text{B} = \min\left(\frac{h \vert \nabla \bm{B} \vert}{\vert\bm{B}\vert},1.0\right),
\end{equation}
where the magnitude of the gradient matrix is computed from the 2-norm \citep{TriccoPrice2013}. This ensures that resistivity is only strong where there are strong gradients in the magnetic field.  The Ohmic diffusion resulting from the artificial resistivity term for a given particle $a$ is given by
\begin{equation}
\label{eq:artificialresis}
\eta_\text{art}^{a} \approx \frac{1}{2}\alpha^{a}_\text{B} v^\text{B}_{\text{sig},a} h_{a}.
\end{equation}
We compute this at each step in the calculation and compare it to the physical diffusion coefficients to ensure that physical resistivity dominates.

Finally, we control the divergence of the magnetic field using the constrained hyperbolic divergence cleaning scheme described in \citet{triccoprice12}. Importantly, this treatment of the magnetic field evolution is completely general as in \citet{PriceTriccoBate2012} and \citet{BateTriccoPrice2014}, unlike the Euler Potentials method used by \citet{pricebate07}.

\subsection{Timestepping}
\label{ssec:num:timestepping}
For non-ideal MHD, the timestep for particle $a$ is constrained by
\begin{equation}
\label{eq:dtni}
\text{d}t_{a} < C_\text{non-ideal}\frac{h_a^2}{|\eta_a|},
\end{equation}
where $\eta_a = \max\left(\eta_{\text{OR},a},\eta_{\text{HE},a},\eta_{\text{AD},a}\right)$ and $C_\text{non-ideal} < 1$ is a positive coefficient analogous to the Courant number.  Wave calculations involving the whistler mode suggest $C_\text{non-ideal} = \frac{1}{2\pi}$ as the optimal value.  This agrees with the stability test in \citet{Bai2014}, who optimally sets $C_\text{non-ideal} = \frac{1}{6}$.  We find that values much larger than this yield unstable results.  

In some cases, this timestep can be considerably smaller than the normal Courant timestep, thus can drastically slow down the simulation.  For the C-shock test presented in \citet{MNKW95} and \citet{WPA2014}, the timestep was 30--40 times shorter for the case that included ambipolar diffusion compared to the case using ideal MHD.  

For the diffusive terms that are parabolic in nature (i.e. Ohmic resistivity and ambipolar diffusion; the Hall effect is hyperbolic), we can relax the stringent condition imposed by (48) by implementing super-timestepping \citep{AlexiadesEtAl96}.  Super-timestepping requires stability at the end of a cycle of $N$ steps rather than at the end of every step.  To implement super-timestepping, we first choose timesteps such that 
\begin{flalign}
\text{d}t =& \min\left(\text{d}t_\text{Courant}, \text{d}t_\text{HE}\right), \label{newdtC}&\\
\text{d}t'_\text{diff} =& \min\left(\text{d}t_\text{OR}, \text{d}t_\text{AD}\right),&
\end{flalign}
and dictate that the simulation must be stable when it has progressed time $\text{d}t$.  If $\text{d}t > \text{d}t'_\text{diff}$, then the number of timesteps required to progress d$t$ is $N = \text{int}\left(\sqrt{\frac{\text{d}t}{k \text{d}t'_\text{diff}}}\right) + 1$ \citep{ckw09}, where $k \le 1$ is a positive scalar; we set $k = 0.9$.  We then reset the diffusive timestep to 
\begin{equation}
\label{eqn:dtdiff}
\text{d}t_\text{diff} = \frac{N^2}{k} \text{d}t.
\end{equation}
The individual sub-steps are then given by 
\begin{equation}
\text{d}\tau_j = \text{d}t_\text{diff}\left[\left(\nu-1\right)\cos\left(\frac{2j-1}{N}\frac{\pi}{2}\right) + \nu + 1\right]^{-1} 
\end{equation}
for $j = 1,...,N$, where $0 < \nu < 1$ is pre-calculated for a given $N$ using the relation in \eqref{eqn:dtdiff}.  The full timestep, $\text{d}t$, is recovered with
\begin{equation}
\label{eq:dtsts}
\text{d}t = \sum^N_{j=1} \text{d}\tau_j= \text{d}t_\text{diff}\frac{N}{2\sqrt{\nu}}\left(\frac{\left(1+\sqrt{\nu}\right)^{2N} - \left(1-\sqrt{\nu}\right)^{2N}}{\left(1+\sqrt{\nu}\right)^{2N} + \left(1-\sqrt{\nu}\right)^{2N}}\right).
\end{equation}
Given the predictor-corrector method used by {\sc Phantom}, we decrease $\text{d}t \rightarrow \text{d}t/2$ if the signal velocity is predicted to increase more than 10 per cent during a given $\text{d}\tau$.  Although decreasing $\text{d}t$ to maintain small changes in the signal velocity can counteract the benefits of super-timestepping, it does provide the required stability over $\text{d}t$ while allowing a decrease in runtime.  In Appendix \ref{app:num:sts}, we discuss super-timestepping for both test cases and our models; further tests of super-timestepping can be found in \citet{CommerconEtAl2011} and \citet{tii13}.

The Hall term is hyperbolic, thus super-timestepping cannot be applied.  Further, we have no explicit treatment of this term (i.e. the minimum timestep is given by Eqn. \ref{newdtC}), thus the Hall effect can cause a considerable slow-down in the simulations.  Under certain circumstances, $\text{d}t_\text{HE}$ can be several hundred or thousand times smaller than the Courant-limited timestep, which essentially results in the premature end of the simulation.

\section{Initial conditions}
\label{sec:ic} 
Our setup is similar to that used in \citetalias{pricebate07}.  We use a spherical cloud of radius $R=4\times~10^{16}$~cm = 0.013~pc, mass $M=1$~M$_{\astrosun}$, and mean density of $\rho_0=7.43\times~10^{-18}$~g~cm$^{-3}$.  The cloud has an initial rotational velocity of $\Omega = 1.77\times 10^{-13}$ rad s$^{-1}$, an initial sound speed of $c_\text{s,0} = 2.19\times 10^4$~cm~s$^{-1}$, and we assume a uniform magnetic field aligned (or anti-aligned) with the axis of rotation, i.e. $B_{0,x} = B_{0,y} = 0$, $B_{0,z} \ne 0$.  The free-fall time is $t_\text{ff}=2.4\times~10^4$~yr, which is the characteristic timescale for this study.

To avoid boundary conditions at the edge of the sphere, the cloud is placed in a uniform, low-density box of edge length $l = 4R = 0.052$~pc, which is in pressure equilibrium with the cloud; the density contrast between the cloud and the surrounding medium is 30:1.  This allows the cloud to be modelled self-consistently, and the large ratio ensures that the surrounding medium will not contribute significantly to the self-gravity of the cloud.  We use quasi-periodic boundary conditions at the edge of the box, in which SPH particles interact hydrodynamically `across the box', but not gravitationally.  

Sink particles \citep{bbp95} are introduced so that we can follow the collapse efficiently after the formation of the first hydrostatic core.  When the maximum gas density surpasses $\rho_\text{crit}~=~10^{-10}$~g~cm$^{-3}$, the densest gas particle is replaced with a sink particle when it and its neighbours within $r_\text{acc} =$~6.7~AU meet a given set of criteria; all the neighbours are immediately accreted onto the sink particle.  Gas which later enters this radius is checked against given criteria to determine if is accreted onto the sink particle.  Sink particles interact with the gas only via gravity and accretion.  Thus, magnetic fields in the central regions are removed and not allowed to feed back on the surrounding material.  While this is a crude approximation, it enables us to perform our study efficiently (\citetalias{pricebate07}).  However, sink particle boundaries with magnetic fields are problematic in SPMHD and a systematic study of alternative approaches would be worthwhile.

The parameters that govern the sink particle (i.e. the critical density, $\rho_\text{crit}$ and the accretion radius, $r_\text{acc}$) must be chosen carefully since these parameters will influence the results.  Objects smaller than the accretion radius are necessarily unresolved, but the density profile around the sink will also vary depending on $r_\text{acc}$ since the sink effectively adds an outflow boundary condition at this radius; this can lead to the sink influencing the gas on scales larger than $r_\text{acc}$  \citep{MachidaInutsukaMatsumoto2014}.

Given our initial conditions and our chosen equation of state, we require at least 30~000 particles to resolve the local Jeans mass throughout the calculation (\citealt{bateburkert97}; PB07).   We present simulations at two different resolutions: 445~000 particles including 302~000 in the sphere, and 1~484~000 particles including 1~004~000 in the sphere.  Thus, this resolution condition is clearly satisfied in our models.  Our primary analysis of a strong magnetic field and \zetaos \ will be performed at the higher resolution, while, for computational efficiency, the remainder of the analyses will be performed at the lower resolution.  Our included resolution studies indicate that our conclusions are independent of resolution.  All particles are set up on a regular close-packed lattice (e.g. \citealp{morris96}).  All undesirable effects initially introduced by the regularity of the lattice are transient and washed out long before the star formation occurs.  

We characterise the magnetic field in terms of the normalised parameter $\mu$, where 
\begin{equation}
\label{eq:masstofluxmu}
\mu \equiv \frac{M/\Phi_\text{B}}{\left(M/\Phi_\text{B}\right)_\text{crit}},
\end{equation}
where ${M}/{\Phi_\text{B}}$ is the mass-to-flux ratio
\begin{equation}
\label{eq:masstoflux}
\frac{M}{\Phi_\text{B}} \equiv \frac{M}{\pi R^2 B},
\end{equation}
and $({M}/{\Phi_\text{B}})_\text{crit}$ is the critical value where magnetic fields prevent gravitational collapse altogether
\begin{equation}
\label{eq:masstofluxcrit}
\left(\frac{M}{\Phi_\text{B}}\right)_\text{crit} = \frac{c_1}{3\pi}\sqrt{ \frac{5}{G} },
\end{equation}
where $M$ is the total mass contained within the cloud, $\Phi_\text{B}$ is the magnetic flux threading the surface of the (spherical) cloud at radius $R$ assuming a uniform magnetic field of strength $B$, $G$ is the gravitational constant and $c_1 \simeq 0.53$ is a parameter numerically determined by \citet{mouschoviasspitzer76}. Observations suggest $\mu \sim 2-10$ in molecular cloud cores \citep[e.g.][]{crutcher99,bourkeetal01,HeilesCrutcher2005}; this value could be even smaller once projection effects are taken into account \citep{LiFangHenningKainulainen2013}.

In this study, we test the effect of the different non-ideal MHD terms, the initial mass-to-flux-ratio, $\mu_0$, the direction of the magnetic field, the cosmic ray ionisation rate, and resolution. Table \ref{table:ic:fiducial} lists these parameters, along with values to be tested.  Unless otherwise stated, the cosmic ionisation rate will be set at \zetaos.  We define models listed as `non-ideal' as containing all three non-ideal MHD terms. 
\begin{table}
\begin{center}
\begin{tabular}{c c}
\hline
Parameter                                        & All values to be tested                                \\
\hline
Non-ideal MHD component             & ideal MHD, non-ideal MHD                         \\
                                                        &Ohmic-only, Hall-only, ambipolar-only          \\
Initial mass-to-flux ratio, $\mu_0$   & 5.0, 7.5, 10.0                                               \\
Direction of the magnetic field         & \bopos, \boneg                                            \\
Cosmic ray ionisation rate, $\zeta$ & $10^{-17}$~s$^{-1}$, $10^{-18}$~s$^{-1}$ \\
Resolution (particles in sphere)       & $\sim$3$\times10^5$, $\sim$$10^6$         \\
\hline
\end{tabular}
\caption{A list of the parameters varied in this study.  The second column lists values of the parameter that we tested.  For our initial conditions, $\mu_0 = 5$, 7.5 and 10 correspond to magnetic field strengths of $B = \{1.63, 1.09, 0.817 \} \times 10^{-4}$ G, respectively.   We define models listed as `non-ideal' as containing all three non-ideal MHD terms.  Unless otherwise stated, the cosmic ionisation rate will be set at \zetaos.}
\label{table:ic:fiducial} 
\end{center}
\end{table}

\section{Results}
\label{sec:results}
\subsection{Ideal MHD}
\label{sec:results_Ideal}
\begin{figure*}
\begin{center}
\includegraphics[width=\textwidth]{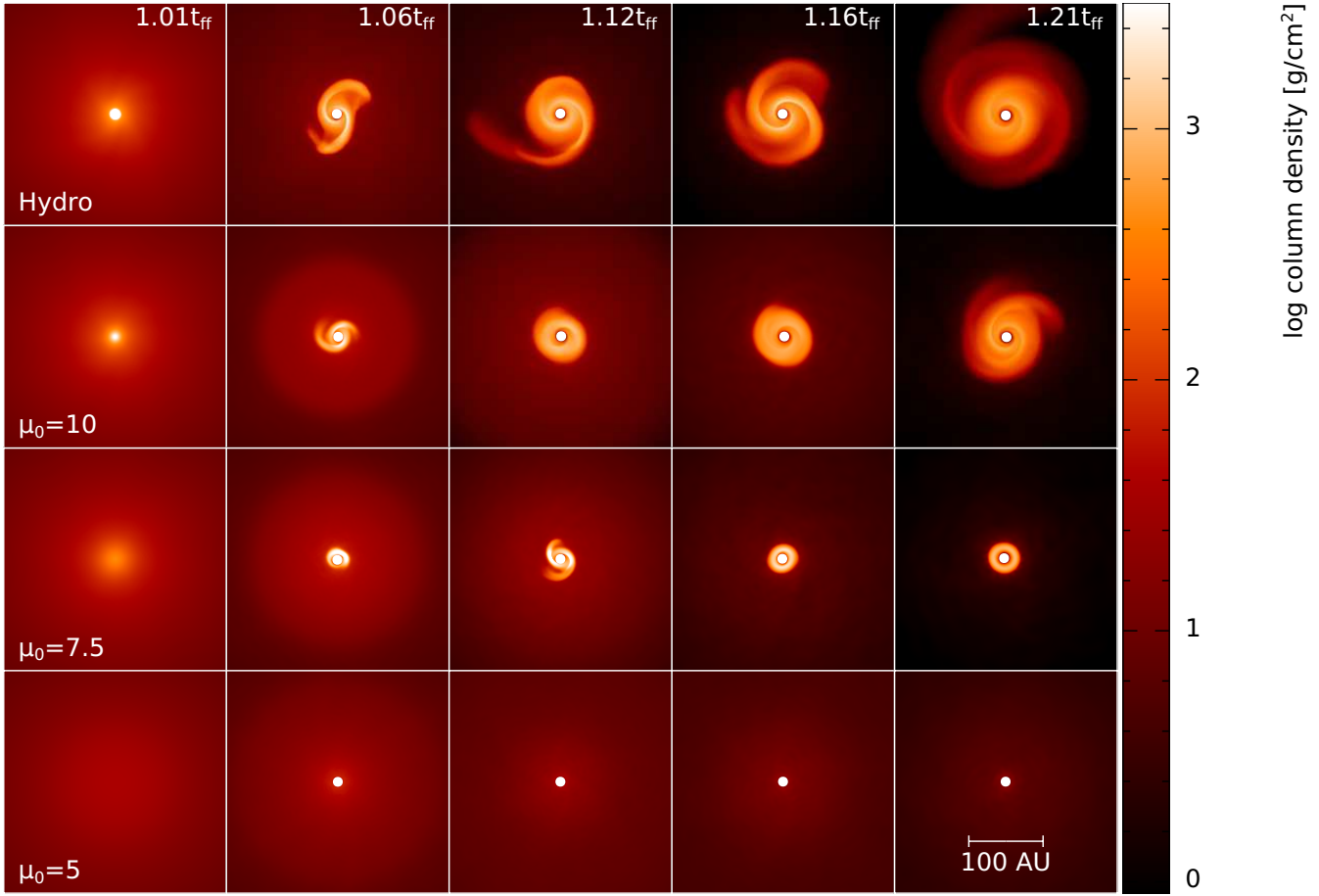}
\caption{Face-on gas column density using ideal MHD.  The initial rotation is counterclockwise and the initial magnetic field is directed out of the page (i.e. \bopos).  Each model is initialised with $\sim$3$\times 10^5$ particles within the sphere.  From left to right, the columns represent snapshots at a given time (in units of the free-fall time, $t_\text{ff} = 2.4\times~10^4$~yr).  The rows represent models with different initial magnetic field strengths given in terms of $\mu_0$ (i.e. the initial mass-to-flux ratio normalised to the critical mass-to-flux ratio).  The top row has no initial magnetic field and the bottom row has the strongest magnetic field (i.e. increasing magnetic field strength corresponds to a decreasing value of $\mu_0$).  The white circles represent the sink particle with the radius of the circle representing the accretion radius of the sink particle.  Each frame is (300~AU)$^2$.  The discs grow in size and mass with time.  At any given time, the models with stronger magnetic fields have smaller and less massive discs than the models with the weaker initial magnetic field.  The hydrodynamic model yields the largest and most massive disc in our entire suite of simulations.}
\label{fig:results:ideal:colxy}
\end{center}
\end{figure*}

For a baseline comparison, we present four simulations of the axisymmetric collapse of a molecular cloud core using ideal MHD.  This is similar, but smaller in scope, to the study presented in \citetalias{pricebate07}.  Fig. \ref{fig:results:ideal:colxy} shows the face-on gas column density profiles for $\mu_0 = \infty$ (i.e. no magnetic fields; labelled as `Hydro' in the figures), 10, 7.5 and 5 at five different times during the collapse; these ratios correspond to magnetic field strengths of $B = \{0, 0.817,1.09,1.63\} \times 10^{-4}$ G, respectively.  There are initially $\sim3\times 10^5$ particles in the gas sphere.

In Fig.~\ref{fig:results:ideal:colxy}, the cloud is initially rotating counterclockwise with the initial magnetic field directed out of the page (i.e. \bopos).  The hydrodynamic model forms a large, massive disc.  This is expected since there is no mechanism (i.e. magnetic fields) to transport the angular momentum out of the system.  For the magnetic models, increasing the initial magnetic field strength (i.e. decreasing $\mu_0$) retards the collapse; this is seen at $t=1.01t_\text{ff}$, where the central density is lower for stronger magnetic fields.  

To perform a quantitative comparison, we first define the disc and the star+disc system.  Gas is `in the disc' if it has density $\rho > \rho_\text{disc,min} = 10^{-13}$ g cm$^{-3}$, which is one order of magnitude above which the gas becomes adiabatic.  The mass of the star+disc system is defined as the mass of the disc plus the mass of the sink particle, which represents the first hydrostatic core.  The radius of the disc is defined as the radius in which 99 per cent mass of the star+disc system is contained \citepalias{pricebate07}.  We caution that the disc characteristics are variable with time since gas is condensing onto it as well as being fed to the sink from it.  Thus, the star+disc system yields a more robust analyse of mass since the mass that accretes onto the disc remains in either the disc or the sink particle.  Despite its temporal variability, the disc properties will be analysed since whether or not a disc forms is the focus of this study. 

\begin{figure*}
\begin{center}
\includegraphics[width=0.76\textwidth]{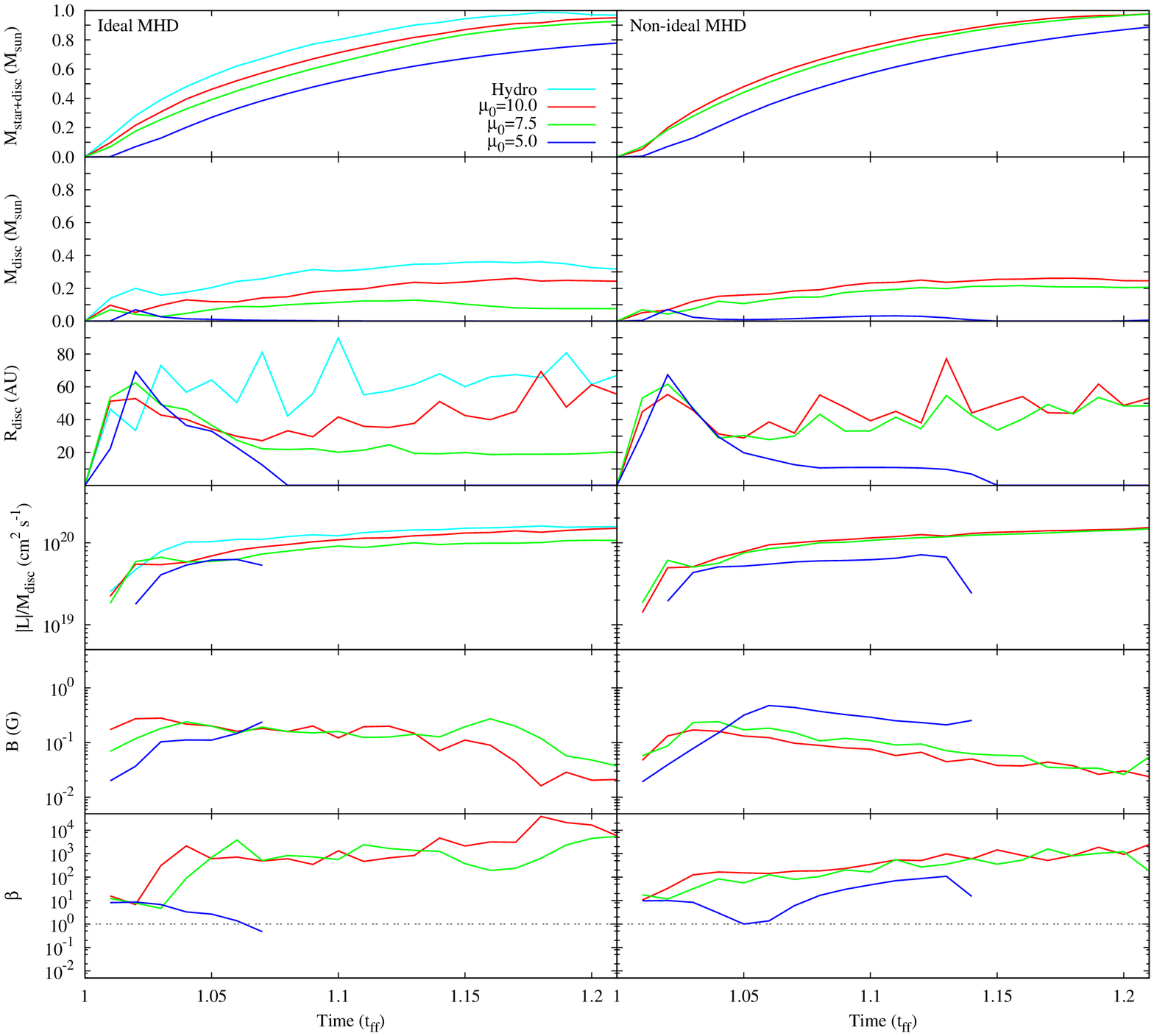}
\caption{Disc properties for the ideal (left) and non-ideal (right) MHD models with $\sim$3$\times 10^5$ particles in the sphere.  \emph{Top to bottom}: Mass of the star+disc system, the disc mass, disc radius, the specific angular momentum of the disc, the average magnetic field, and the average plasma beta (this frame includes a reference line at $\beta=1$, where gas and magnetic pressure are equal).  The disc is defined by $\rho > \rho_\text{disc,min}~=~10^{-13}$~g~cm$^{-3}$, which is one order of magnitude above which the gas becomes adiabatic.  The disc radius is defined as the radius which contains 99 per cent of the mass of the star+disc system.  The average magnetic field and plasma beta are averages over all of the particles in the disc.  The masses of the disc and star+disc system, as well as the disc radius and specific angular momentum, decrease for increasing magnetic field strength.  The hydrodynamic model yields the largest and most massive disc.  Magnetic fields counteract the gravitational collapse, thus, for both ideal and non-ideal MHD, stronger magnetic fields yield smaller and less massive discs.  For the models with $\mu_0=5$,  all traces of the disc have been erased by $t\approx1.08t_\text{ff}$ and $\approx1.15t_\text{ff}$ for the ideal and non-ideal MHD models, respectively.}
\label{fig:results:inim:discMR}
\end{center}
\end{figure*}

The left-hand column of Fig.~\ref{fig:results:inim:discMR} shows the mass of the star+disc system and important disc properties, including mass, radius, specific angular momentum, average magnetic field, and plasma beta.  
The star+disc system forms at $t~\approx1.01t_\text{ff}$.  For the hydrodynamic model, 97 per cent of the mass that was initially in the gas cloud resides in the star+disc system by $t=1.21t_\text{ff}$ and 32 per cent of that mass resides in the disc.  When magnetic fields are included, angular momentum is efficiently transported outwards, so the specific angular momentum in the disc decreases as the initial magnetic field strength is increased.  As the initial magnetic field strength is increased, the disc radius and the masses of the disc and star+disc system decrease.  For $\mu_0 = 5$, the majority of the high-density material is converted into the sink particle during its formation, and the remaining high-density material is quickly accreted; by $t\approx1.08t_\text{ff}$, all evidence of the disc has been erased.  The non-smooth evolution of the disc radii is caused by gravitational instabilities that trigger transient features in the disc, including spiral arms.  These are also pronounced since the disc characteristics are typically calculated at intervals of d$t= 0.01t_\text{ff}$.

This analysis is similar to the results in fig.~4 of \citetalias{pricebate07}.  While the same trends are observed, the quantitative results are different.  This is a result of an error in the initial conditions of \citetalias{pricebate07}, where the sound speed was not defined in the low-density background.  This resulted in a slower collapse, and less gas reaching $\rho > \rho_\text{disc,min}$ since the gas in the initial cloud was not pressure-confined.  Hence, they reported lower disc masses and radii.  We were able to reproduce their values by simulating a model where our gas cloud was not in a pressure confined medium.  

The relationship between magnetic field strength and density in these models is not as well-defined as in Sections~\ref{ssec:num:ionise} and \ref{ssec:num:conduct}, where we described the ionisation and conductivities; there, the magnetic field was defined as in \eqref{eq:Btest}.  In these models, the relationship is typically $B \propto \rho^p$, where $p\in\left[\frac{1}{2},\frac{2}{3}\right]$.  This agrees with the range presented in \citet{TsukamotoEtAl2015}.  For the three magnetic models, the average magnetic field in the disc is similar; recall, though, that the these models had initial magnetic field strengths that differed only by a factor of two.    

In summary, hydrodynamical collapses result in large, massive discs while magnetohydrodynamical collapses hinder or suppress the formation of discs, with smaller discs forming in simulations with stronger initial magnetic fields -- assuming a disc forms at all.  In agreement with (e.g.) \citet{als03}, \citetalias{pricebate07}, \citet{mellonli08}, and \citet{hennebellefromang08}, this demonstrates the magnetic braking catastrophe.

 \subsubsection{Resolution}
\begin{figure}
\begin{center}
\includegraphics[width=\columnwidth]{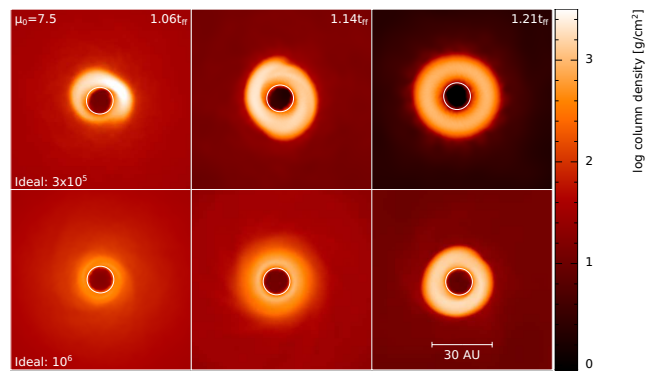}
\caption{Face-on column density as in Fig.~\ref{fig:results:ideal:colxy} but for ideal MHD at two different resolutions and zoomed in to (90 AU)$^2$; both use $\mu_0=7.5$.  The open circles represent the sink particle with the radius of the circle representing the accretion radius of the sink particle.  At both resolutions, disc masses and radii are similar.}
\label{fig:results:resolution:colxy:ideal}
\end{center}
\end{figure}

Fig.~\ref{fig:results:resolution:colxy:ideal} shows a comparison of the discs formed at resolutions of $\sim$3$\times~10^5$ particles in the collapsing sphere (top row) and $\sim$$10^6$ particles (bottom row) using  $\mu_0 = 7.5$.  This magnetic field strength was used so that disc characteristics could be compared.  The $\sim$$10^6$ particle model took $\sim$3.5 times longer to run, which is reasonable given the increase in resolution.  

The two resolutions follow the same general trend, with large discs forming.  For $1.10~\lesssim~t/t_\text{ff}~\lesssim~1.21$, the  star+disc mass, disc mass and disc radius typically differ by less than 20 per cent.  Thus, these results are relatively robust to the resolution increase presented here.

\subsection{Ideal MHD --- outflows}
\begin{figure*}
\begin{center}
\includegraphics[width=\textwidth]{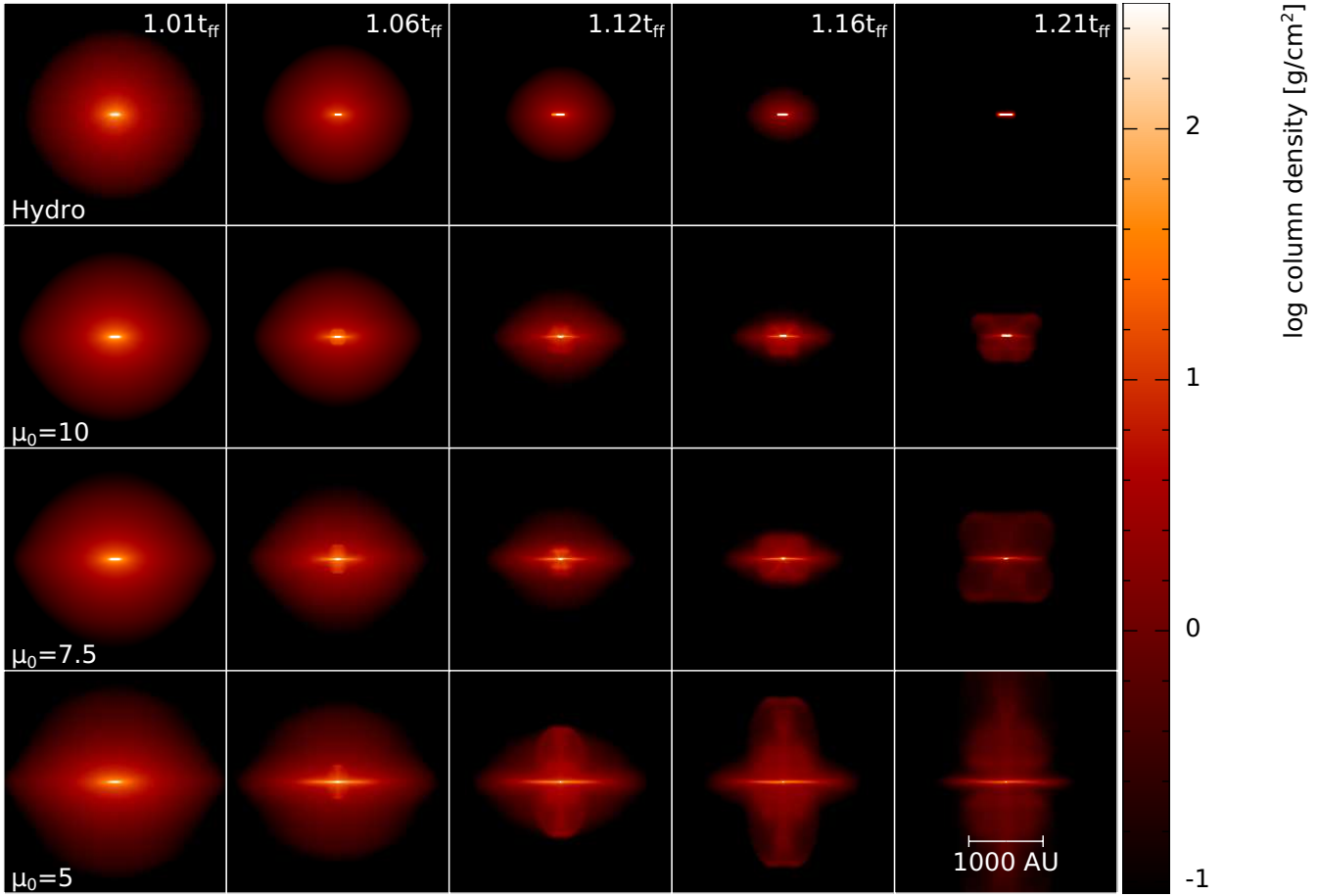}
\caption{Edge-on gas column density using ideal MHD and zoomed out to (3000~AU)$^2$ and using a density range shifted down by a factor of ten to visualise the full extent of the outflows launched shortly after the collapse ($t\approx1.02t_\text{ff}$) in the magnetic models.  The models with stronger magnetic fields have faster and more collimated outflows.}
\label{fig:results:ideal:colxz}
\end{center}
\end{figure*}

Fig.~\ref{fig:results:ideal:colxz} shows the edge-on column density for the ideal MHD calculations.  We see that the models with magnetic fields launch bipolar outflows shortly after the collapse of the core, in agreement with \citet{Tomisaka1998,Tomisaka2002}, \citet{TomisakaMachidaMatsumoto2004}, \citet{MachidaTomisakaMatsumoto2004,machidaetal06,machidaetal08}, \citet{hennebellefromang08}, \citet{commerconetal10}, \citet{burzleetal11a}, and \citet{PriceTriccoBate2012}; stronger and more collimated outflows are launched in models with stronger magnetic fields.  Some numerical asymmetries are visible at $t \gtrsim 1.12t_\text{ff}$ in the $\mu_0 = 10$ and 7.5 models, which are caused by a lack of momentum conservation in the second term of \eqref{eq:spmhdmom}.   Angular momentum, however, is conserved within two per cent until sink formation.  We have verified that the loss in conservation is independent of whether or not a sink particle is inserted.  We also find that in ideal MHD, in agreement with previous authors, the outflow properties are resolution dependent.  Hence, while a qualitative study of the outflows is useful, we urge caution regarding any quantitative properties. 


\subsection{Non-ideal MHD}
\label{sec:results_NI}

\begin{figure*}
\begin{center}
\includegraphics[width=\textwidth]{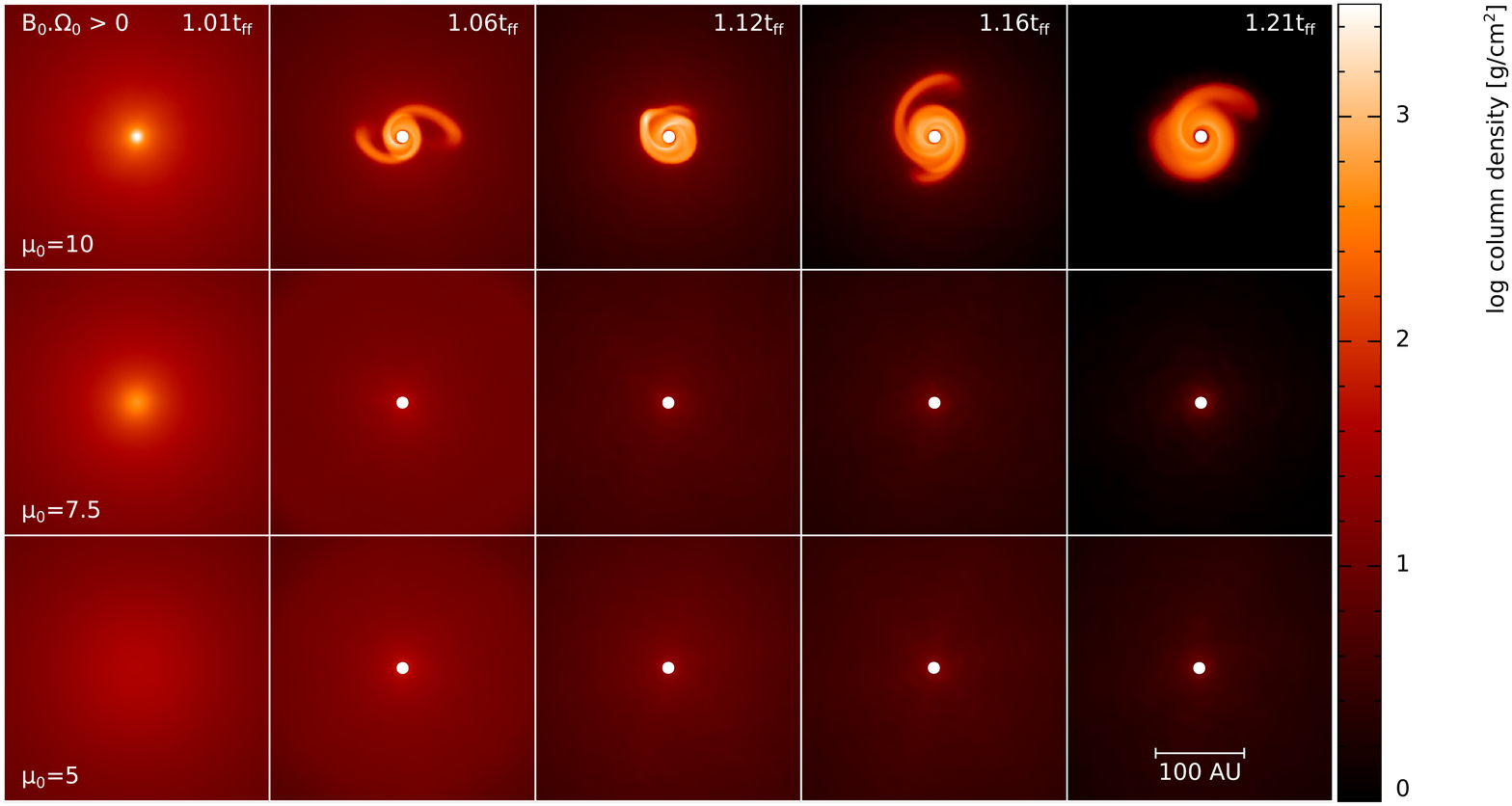}
\includegraphics[width=\textwidth]{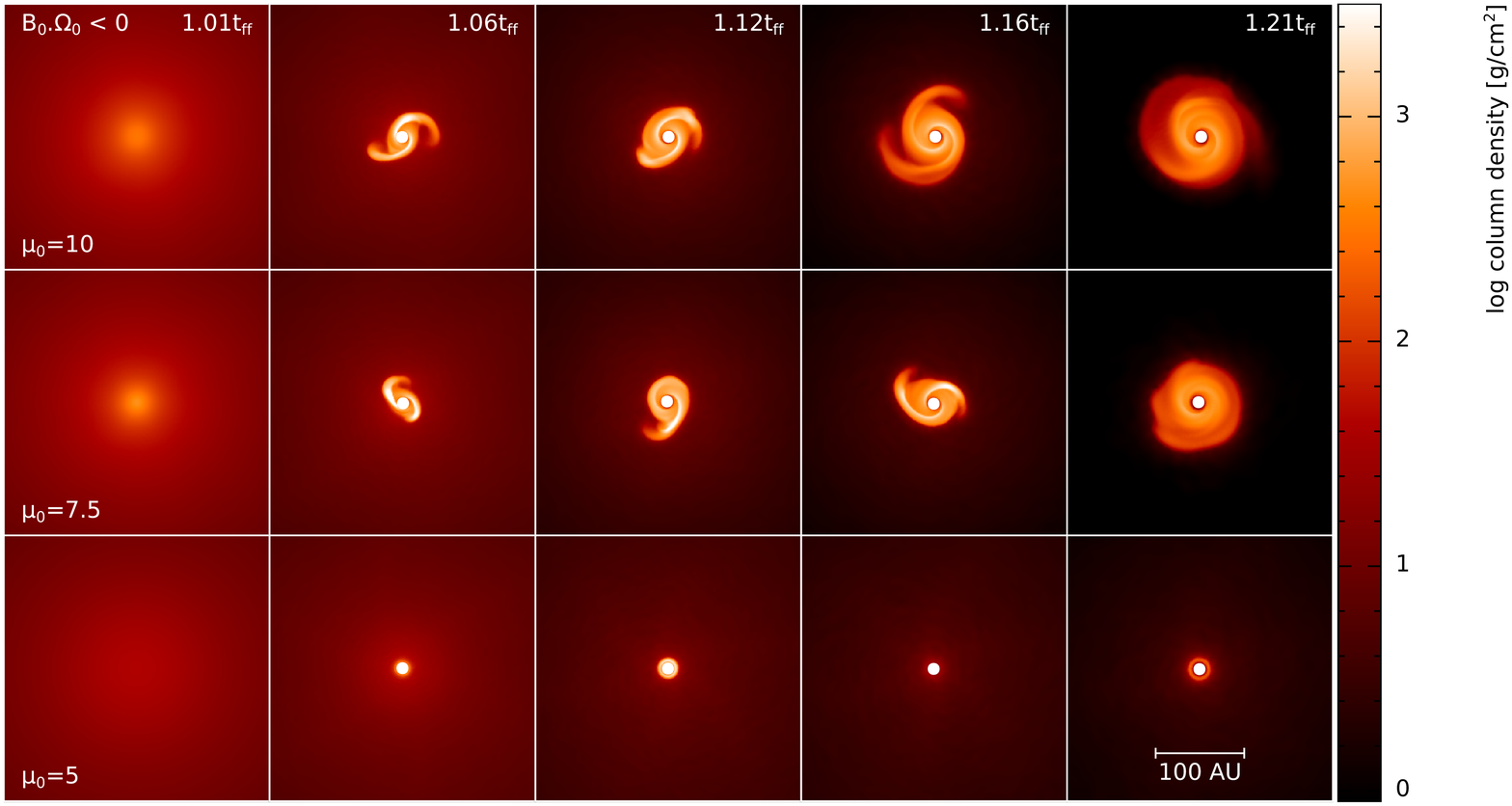}
\caption{Face-on gas column density, as in Fig.~\ref{fig:results:ideal:colxy} but for non-ideal MHD including the effect of Ohmic resistivity, the Hall effect and ambipolar diffusion.  The top plot has the magnetic field initialised with \bopos, and the bottom plot with \boneg.  Compared to ideal MHD, disc sizes are smaller for \bopos, but larger for \boneg.  This indicates that the Hall effect is the most important non-ideal MHD term for disc formation.}
\label{fig:results:ni:colxy}
\end{center}
\end{figure*}

Fig.~\ref{fig:results:ni:colxy} shows the face-on column density plots for three non-ideal MHD models using initial magnetic field strengths of $\mu_0 = 10$, 7.5 and 5; this figure is directly comparable to the bottom three rows in Fig.~\ref{fig:results:ideal:colxy}.  Since the sign of \bo \ is important for the Hall effect, the top and bottom panels have the magnetic field initialised with \bopos \ and $<0$, respectively.

Of the \bopos \ models, only the $\mu_0 = 10$ model forms a disc.  At $t = 1.21t_\text{ff}$, the non-ideal MHD model has a star+disc mass that is 4.2 per cent more massive and a disc that is 10 per cent less massive than its ideal MHD counterpart.  The radii of the two discs differ by less than 30 per cent; at any given time throughout the evolution, there is up to 60 per cent difference in radius, although it varies which model has the larger disc.  

Discs form in all three \boneg \ models, and their characteristics are plotted in the right-hand column of Fig.~\ref{fig:results:inim:discMR}.  As with the ideal MHD models, increasing the initial magnetic field strength decreases the star+disc and disc masses and the radius of the disc, although the effect is not as pronounced as in the ideal MHD suite.  Since a disc does not form in the ideal MHD model with $\mu_0=5$, it is reasonable to only compare star+disc masses.  At the remaining two magnetic field strengths, the non-ideal MHD models have larger disc masses and radii, and the specific angular momentum is similar or slightly larger.  The ideal MHD models have stronger magnetic fields in the disc; this is expected given the inclusion of the two dissipative terms in the non-ideal MHD models.  On average, gas pressure dominates the magnetic pressure in the disc.  

\subsubsection{Resolution}

As with our ideal MHD simulations, we analyse the effect of increasing the resolution from  $\sim$3$\times 10^5$ particles in the initial gas cloud to $\sim$$10^6$ particles.  Given the $h^2$ dependence that the smoothing length has on the non-ideal MHD timestep, the increase in runtime is considerable for the models that form discs and include the Hall effect (since super-timestepping cannot be used).  It takes the non-ideal MHD model with $\mu_0 = 5$, \boneg \ and $\sim$$10^6$ particles $\sim$19 times longer to reach $t = 1.15t_\text{ff}$ than its  $\sim$3$\times 10^5$ particle counterpart; this is the time when the disc dissipates in the $\sim$3$\times 10^5$  model.  For comparison, it takes the \boneg \ model with $\sim$$10^6$ particles \  $\sim$6.8 times longer to reach $t = 1.21t_\text{ff}$ than its $\sim$3$\times 10^5$ counterpart.

\begin{figure}
\begin{center}
\includegraphics[width=\columnwidth]{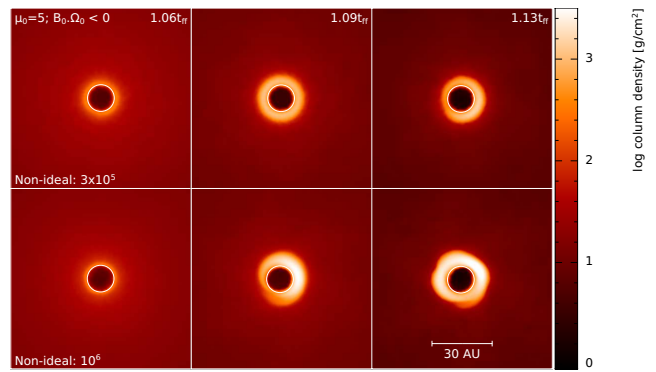}
\caption{Resolution study, as in Fig.~\ref{fig:results:resolution:colxy:ideal}, but for non-ideal MHD with $\mu_0 = 5$ and \boneg.   For non-ideal MHD, increasing the resolution decreases the mass of the star+disc system by only $\sim5$ per cent.}
\label{fig:results:resolution:colxy:ni}
\end{center}
\end{figure}

\begin{figure}
\begin{center}
\includegraphics[width=\columnwidth]{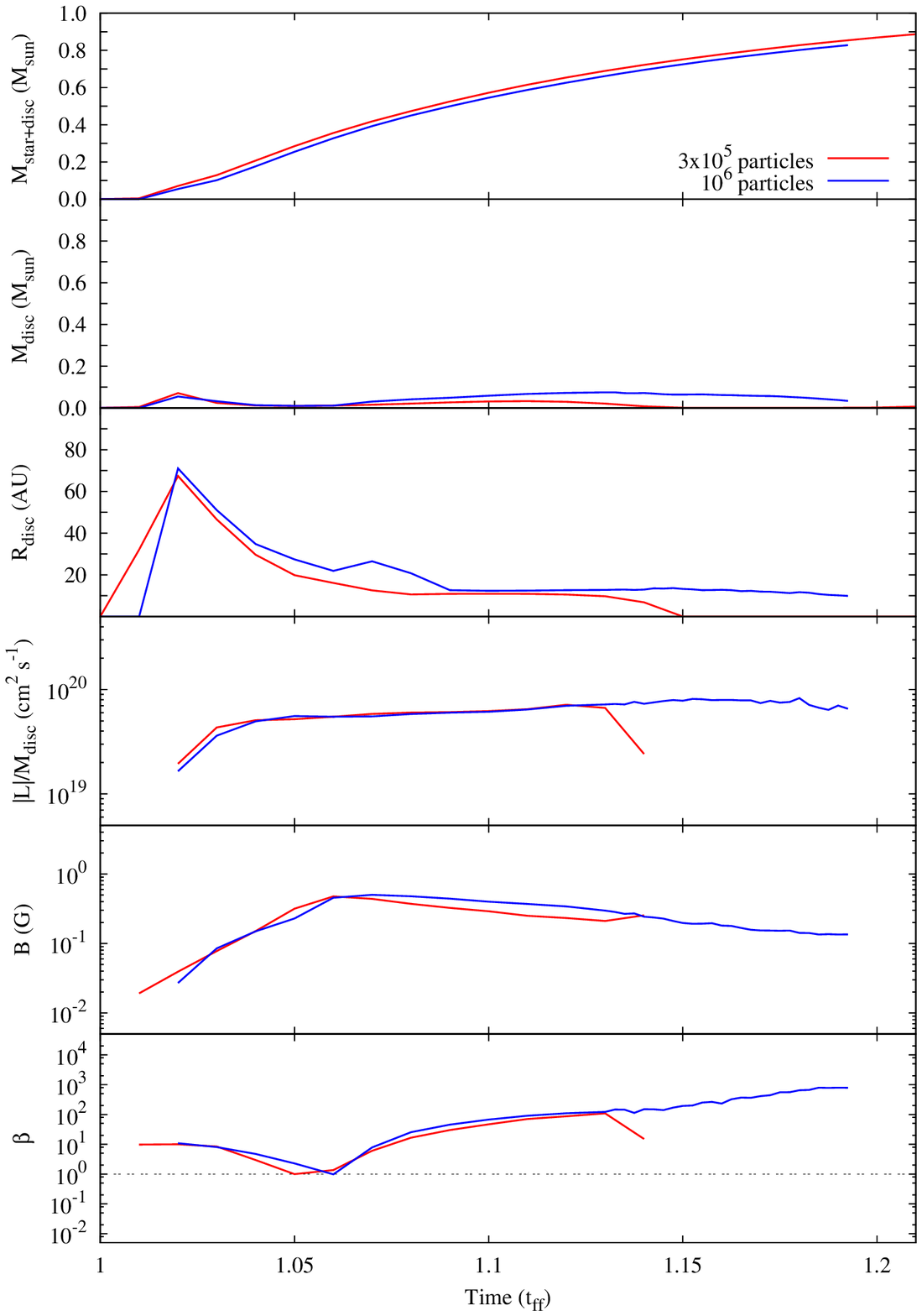}
\caption{As in Fig.~\ref{fig:results:inim:discMR} but for the non-ideal MHD models with \boneg \ and $\mu_0 = 5$ at resolutions of \lores \ and \hires \ particles in the sphere.  At $t = 1.15t_\text{ff}$, the star+disc system masses differ by less than $\sim$5 per cent.  The \hires \ model has only evolved to $t \approx 1.18t_\text{ff}$.}
\label{fig:results:nimohaC:discMR}
\end{center}
\end{figure}

Fig. \ref{fig:results:resolution:colxy:ni} shows the face-on gas column densities for the non-ideal MHD model with $\mu_0 = 5$ and \boneg, and Fig.~\ref{fig:results:nimohaC:discMR} shows the disc characteristics.
Increasing the resolution for the non-ideal MHD models has a minimal effect on the disc over the time of analysis ($t \le 1.15t_\text{ff}$; i.e. the life of the disc in the $\sim$3$\times 10^5$  model).  By increasing the resolution, the mass of the star+disc system decreases only by $\sim$5 per cent.  The high-resolution disc is more massive within a factor of two than its counterpart, and the evolution indicates that it will not dissipate.  

Our \lores \ particle models meet the resolution criteria set out by \citet{bateburkert97} (c.f. Section \ref{sec:ic}), and our brief resolution study indicates that our results agree at both resolutions.  Thus, to save computational costs of the \boneg \ models with weaker magnetic fields, the bottom panel in Fig.~\ref{fig:results:ni:colxy} shows the lower resolution models.  For consistency, we thus present all the models in Sections~\ref{sec:results_Ideal} and \ref{sec:results_NI} at the lower resolution.  The remainder of this study is performed using the $\sim$$10^6$ particle models, with the exception of our discussion of the cosmic ionisation rate.  Note that the non-ideal MHD model with $\mu_0=5$  and \boneg \ has only evolved to $t \approx 1.18t_\text{ff}$.

\subsection{Non-ideal MHD --- outflows}
\label{sec:results_NIo}
\begin{figure*}
\begin{center}
\includegraphics[width=\textwidth]{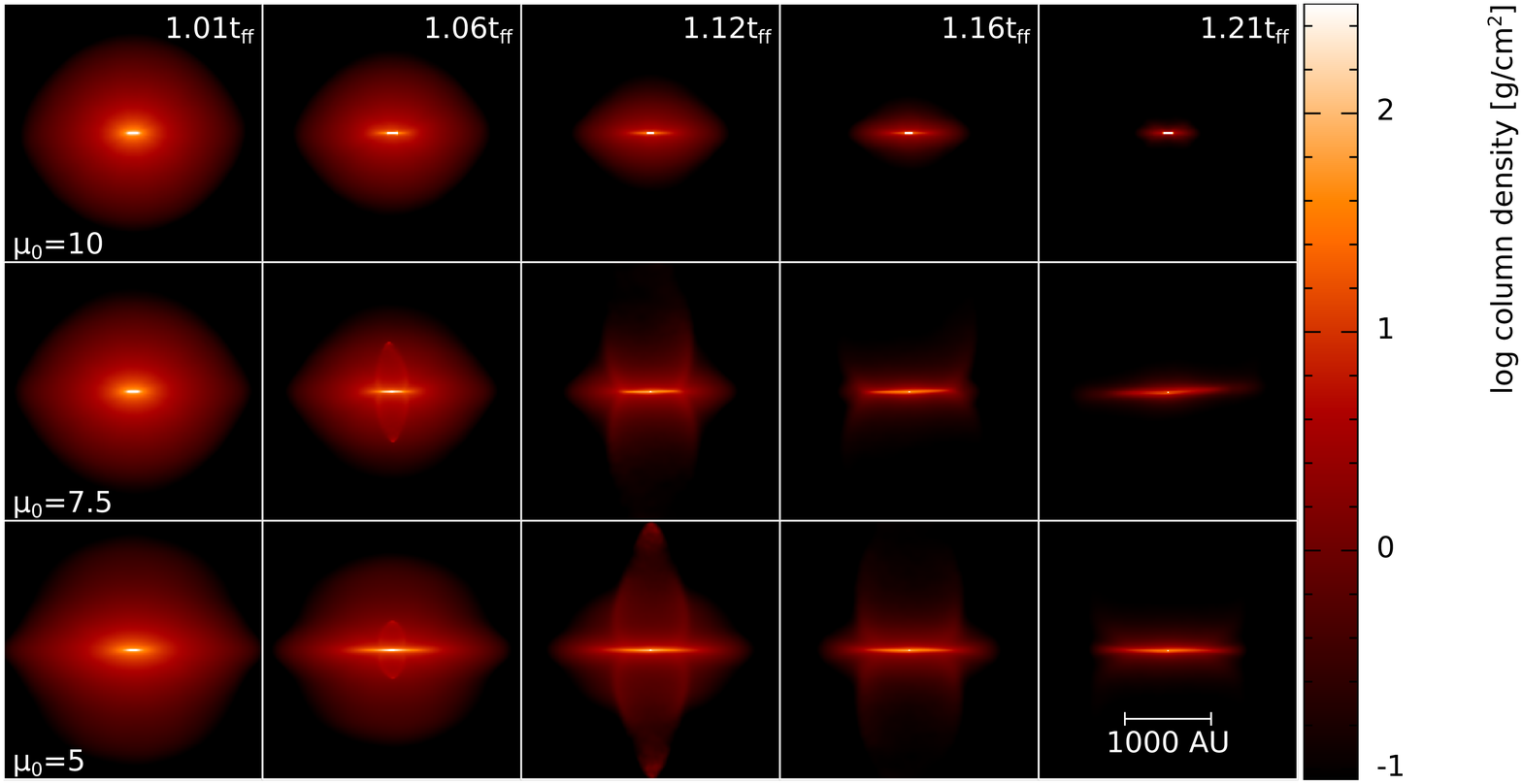}
\includegraphics[width=\textwidth]{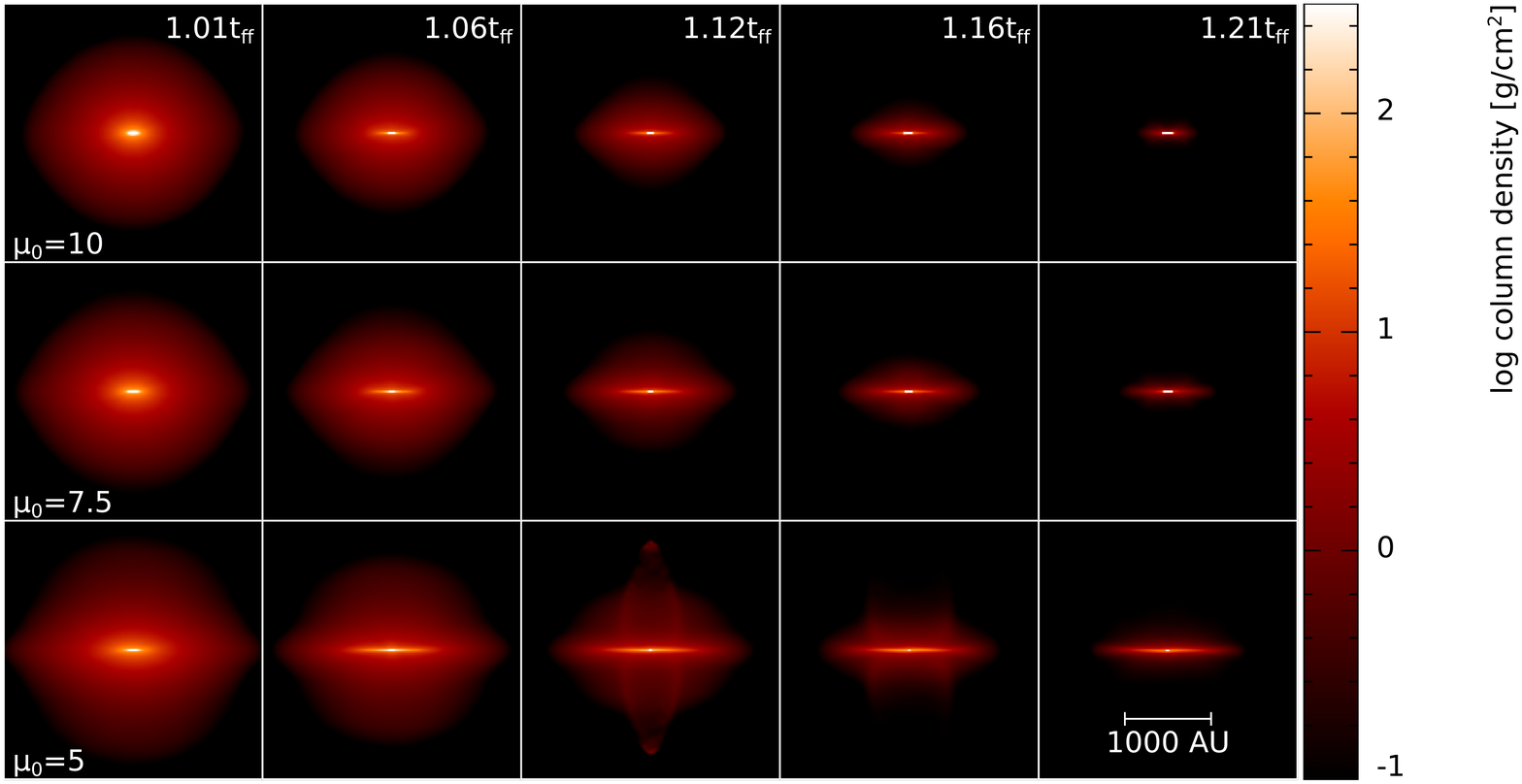}
\caption{Edge-on gas column density, as in Fig.~\ref{fig:results:ideal:colxz} but for non-ideal MHD including the effect of Ohmic resistivity, the Hall effect and ambipolar diffusion.  The top plot has the magnetic field initialised with \bopos, and the bottom plot with \boneg.  Outflows form in the calculations that form small discs.}
\label{fig:results:ni:colxz}
\end{center}
\end{figure*}

Fig.~\ref{fig:results:ni:colxz} shows the edge-on column density for the non-ideal MHD calculations, showing the models with \bopos \ and \boneg \ in the top and bottom plots, respectively.
The most interesting aspect is that outflows appear to anticorrelate with the presence of discs.  That is, outflows carry away angular momentum, which hinders the formation of discs.  This is counterintuitive since one would normally expect outflows to be launched from a disc.  Here, as in \citet{PriceTriccoBate2012}, the outflows are powered by a rotating, sub-Keplerian flow, and carry away sufficient angular momentum to prevent the formation of a Keplerian disc.  Non-ideal MHD, in general, appears to suppress the formation of outflows.  This is quantified further in Section~\ref{sec:results_outflows}, where we discuss the influence of individual non-ideal MHD terms.  

\subsection{Which non-ideal MHD terms are important?}
\label{sec:results_NIcomp}

\begin{figure*}
\begin{center}
\includegraphics[width=\textwidth]{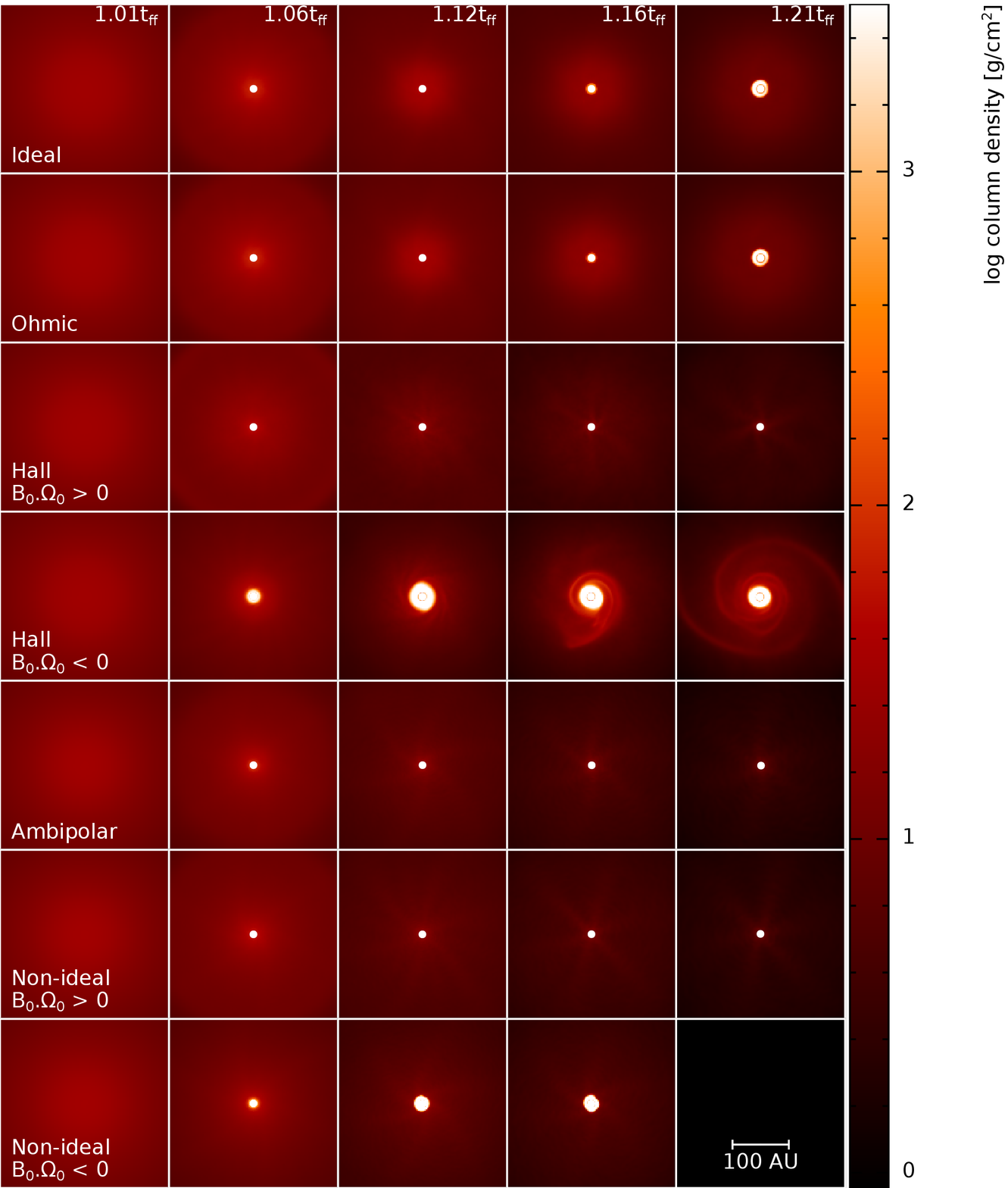}
\caption{As in Fig.~\ref{fig:results:ideal:colxy} but for ideal MHD, Ohmic-only, Hall-only, ambipolar-only and non-ideal MHD models, using $\mu_0=5$ and $\sim$$10^6$ particles in the sphere.  The Hall effect is sensitive to the sign of \bo, thus models including the Hall effect are modelled using both orientations of the initial magnetic field; all other models are insensitive to the sign of \bo \ thus use \bopos.  Small discs form at late times in the ideal MHD and Ohmic-only models.  In the Hall-only and non-ideal MHD models, $r \approx 38$ and $13$~AU disc exists by $t = 1.15t_\text{ff}$, respectively. The non-ideal MHD model with $\mu_0=5$  and \boneg \ has only evolved to $t \approx 1.18t_\text{ff}$.}
\label{fig:results:all:colxy}
\end{center}
\end{figure*}

\begin{figure*}
\begin{center}
\includegraphics[width=\textwidth]{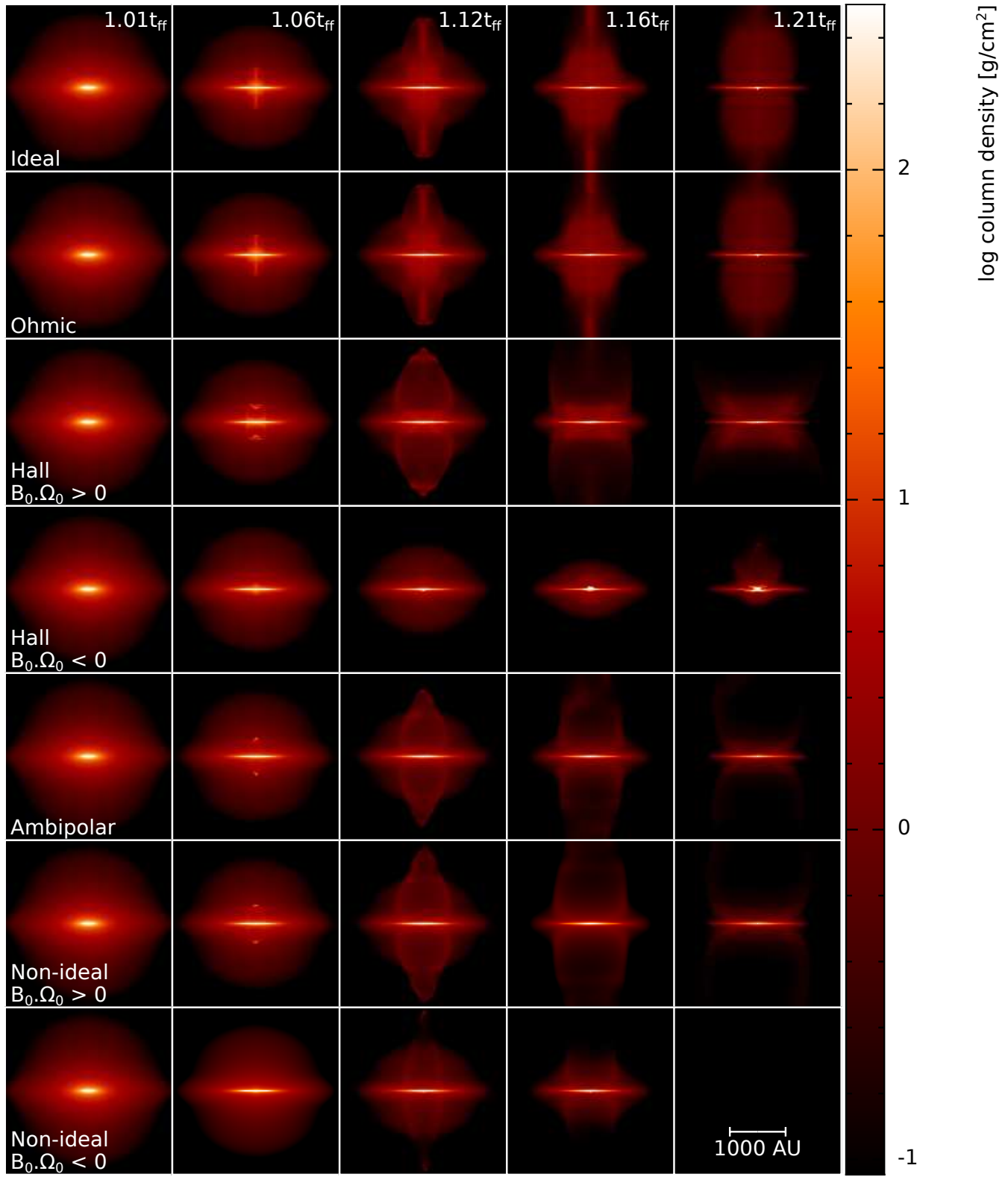}
\caption{As in Fig.~\ref{fig:results:all:colxy} but showing the edge-on gas column density.  Strong outflows correspond to models with small or no discs, and the outflow is more collimated for more ideal models (i.e. ideal MHD and Ohmic-only models).}
\label{fig:results:all:colxz}
\end{center}
\end{figure*}

The results in Section \ref{sec:results_NI} clearly show the importance of including non-ideal MHD.  To determine the specific effect of each term, we model the collapse using only one non-ideal MHD term at a time, and using both signs of \bo \ in models that include the Hall effect.  These models represent contrived and idealised situations since the physical environment dictates the importance of each term, thus these terms cannot be selected \emph{a priori}.  However, these models will allow us to determine the impact each effect has, as well as to compare our results to those in the literature.  Figs.~\ref{fig:results:all:colxy} and \ref{fig:results:all:colxz} show the the face-on and edge-on gas column density, respectively, for the ideal MHD, Ohmic-only, Hall-only (for \bopos \ and $<0$), ambipolar-only, and non-ideal MHD (for \bopos \ and $<0$) models.  All simulations use $\sim$$10^6$ particles in the sphere and $\mu_0 = 5$, since this is the magnetic field strength in our suite that it is most comparable to observed magnetic field strengths

\begin{figure*}
\begin{center}
\includegraphics[width=\columnwidth]{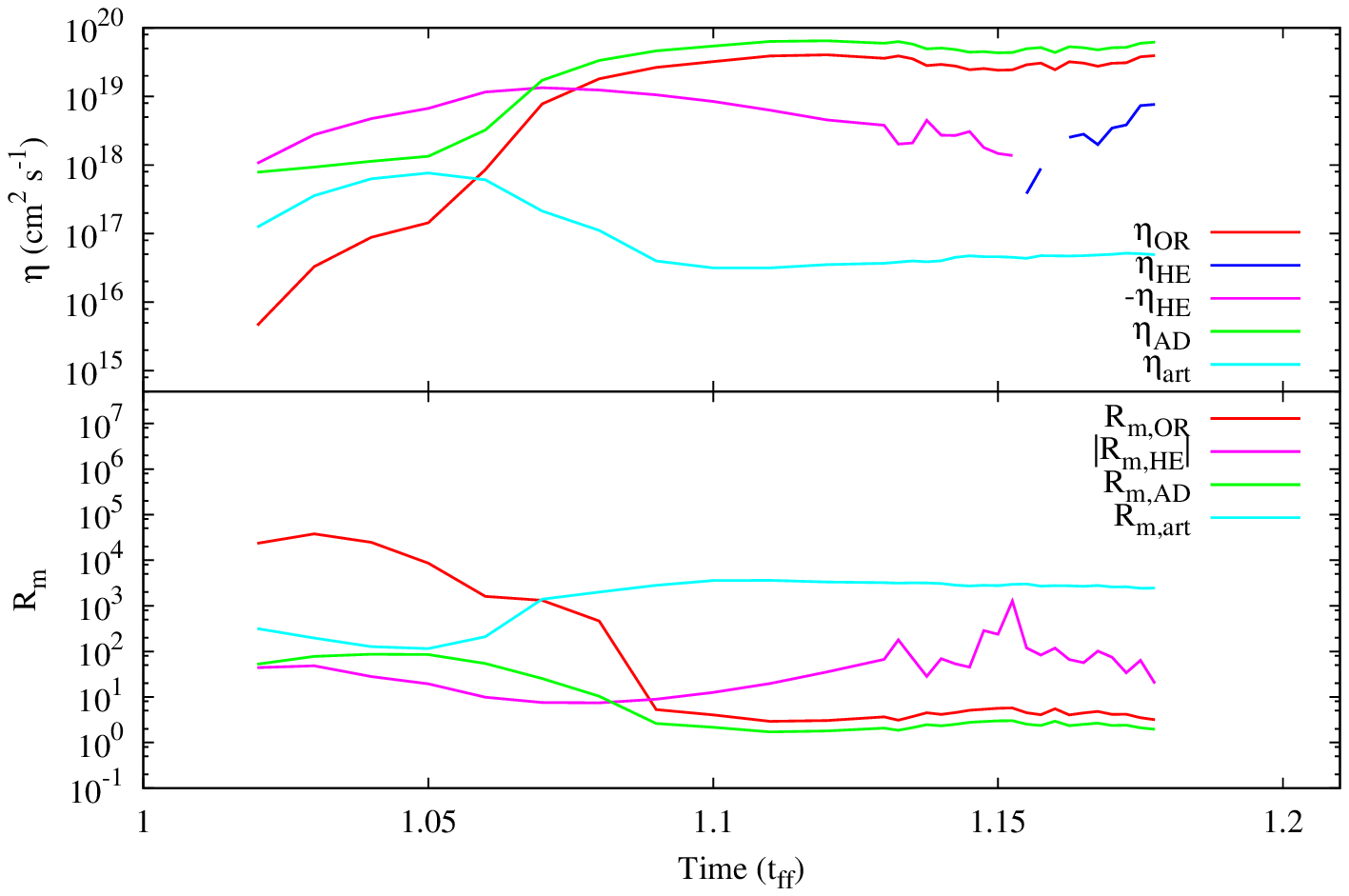}
\includegraphics[width=\columnwidth]{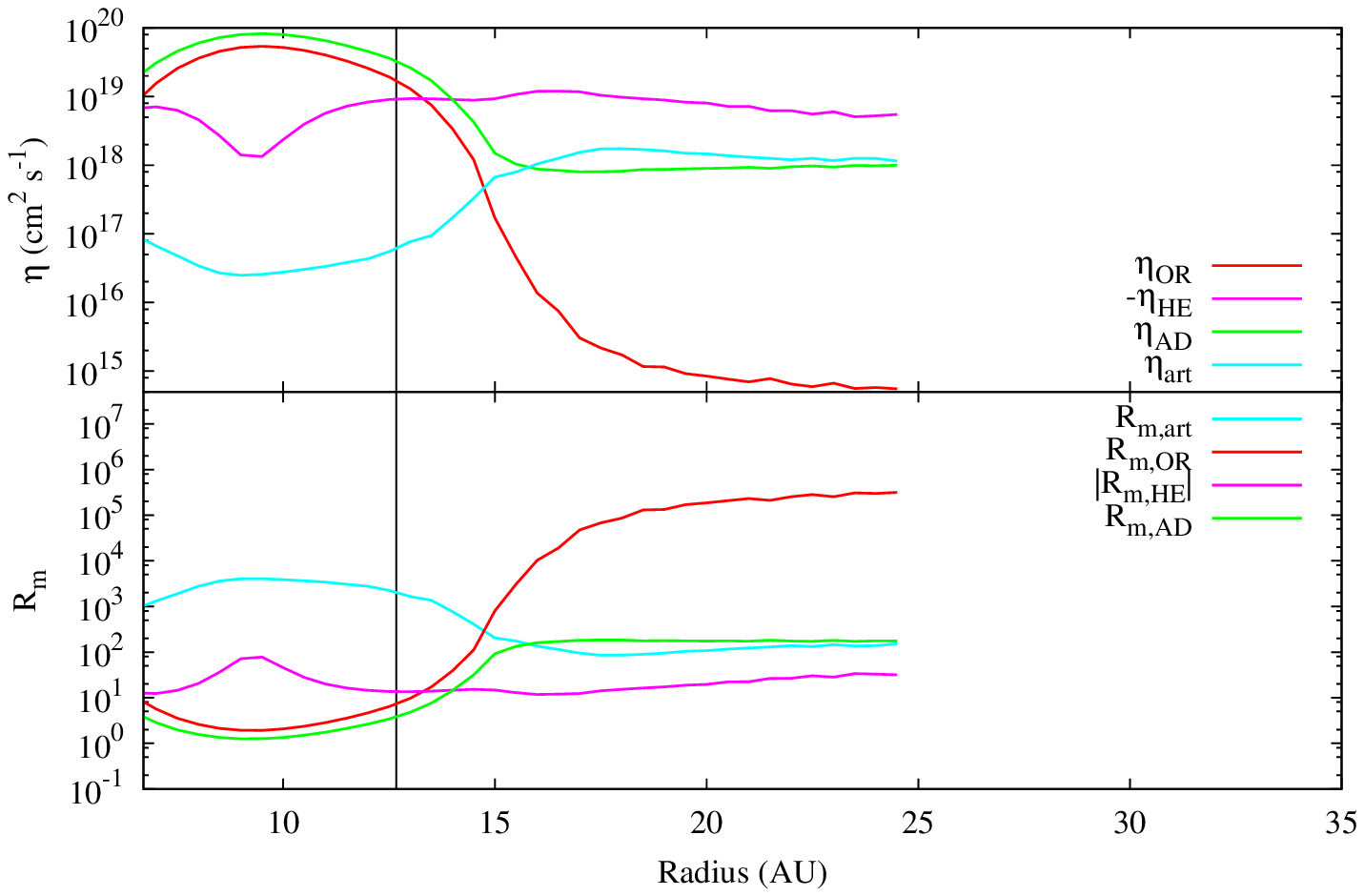}
\caption{Non-ideal MHD coefficients, $\eta$ (top panels) and magnetic Reynolds numbers, $R_\text{m} = v_\text{C}r/\eta$ (bottom panels) for the non-ideal MHD model with $\mu_0 = 5$ and \boneg.  The left-hand panel shows the average values in the disc, and the right-hand panel shows the radial profile at $t = 1.12t_\text{ff}$.  Horizontal axes are chosen for consistency with other plots in this paper.  The vertical line in the right-hand panel corresponds to the defined radius of the disc at that time.  Shortly after the formation of the disc, the Hall effect is the dominant term, but as the disc begins to dissipate, the dissipative terms begin to dominate.  The value of $R_\text{m} \gtrsim 1$ indicates that the diffusion terms are important in the disc.}
\label{fig:results:etaR:C}
\end{center}
\end{figure*}

At this resolution and magnetic field strength, discs form in the ideal MHD and Ohmic-only models at $t \approx 1.17t_\text{ff}$; by $t = 1.21t_\text{ff}$, these discs have grown to $r \approx 14$ and 13~AU, respectively.  The ambipolar-only model, and the models that include the Hall effect with \bopos \ fail to form discs.  Large discs are formed when the Hall effect is included and the gas is initialised with \boneg.

Qualitatively, the Hall effect with \boneg \ appears to be the most important non-ideal MHD term.  Quantitatively, the importance of each non-ideal term in the disc can be analysed using the non-ideal MHD model with \boneg.  The left-hand panel of Fig.~\ref{fig:results:etaR:C} shows the average coefficients of the three non-ideal MHD terms and the artificial resistivity, $\eta$, and magnetic Reynolds number, $R_\text{m}$, in the disc at any given time; the right-hand panel shows the radial profile of these terms at $t = 1.12t_\text{ff}$.  

Shortly after the formation of the disc, the Hall effect is the dominant term, and ambipolar diffusion is comparable to the artificial resistivity.  As the system evolves, ambipolar diffusion becomes the dominant term, and the Hall effect becomes less important.  Ohmic resistivity becomes important as the density of the disc increases.  At the snapshot of $t = 1.12t_\text{ff}$, ambipolar diffusion and Ohmic resistivity dominate the Hall effect within the defined disc.  When considering earlier times (e.g. $1.06t_\text{ff}$), the radial profile is approximately constant with radius, with $|\bar{\eta}_\text{HE}| > \bar{\eta}_\text{art} > \bar{\eta}_\text{AD} > \bar{\eta}_\text{OR}$.

A magnetic Reynolds number, $R_\text{m} = v_\text{C}r/\eta$, where $v_\text{C} = \sqrt{GM(r)/r}$ is the circular velocity at radius $r$ and $M(r)$ is the enclosed star+disc mass, is calculated for each physical and artificial resistivity.  For all time, at least one Reynolds number from a physical terms is lower than the Reynolds number from artificial resistivity.  This indicates that the physical resistivity is more important than artificial resistivity.  Moreover, the value of $R_\text{m} \gtrsim 1$ indicates that the diffusion terms are important in the disc.  

\subsection{Effect of non-ideal MHD terms --- outflows}
\label{sec:results_outflows}
All models that include magnetic fields launch bipolar outflows.  Increasing the magnetic field strength (i.e. Sections~\ref{sec:results_Ideal} and \ref{sec:results_NI}) yields faster and more collimated outflows, which are sustained over a longer period of time. 

At $\mu_0 = 5$ (c.f. Figs.~\ref{fig:results:all:colxy} and \ref{fig:results:all:colxz}), both the ideal MHD and Ohmic-only models launch strong, collimated outflows.  As the systems evolve, the base of the outflow broadens, resulting in a less collimated system, although the remnant of the collimated outflow persists.  By the time the disc forms at $t \approx 1.17t_\text{ff}$, the outflow has been well-established and is in the process of weakening.  When the Hall effect or ambipolar diffusion are included for \bopos, broad outflows form; unlike the ideal MHD and Ohmic-only models, there is no collimated central outflow.  Thus, in models that do not form discs, strong bipolar outflows form, and the Hall effect and ambipolar diffusion prevent collimation.  

A large disc forms in the Hall-only model with \boneg, however the outflow is almost completely suppressed.  A weak outflow is launched at $t \approx 1.05t_\text{ff}$, but dissipates by $\approx 1.08t_\text{ff}$.   A small dense disc forms in the non-ideal MHD model with \boneg, and a weak outflow is launch at $t \approx 1.07t_\text{ff}$.  The outflow continues to expand as the system evolves, however it is never collimated and never becomes as broad as in the \bopos \ models.  

As discussed in Section~\ref{sec:results_NIo}, we find that the presence of a collimated outflow is anticorrelated to the presence of a large disc.  
\subsection{Effect of magnetic field direction}
\label{sec:results_NIB}

The direction of the magnetic field with respect to the rotation vector (i.e. the sign of \bo) has a profound impact.  This can be seen by comparing third to the fourth row (Hall-only), and the seventh to the eighth row (non-ideal MHD) of Figs.~\ref{fig:results:all:colxy} and \ref{fig:results:all:colxz}; the bottom row in each pair is initialised with \boneg.  Using \boneg, an $r\approx38$~AU disc forms in the Hall-only model, and an $r\approx 13$~AU disc forms in the non-ideal MHD model by $t = 1.15t_\text{ff}$; the Hall-only model has its maximum disc radius at this time. Thus, at a magnetic field strength of $\mu_0 = 5$, discs can only be formed if the Hall effect is included and \boneg. 

\begin{figure}
\begin{center}
\includegraphics[width=\columnwidth]{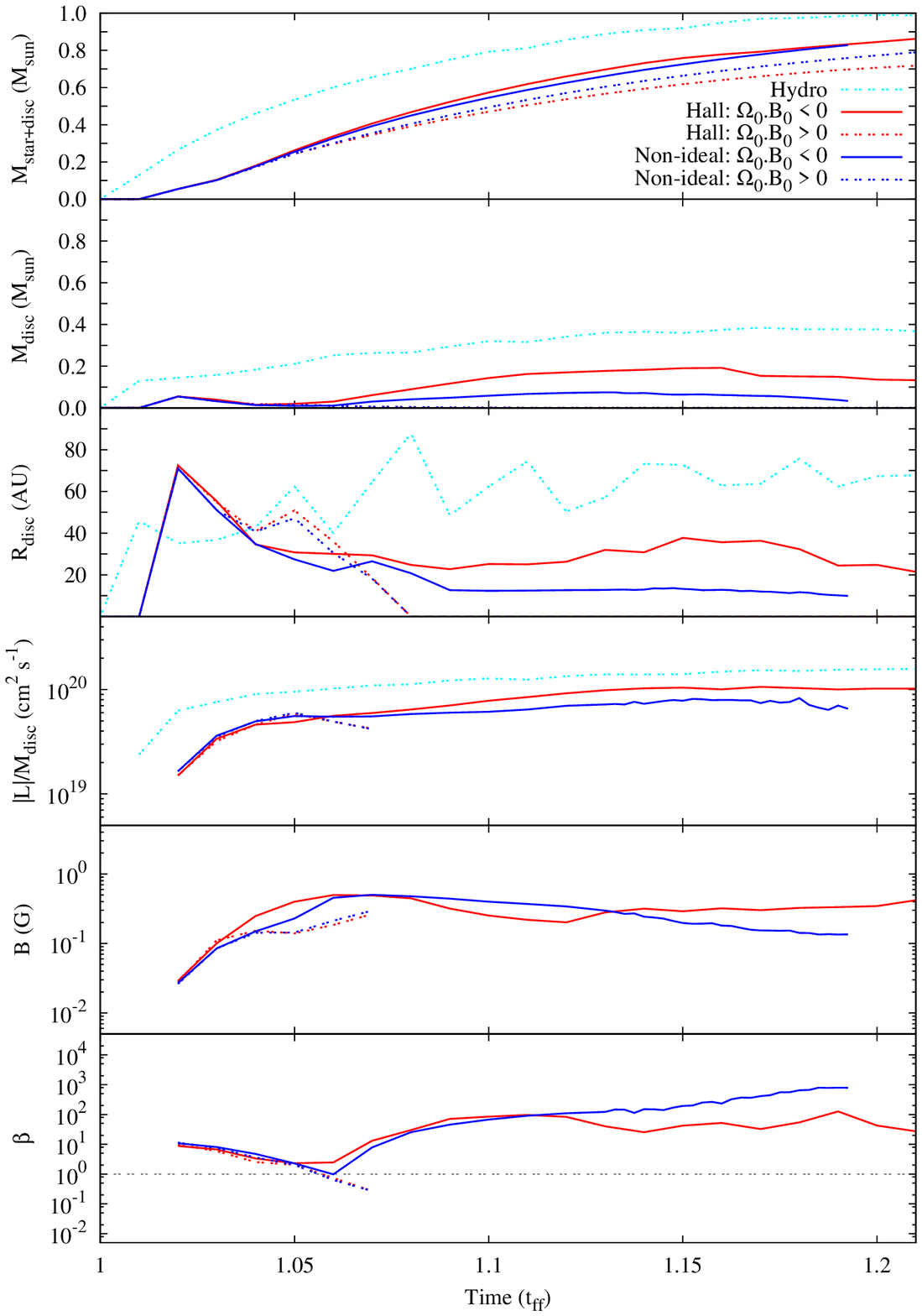}
\caption{Disc properties as in Fig.~\ref{fig:results:inim:discMR}, but for Hall-only (red) and non-ideal MHD (blue) models with \boneg \ (solid lines) and \bopos \ (dashed lines) for $\mu_0=5$ and \hires \ particles.  The properties for the hydrodynamical model are included for reference.  In both cases, initialising the magnetic field such that \boneg \  permits a disc to form, and yields a larger star+disc mass than their  \bopos \  counterpart.  The non-ideal MHD model with $\mu_0=5$  and \boneg \ has only evolved to $t \approx 1.18t_\text{ff}$.}
\label{fig:results:negB:discMR}
\end{center}
\end{figure}

Fig.~\ref{fig:results:negB:discMR} shows the masses and sizes of these discs (along with the limited information from their \bopos \ counterparts).
In all five models in this figure, a sink particle is formed at $t\approx 1.025t_\text{ff}$.  By $t~\approx~1.07t_\text{ff}$, the disc disappears in the Hall-only and non-ideal MHD models with \bopos \  as the remainder of the high-density ($\rho > \rho_\text{disc,min}$) material is accreted onto the sink particle.  In these models, a true `disc' may never have formed, and the reported disc properties are for the high-density material that satisfies our chosen definition of `disc'.  Thus, at this magnetic field strength, there are no discs with \bopos \  to which we can compare.  

The star+disc masses in the \boneg \ models are $\sim$23 and 9 per cent more massive than their \bopos \  counterparts for the Hall-only and non-ideal MHD models, respectively, at $t = 1.15t_\text{ff}$.  At $t = 1.21t_\text{ff}$, the Hall-only model with \boneg \ has a large disc, which is $\sim$64 and $\sim$68 per cent smaller in mass and radius,  respectively, than the hydrodynamic disc.

Both \boneg \ models have magnetic field strengths and plasma beta's that differ by less than a factor of two.  Thus, the Hall effect is the non-ideal MHD term that is primarily responsible for preventing the transport of angular momentum to allow the disc to grow.  

\begin{figure}
\begin{center}
\includegraphics[width=\columnwidth]{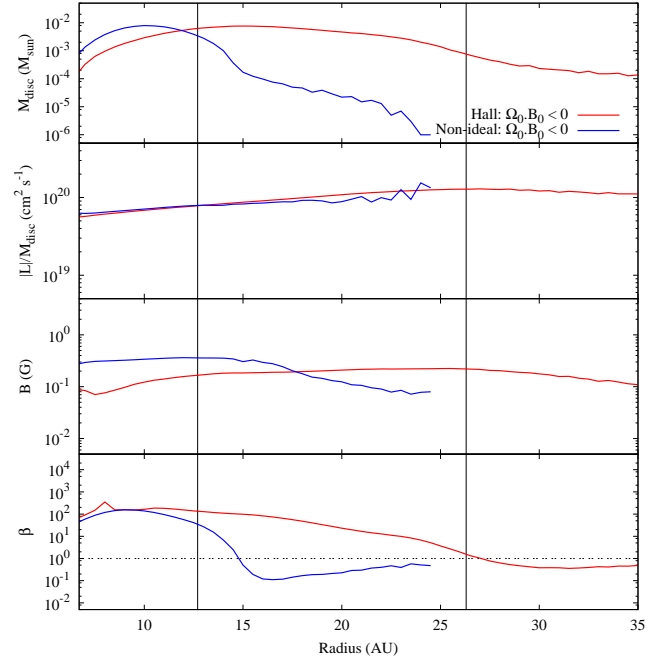}
\caption{Radial profile of the disc at $t=1.12t_\text{ff}$ for the Hall-only (red) and non-ideal MHD (blue) models with \boneg, for $\mu_0=5$ and \hires \ particles.  The profile includes all the gas particles with density $\rho > \rho_\text{disc,min}$, and starts at $r = 6.7$~AU, which is the sink radius. The vertical lines at $\sim$13 and 26~AU represent the defined radii of the discs in the non-ideal MHD and Hall-only models, respectively.  \emph{From top to bottom}: Disc mass, specific angular momentum, magnetic field strength, and plasma beta.}
\label{fig:results:negB:vR}
\end{center}
\end{figure}

Although average trends appear similar between both models, the radial structure of both discs differs, as shown in Fig.~\ref{fig:results:negB:vR} at a snapshot at $t = 1.12t_\text{ff}$.
When considering the magnetic field profile, the Hall-only model has a maximal magnetic field strength at $r\approx26$~AU, which does not correspond to the radius of the maximum mass.  Thus, the Hall effect traps the magnetic field at a larger radius, which is near the outer edge of the disc.  Interior to this, and ignoring the slight rise in magnetic field strength near the sink particle, the maximum plasma beta of $\beta \approx 184$ is at $r \approx 11$~AU, corresponding to a weaker magnetic field.

When all three non-ideal MHD terms are present, the dissipative processes diffuse the magnetic field, and the maximum field strength is reduced.  However, these processes also diffuse the field inwards, so the magnetic field for the non-ideal MHD model is stronger than the Hall-only model for the inner $r\approx$18~AU.  Unlike the Hall-only model, the maximum mass and plasma  beta occurs in the non-ideal MHD model at $r\approx 9$~AU; this radius also corresponds to the maximum non-ideal coefficients, $\eta$ (c.f. the left-hand panel of Fig.~\ref{fig:results:etaR:C}).   Within the defined disc, the magnetic field strength differs by less than 7 per cent, but the magnetic pressure is less important with respect to the gas pressure. 

\begin{figure}
\begin{center}
\includegraphics[width=0.8\columnwidth]{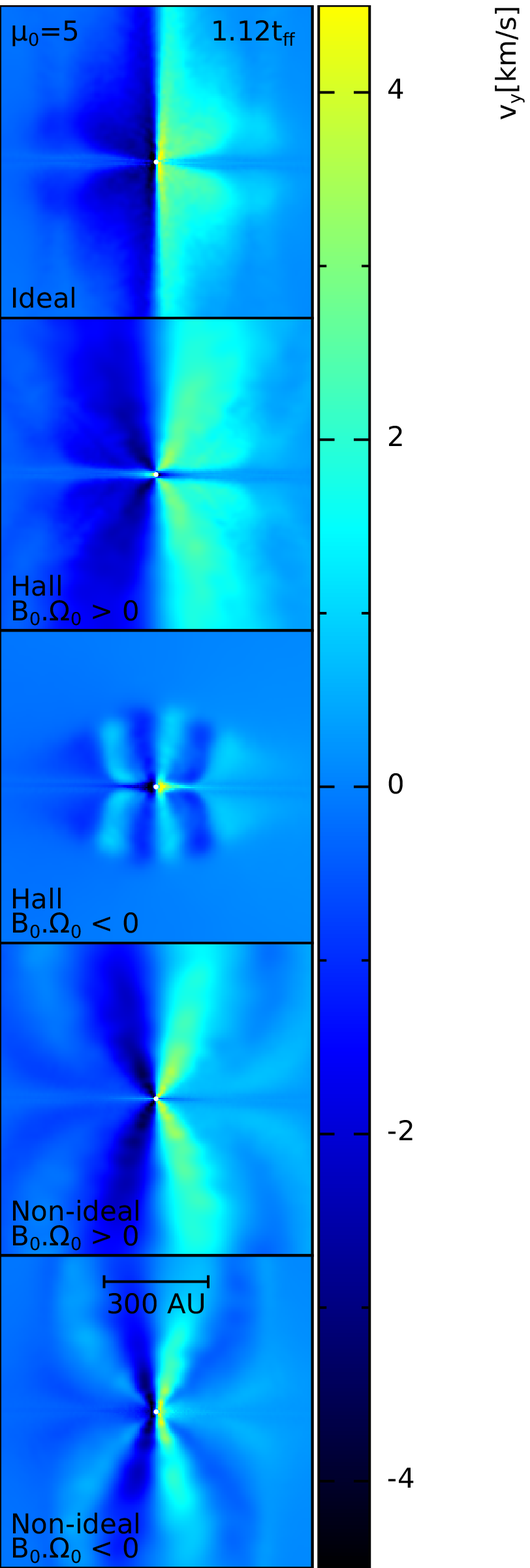}
\caption{The velocity perpendicular to a slice through the outflow (i.e. $v_\text{y}$) for five models with $\mu_0 = 5$ at $t=1.12t_\text{ff}$.  Each frame is (900 AU)$^2$, which is smaller than in Fig.~\ref{fig:results:all:colxz} so that details around the first hydrostatic core can be seen.  The Hall-only model the \boneg \ has a weak bipolar outflow, but forms a counter-rotating envelope.  A weak counter-rotating envelope also exists in the non-ideal MHD model with \boneg.  None of the other models develop a counter-rotating envelope.  The Ohmic-only and ambipolar-only models are very similar to the ideal and non-ideal MHD (\bopos) models, respectively.}
\label{fig:results:Vely}
\end{center}
\end{figure}

The previous analysis has focused on the formation of the disc, however, the surrounding gas is also affected by the processes and parameters, as suggested by Fig.~\ref{fig:results:all:colxz}.  Fig.~\ref{fig:results:Vely} shows the velocity perpendicular to a slice through the outflow (i.e. $v_\text{y}$) at $t = 1.12t_\text{ff}$ for the ideal MHD model, and the Hall-only and non-ideal MHD models for both \bopos \ and $<0$.
In all models, the gas is initially rotating counterclockwise, and in the \bopos \ models, it continues to do so  as the system evolves.  The velocity structure of the ideal MHD model traces the established collimated outflow and the young broad outflow (c.f. Fig.~\ref{fig:results:all:colxz}).  The Hall-only and non-ideal MHD models with \bopos \ have a large opening angle, which corresponds to the broad outflow.  

When the Hall effect is included in the \boneg \ models, the large angular momentum in the disc results in a decrease in the angular momentum of the gas in the cloud from conservation laws, and causes a counter-rotating envelope to form.  This can be clearly seen for the Hall-only model, where a counter-rotating envelope exists at a radius of $r \in (90,150)$~AU from the rotation axis.  The non-disc material interior to this is slowly rotating, which is distinct from the remaining models.  A weak counter-rotating envelope also exists in the non-ideal MHD models at $r \gtrsim 150$~AU.

\subsubsection{Comparison to other works}
Given the numerical difficulty associated with the Hall effect, it has been previously ignored in disc collapse simulations, with the exception of \citet{TsukamotoEtAl2015b}, who independently performed similar collapse simulations while this study was being undertaken.  

In the shearing box simulations of \citet{Bai2014,Bai2015}, their \boneg \ models reduce the horizontal magnetic field which results in negligible magnetic braking.  This would result in the formation of large discs, similar to the results obtained in our simulations.  

Using self-similar calculations, \citet{BraidingWardle2012sf} concluded that the Hall effect is important in determining whether a disc forms, its size, density and rotational profiles.  For \bopos, their solutions yield a decrease in surface density of the disc compared to a disc without the Hall effect.  For \boneg, they determined that the surface density of the disc would be increased to realistic values for protostellar discs.  They also show that a large amount of Hall diffusion is not required to create the asymmetry caused by the sign of \bo \ to be observable.  While we were unable to control the strength of $\eta_\text{Hall}$ since it is self-consistently calculated, we did achieve noticeable differences in the non-ideal MHD models even when the Hall effect had a secondary effect to ambipolar diffusion.

Similar to this study, \citet{TsukamotoEtAl2015b} modelled the collapse of a uniform sphere of gas of $M = 1$ M$_\odot$ with $\mu_0 = 4$.  These models included the Hall effect.  When their cloud was initialised with \bopos, the result was a disc of $r < 1$~AU, while for \boneg, the resulting disc had a radius of $r > 20$~AU.  This is consistent with our results, where our Hall-only models yielded no disc and an $r \approx 38$~AU disc by $t = 1.15t_\text{ff}$ for the two different orientations, respectively.  Given our sink radius of $r = 6.7$~AU, we are unable to resolve discs smaller than that; given our slightly weaker initial magnetic field, we expect a larger disc in our \boneg \ model.  Further, their Hall-only model with \boneg \ forms a counter-rotating envelope similar in size and velocity as ours.  

\subsection{Effect of the cosmic ray ionisation rate}
\label{sec:results_NIzeta}
As discussed in \S \ref{ssec:num:ionise}, the cosmic ray ionisation rate, $\zeta$, is one of the few free parameters in our algorithm.  To test the effect of this parameter, we decrease it by a factor of ten from our fiducial value of \zetaos \ to make the simulations `more non-ideal.'  This decrease in $\zeta$ causes a decrease in d$t_\text{non-ideal}$, and the runtime is substantially increased; the exact slow-down is model-dependent.  Given the slow-down in the models that form discs, the results in this section are complied from the \lores \ particle models.  

In the non-ideal MHD model with \bopos, $\zeta$ has negligible effect, and neither model produces a disc.  The model with \zetaoe \ has a runtime $\sim$3.25 times longer than its fiducial-$\zeta$ counterpart.

\begin{figure}
\begin{center}
\includegraphics[width=\columnwidth]{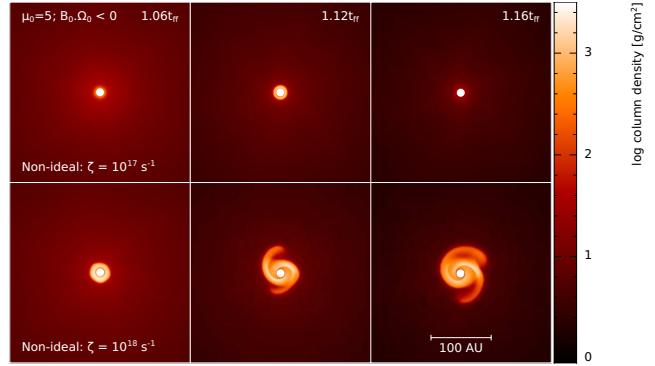}
\caption{As in Fig.~\ref{fig:results:ideal:colxy}, but for two different cosmic ray ionisation rates, $\zeta$, for the non-ideal MHD model with $\mu_0=5$, \boneg, and \lores \ particles initially in the sphere.  For this model, decreasing $\zeta$ allows a large disc to form and persist for the duration of the simulation; for the larger value of $\zeta$, the disc dissipates by $t \approx 1.15t_\text{ff}$.}
\label{fig:results:zetaQ:colxy}
\end{center}
\end{figure}

\begin{figure}
\begin{center}
\includegraphics[width=\columnwidth]{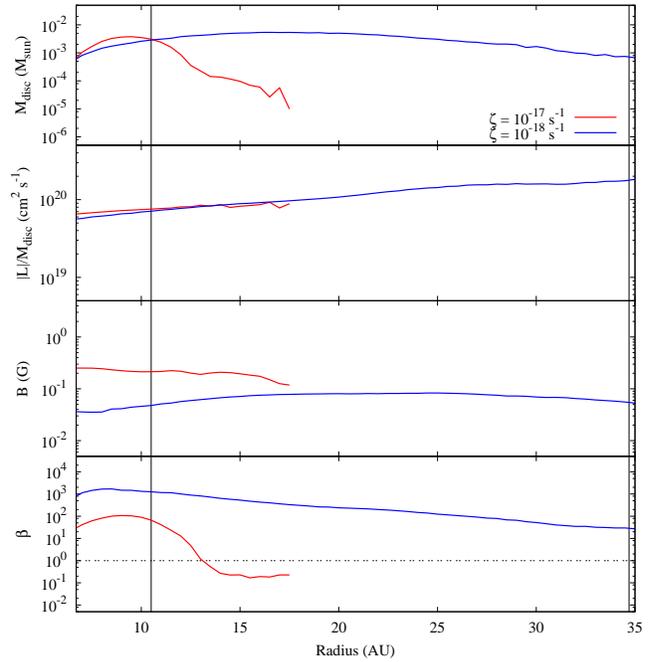}
\caption{As in Fig.~\ref{fig:results:negB:vR}, but for two different cosmic ray ionisation rates, $\zeta$, for the non-ideal MHD model with $\mu_0=5$ and \boneg.  The vertical lines at $r \approx $11 and 35~AU represent the defined radii of the discs using $\zeta = 10^{-17}$ and \dexoe, respectively.}
\label{fig:results:zeta:mmlb}
\end{center}
\end{figure}

For \boneg, a disc forms for both values of $\zeta$.  Fig.~\ref{fig:results:zetaQ:colxy} shows the face-on column density for the non-ideal MHD models using $\zeta = 10^{-17}$ (top) and \dexoe \ (bottom); both models use $\mu_0=5$.  
Decreasing the value of $\zeta$ in this model increases the runtime by a factor of $\sim$30, and d$t_\text{Hall}$ is typically the limiting timestep after a free-fall time.  The resulting disc is larger and more stable (i.e. survives for a longer period of time) for the \zetaoe \ model.  Fig.~\ref{fig:results:zeta:mmlb} shows the disc properties at a snapshot at $t=1.12t_\text{ff}$.
In both models, the magnetic field strength in approximately constant, but is $\sim$3 times higher in the \zetaos \ model.  The maximum plasma beta is also $\sim$15 times higher in the \zetaoe \ model, indicating weaker magnetic fields.

This analysis was intentionally performed using a model with the Hall effect and \boneg \ since a disc forms.  Tests show that the ambipolar-only model is insensitive to the value of the cosmic ionisation rate.  Thus, the physical differences are insensitive to the precise value of $\zeta$, with the exception of models that include the Hall effect which are initialised with \boneg.

\subsection{Comparison to other works}

The studies presented in previous works, such as \citet{LiKrasnopolskyShang2011}, \citet{TomidaOkuzumiMachida2015} and \citet{TsukamotoEtAl2015}, are complementary to this study.  Caution must be used, however, when comparing the studies to one another and to ours since there are differences in initial conditions, physical and numerical processes.  None of these studies included the Hall effect.  Regardless of the differences between the studies, a qualitative comparison is useful.  

\citet{LiKrasnopolskyShang2011} ran a suite of 2D axisymmetric simulations on a polar grid to model the collapse of a 1~M$_{\astrosun}$ gas cloud of initially uniform density.  Their inner radial boundary is an outflow boundary condition set at 6.7~A.U, which is the same radius as our sink particles.  They varied several parameters including magnetic field strength, cosmic ionisation rate, the grain model and which non-ideal MHD terms were included.  In strong magnetic fields, they found that discs did not form, even when Ohmic resistivity and ambipolar diffusion were included.  For weak magnetic fields, they found small discs normally formed at early times, but later dissipated due to powerful outflows.

Both \citet{TomidaOkuzumiMachida2015} and \citet{TsukamotoEtAl2015} modelled the collapse of a gas cloud in 3D, and both studies compared three models using a strong magnetic field, $\mu_0 \approx 4$: ideal MHD, Ohmic-only, Ohmic+ambipolar.  \citet{TomidaOkuzumiMachida2015} initialised their gas cloud as an unstabilised, rotating 1~M$_{\astrosun}$ Bonner-Ebert sphere; their simulations are performed on a nested grid, and includes radiative transfer.  \citet{TsukamotoEtAl2015} used a radiative SPMHD code without sink particles, initialising their 1~M$_{\astrosun}$ gas cloud as an isothermal uniform gas sphere.  Both of these models used a different initial radius than our cloud, thus the absolute initial magnetic field strength is different in each study.
 
Qualitatively, the studies agree that strong magnetic fields in ideal MHD models efficiently transport angular momentum and prevent the formation of a disc.  By adding Ohmic resistivity, \citet{TomidaOkuzumiMachida2015} form a small disc simultaneous with the formation of the protostellar core.  When ambipolar diffusion is also included, the disc forms prior to the formation of the protostellar core, and at the end of the first hydrostatic core phase, the disc radius is $r\approx5$~AU.  They predict their disc will continue to grow, although it is already massive enough to form non-axisymmetric structures and possibly fragment.  \citet{TsukamotoEtAl2015} form an $r \approx 1$~AU disc just after the formation of the protostellar core in both of their models that include non-ideal MHD terms.   

Caution must be used when comparing our models to previously published studies, since our sink particle has a radius of 6.7~AU, thus we are unable to track discs smaller than this; moreover, discs with radii only slightly larger than this may be artificially created as a result of the sink particle's interaction with the nearby gas.  Our definition of `in the disc' (where the only requirement is $\rho_\text{gas} > \rho_\text{disc,min}$) likely allows for different size of discs to be reported for similar simulations.  After the formation of the sink particle at in our $\mu_0 = 5$ models (excluding the Hall-only and non-ideal MHD models that are initialised with \boneg), the high-density gas is accreted onto the sink particle, leaving the systems devoid of any disc-like object.  This lack of disc agrees with the other studies.  In the high resolution version of our Ohmic-only model, a disc does reform prior to the end of the simulation at $t=1.21t_\text{ff}$.  The disc mass is low, comprising $\sim$6 per cent of the star+disc system mass.

\section{Summary and conclusion}
\label{sec:discussion}

We have presented a suite of simulations studying the effect of non-ideal MHD on the formation of circumstellar discs.  Our models were initialised as a 1~M$_{\astrosun}$, spherically symmetric, rotating molecular cloud core with a magnetic field threaded vertically through it.  We followed the gravitational collapse of the core until shortly after the formation of the first hydrostatic core using a barotropic equation of state.  We tested the impact of each non-ideal MHD term (Ohmic resistivity, Hall effect, and ambipolar diffusion) both independently and together.  We further tested the effect of the initial mass-to-flux ratio $\mu_0$, the direction of the seeded vertical magnetic field with respect to the axis of rotation, the cosmic ray ionisation rate $\zeta$, and resolution.  All of the simulations were performed using the SPMHD code {\sc Phantom}, including self-gravity.  Our key results are as follows:
\begin{enumerate}

\item \emph{Ideal MHD}:  We performed simulations using ideal MHD with $\mu_0 = \infty$, 10, 7.5 and 5.  As in \citet{pricebate07}, stronger magnetic fields (smaller values of $\mu_0$) yielded smaller and less massive discs.  A bipolar outflow was launched shortly after a free-fall time in the magnetic models, with faster and more collimated outflows being launched in the models with stronger magnetic fields.

\item \emph{Non-Ideal MHD}:  For $\mu_0 = 10$, 7.5 and 5, we modelled the collapse using all three non-ideal MHD terms.  For \bopos, the non-ideal MHD models yielded smaller discs than their ideal MHD counterparts, but yielded larger discs when initialised with \boneg.  For $\mu_0 = 5$, we modelled the collapse using each effect separately, and used both signs of \bo \ for models that included the Hall effect.   Discs did not form in any of the models initialised with \bopos.  At all magnetic field strengths, the evolution of the system is dependent on the Hall effect and the sign of \bo.  

\item \emph{Outflows}: We found an anticorrelation between the size and speed of the outflow and the size of the disc. That is, outflows carry away angular momentum and this hinders the formation of discs.  

\item \emph{Direction of the magnetic field}:  We modelled the Hall-only and non-ideal MHD models using both signs of \bo.  For \boneg, the Hall effect resisted the momentum transport, and large discs formed.  In the Hall-only model with $\mu_0 = 5$, an $r\approx38$~AU disc formed.  In the non-ideal MHD model, an $r\approx$13~AU disc formed.  In both models, counter-rotating envelopes formed at $r \in (90,150)$ and $\gtrsim 150$~AU, respectively.

\item \emph{Cosmic ray ionisation rate}:  Our models are relatively insensitive to the cosmic ray ionisation rate.  The exceptions are the models that include the Hall effect and are initialised with \boneg, in which larger and more massive discs form in the models with lower cosmic ionisation rates.

\end{enumerate}

In the presence of strong magnetic fields, Ohmic resistivity and ambipolar diffusion cannot solve the magnetic braking catastrophe; as in ideal MHD models, large discs cannot form.  However, the Hall effect with \boneg \ can solve the magnetic braking catastrophe, allowing rotationally supported discs to form.  Thus, the direction of the magnetic field with respect to the rotation axis is important (e.g. \citealp{wardleng99}), as found in \citet{KrasnopolskyLiShang2011} and \citet{BraidingWardle2012accretion} and confirmed in our study.  Larger and more massive discs form for \boneg, while catastrophic magnetic braking may occur for \bopos.  Our results agree with the conclusions found in \citet{TsukamotoEtAl2015b}.

 As discussed in \citet{BraidingWardle2012sf}, targeted observations with telescopes such as ALMA should be able to determine the direction of the magnetic field with respect to the rotation axis of discs around newly forming stars.  Our results on the Hall effect suggest that a correlation between \bo \ and disc size should be observed.  

\section*{Acknowledgements}
We thank the referee for a prompt and thorough review, whose comments resulted in the improved quality of this manuscript.  This work was funded by an Australian Research Council (ARC) Discovery Projects Grant (DP130102078), including a Discovery International Award funding MRB's sabbatical at Monash.  DJP is funded by ARC Future Fellowship FT130100034. MRB also acknowledges support from the European Research Council under the European Community's Seventh Framework Programme (FP7/2007- 2013 grant agreement no. 339248).  We would like to thank Mark Wardle for helpful discussions and clarifications.   This work was supported by resources on the gSTAR national facility at Swinburne University of Technology. gSTAR is funded by Swinburne and the Australian Government's Education Investment Fund.  For the column density figures, we used {\sc splash} \citep{price07}. 

\bibliography{WPBbib}

\begin{thebibliography}{}

\bibitem[\protect\citeauthoryear{{Alexiades}, {Amiez} \& {Gremaud}}{{Alexiades}
  et~al.}{1996}]{AlexiadesEtAl96}
{Alexiades} V.,  {Amiez} G.,    {Gremaud} P.-A.,  1996, Commun. Numer. Meth.
  Eng., 12, 31

\bibitem[\protect\citeauthoryear{{Allen}, {Li} \& {Shu}}{{Allen}
  et~al.}{2003}]{als03}
{Allen} A.,  {Li} Z.-Y.,    {Shu} F.~H.,  2003, \apj, 599, 363

\bibitem[\protect\citeauthoryear{{Asplund}, {Grevesse}, {Sauval} \&
  {Scott}}{{Asplund} et~al.}{2009}]{AsplundEtAl2009}
{Asplund} M.,  {Grevesse} N.,  {Sauval} A.~J.,    {Scott} P.,  2009, \araa, 47,
  481

\bibitem[\protect\citeauthoryear{{Bai}}{{Bai}}{2014}]{Bai2014}
{Bai} X.-N.,  2014, \apj, 791, 137

\bibitem[\protect\citeauthoryear{{Bai}}{{Bai}}{2015}]{Bai2015}
{Bai} X.-N.,  2015, \apj, 798, 84

\bibitem[\protect\citeauthoryear{{Bate}, {Bonnell} \& {Price}}{{Bate}
  et~al.}{1995}]{bbp95}
{Bate} M.~R.,  {Bonnell} I.~A.,    {Price} N.~M.,  1995, \mnras, 277, 362

\bibitem[\protect\citeauthoryear{{Bate} \& {Burkert}}{{Bate} \&
  {Burkert}}{1997}]{bateburkert97}
{Bate} M.~R.,  {Burkert} A.,  1997, \mnras, 288, 1060

\bibitem[\protect\citeauthoryear{{Bate}, {Tricco} \& {Price}}{{Bate}
  et~al.}{2014}]{BateTriccoPrice2014}
{Bate} M.~R.,  {Tricco} T.~S.,    {Price} D.~J.,  2014, \mnras, 437, 77

\bibitem[\protect\citeauthoryear{{B{\o}rve}, {Omang} \& {Trulsen}}{{B{\o}rve}
  et~al.}{2001}]{bot01}
{B{\o}rve} S.,  {Omang} M.,    {Trulsen} J.,  2001, \apj, 561, 82

\bibitem[\protect\citeauthoryear{{B{\o}rve}, {Omang} \& {Trulsen}}{{B{\o}rve}
  et~al.}{2004}]{bot04}
{B{\o}rve} S.,  {Omang} M.,    {Trulsen} J.,  2004, \apjs, 153, 447

\bibitem[\protect\citeauthoryear{{Bourke}, {Myers}, {Robinson} \&
  {Hyland}}{{Bourke} et~al.}{2001}]{bourkeetal01}
{Bourke} T.~L.,  {Myers} P.~C.,  {Robinson} G.,    {Hyland} A.~R.,  2001, \apj,
  554, 916

\bibitem[\protect\citeauthoryear{{Braiding} \& {Wardle}}{{Braiding} \&
  {Wardle}}{2012a}]{BraidingWardle2012sf}
{Braiding} C.~R.,  {Wardle} M.,  2012a, \mnras, 422, 261

\bibitem[\protect\citeauthoryear{{Braiding} \& {Wardle}}{{Braiding} \&
  {Wardle}}{2012b}]{BraidingWardle2012accretion}
{Braiding} C.~R.,  {Wardle} M.,  2012b, \mnras, 427, 3188

\bibitem[\protect\citeauthoryear{{B{\"u}rzle}, {Clark}, {Stasyszyn}, {Dolag} \&
  {Klessen}}{{B{\"u}rzle} et~al.}{2011}]{burzleetal11a}
{B{\"u}rzle} F.,  {Clark} P.~C.,  {Stasyszyn} F.,  {Dolag} K.,    {Klessen}
  R.~S.,  2011, \mnras, 417, L61

\bibitem[\protect\citeauthoryear{{Choi}, {Kim} \& {Wiita}}{{Choi}
  et~al.}{2009}]{ckw09}
{Choi} E.,  {Kim} J.,    {Wiita} P.~J.,  2009, \apjs, 181, 413

\bibitem[\protect\citeauthoryear{{Commer{\c c}on}, {Hennebelle}, {Audit},
  {Chabrier} \& {Teyssier}}{{Commer{\c c}on} et~al.}{2010}]{commerconetal10}
{Commer{\c c}on} B.,  {Hennebelle} P.,  {Audit} E.,  {Chabrier} G.,
  {Teyssier} R.,  2010, \aap, 510, L3

\bibitem[\protect\citeauthoryear{{Commer{\c c}on}, {Teyssier}, {Audit},
  {Hennebelle} \& {Chabrier}}{{Commer{\c c}on}
  et~al.}{2011}]{CommerconEtAl2011}
{Commer{\c c}on} B.,  {Teyssier} R.,  {Audit} E.,  {Hennebelle} P.,
  {Chabrier} G.,  2011, \aap, 529, A35

\bibitem[\protect\citeauthoryear{{Crutcher}}{{Crutcher}}{1999}]{crutcher99}
{Crutcher} R.~M.,  1999, \apj, 520, 706

\bibitem[\protect\citeauthoryear{{Dapp} \& {Basu}}{{Dapp} \&
  {Basu}}{2010}]{DappBasu2010}
{Dapp} W.~B.,  {Basu} S.,  2010, \aap, 521, L56

\bibitem[\protect\citeauthoryear{{Dapp}, {Basu} \& {Kunz}}{{Dapp}
  et~al.}{2012}]{DappBasuKunz012}
{Dapp} W.~B.,  {Basu} S.,    {Kunz} M.~W.,  2012, \aap, 541, A35

\bibitem[\protect\citeauthoryear{{Draine}}{{Draine}}{1980}]{Draine80}
{Draine} B.~T.,  1980, \apj, 241, 1021

\bibitem[\protect\citeauthoryear{{Duffin} \& {Pudritz}}{{Duffin} \&
  {Pudritz}}{2009}]{duffinpudritz09}
{Duffin} D.~F.,  {Pudritz} R.~E.,  2009, \apjl, 706, L46

\bibitem[\protect\citeauthoryear{{Dunham}, {Chen}, {Arce}, {Bourke}, {Schnee}
  \& {Enoch}}{{Dunham} et~al.}{2011}]{dunhametal11}
{Dunham} M.~M.,  {Chen} X.,  {Arce} H.~G.,  {Bourke} T.~L.,  {Schnee} S.,
  {Enoch} M.~L.,  2011, \apj, 742, 1

\bibitem[\protect\citeauthoryear{{Falle}}{{Falle}}{2003}]{Falle2003}
{Falle} S.~A.~E.~G.,  2003, \mnras, 344, 1210

\bibitem[\protect\citeauthoryear{{Fujii}, {Okuzumi} \& {Inutsuka}}{{Fujii}
  et~al.}{2011}]{foi11}
{Fujii} Y.~I.,  {Okuzumi} S.,    {Inutsuka} S.-i.,  2011, \apj, 743, 53

\bibitem[\protect\citeauthoryear{{Gafton} \& {Rosswog}}{{Gafton} \&
  {Rosswog}}{2011}]{gaftonrosswog11}
{Gafton} E.,  {Rosswog} S.,  2011, \mnras, 418, 770

\bibitem[\protect\citeauthoryear{{Galli}, {Lizano}, {Shu} \& {Allen}}{{Galli}
  et~al.}{2006}]{gallietal06}
{Galli} D.,  {Lizano} S.,  {Shu} F.~H.,    {Allen} A.,  2006, \apj, 647, 374

\bibitem[\protect\citeauthoryear{{Heiles} \& {Crutcher}}{{Heiles} \&
  {Crutcher}}{2005}]{HeilesCrutcher2005}
{Heiles} C.,  {Crutcher} R.,  2005, in {Wielebinski} R.,  {Beck} R.,  eds,
  Cosmic Magnetic Fields Vol.~664 of Lecture Notes in Physics, Berlin Springer
  Verlag, {Magnetic Fields in Diffuse HI and Molecular Clouds}.
p.~137

\bibitem[\protect\citeauthoryear{{Hennebelle} \& {Ciardi}}{{Hennebelle} \&
  {Ciardi}}{2009}]{hennebelleciardi09}
{Hennebelle} P.,  {Ciardi} A.,  2009, \aap, 506, L29

\bibitem[\protect\citeauthoryear{{Hennebelle} \& {Fromang}}{{Hennebelle} \&
  {Fromang}}{2008}]{hennebellefromang08}
{Hennebelle} P.,  {Fromang} S.,  2008, \aap, 477, 9

\bibitem[\protect\citeauthoryear{{Joos}, {Hennebelle}, {Ciardi} \&
  {Fromang}}{{Joos} et~al.}{2013}]{JoosEtAl2013}
{Joos} M.,  {Hennebelle} P.,  {Ciardi} A.,    {Fromang} S.,  2013, \aap, 554,
  A17

\bibitem[\protect\citeauthoryear{{Keith} \& {Wardle}}{{Keith} \&
  {Wardle}}{2014}]{keithwardle14}
{Keith} S.~L.,  {Wardle} M.,  2014, \mnras, 440, 89

\bibitem[\protect\citeauthoryear{{Krasnopolsky}, {Li} \&
  {Shang}}{{Krasnopolsky} et~al.}{2010}]{kls10}
{Krasnopolsky} R.,  {Li} Z.-Y.,    {Shang} H.,  2010, \apj, 716, 1541

\bibitem[\protect\citeauthoryear{{Krasnopolsky}, {Li} \&
  {Shang}}{{Krasnopolsky} et~al.}{2011}]{KrasnopolskyLiShang2011}
{Krasnopolsky} R.,  {Li} Z.-Y.,    {Shang} H.,  2011, \apj, 733, 54

\bibitem[\protect\citeauthoryear{{Larson}}{{Larson}}{1969}]{Larson1969}
{Larson} R.~B.,  1969, \mnras, 145, 271

\bibitem[\protect\citeauthoryear{{Lewis}, {Bate} \& {Price}}{{Lewis}
  et~al.}{2015}]{LewisBatePrice2015}
{Lewis} B.~T.,  {Bate} M.~R.,    {Price} D.~J.,  2015, \mnras, 451, 288

\bibitem[\protect\citeauthoryear{{Li}, {Fang}, {Henning} \& {Kainulainen}}{{Li}
  et~al.}{2013}]{LiFangHenningKainulainen2013}
{Li} H.-b.,  {Fang} M.,  {Henning} T.,    {Kainulainen} J.,  2013, \mnras, 436,
  3707

\bibitem[\protect\citeauthoryear{{Li}, {Krasnopolsky} \& {Shang}}{{Li}
  et~al.}{2011}]{LiKrasnopolskyShang2011}
{Li} Z.-Y.,  {Krasnopolsky} R.,    {Shang} H.,  2011, \apj, 738, 180

\bibitem[\protect\citeauthoryear{{Li}, {Krasnopolsky}, {Shang} \& {Zhao}}{{Li}
  et~al.}{2014}]{lietal14}
{Li} Z.-Y.,  {Krasnopolsky} R.,  {Shang} H.,    {Zhao} B.,  2014, \apj, 793,
  130

\bibitem[\protect\citeauthoryear{{Lindberg}, {J{\o}rgensen}, {Brinch},
  {Haugb{\o}lle}, {Bergin}, {Harsono}, {Persson}, {Visser} \&
  {Yamamoto}}{{Lindberg} et~al.}{2014}]{LindbergEtAl2014}
{Lindberg} J.~E.,  {J{\o}rgensen} J.~K.,  {Brinch} C.,  {Haugb{\o}lle} T.,
  {Bergin} E.~A.,  {Harsono} D.,  {Persson} M.~V.,  {Visser} R.,    {Yamamoto}
  S.,  2014, \aap, 566, A74

\bibitem[\protect\citeauthoryear{{Liu}, {Goree}, {Nosenko} \& {Boufendi}}{{Liu}
  et~al.}{2003}]{LiuEtAl2003}
{Liu} B.,  {Goree} J.,  {Nosenko} V.,    {Boufendi} L.,  2003, Physics of
  Plasmas, 10, 9

\bibitem[\protect\citeauthoryear{{Mac Low}, {Norman}, {Konigl} \&
  {Wardle}}{{Mac Low} et~al.}{1995}]{MNKW95}
{Mac Low} M.-M.,  {Norman} M.~L.,  {Konigl} A.,    {Wardle} M.,  1995, \apj,
  442, 726

\bibitem[\protect\citeauthoryear{{Machida}, {Inutsuka} \&
  {Matsumoto}}{{Machida} et~al.}{2008}]{MachidaInutsukaMatsumoto2008}
{Machida} M.~N.,  {Inutsuka} S.-i.,    {Matsumoto} T.,  2008, \apj, 676, 1088

\bibitem[\protect\citeauthoryear{{Machida}, {Inutsuka} \&
  {Matsumoto}}{{Machida} et~al.}{2011}]{MachidaInutsukaMatsumoto2011}
{Machida} M.~N.,  {Inutsuka} S.-I.,    {Matsumoto} T.,  2011, \pasj, 63, 555

\bibitem[\protect\citeauthoryear{{Machida}, {Inutsuka} \&
  {Matsumoto}}{{Machida} et~al.}{2014}]{MachidaInutsukaMatsumoto2014}
{Machida} M.~N.,  {Inutsuka} S.-i.,    {Matsumoto} T.,  2014, \mnras, 438, 2278

\bibitem[\protect\citeauthoryear{{Machida}, {Matsumoto}, {Hanawa} \&
  {Tomisaka}}{{Machida} et~al.}{2006}]{machidaetal06}
{Machida} M.~N.,  {Matsumoto} T.,  {Hanawa} T.,    {Tomisaka} K.,  2006, \apj,
  645, 1227

\bibitem[\protect\citeauthoryear{{Machida}, {Tomisaka} \&
  {Matsumoto}}{{Machida} et~al.}{2004}]{MachidaTomisakaMatsumoto2004}
{Machida} M.~N.,  {Tomisaka} K.,    {Matsumoto} T.,  2004, \mnras, 348, L1

\bibitem[\protect\citeauthoryear{{Machida}, {Tomisaka}, {Matsumoto} \&
  {Inutsuka}}{{Machida} et~al.}{2008}]{machidaetal08}
{Machida} M.~N.,  {Tomisaka} K.,  {Matsumoto} T.,    {Inutsuka} S.-i.,  2008,
  \apj, 677, 327

\bibitem[\protect\citeauthoryear{{Masunaga} \& {Inutsuka}}{{Masunaga} \&
  {Inutsuka}}{2000}]{MasunagaInutsuka2000}
{Masunaga} H.,  {Inutsuka} S.-i.,  2000, \apj, 531, 350

\bibitem[\protect\citeauthoryear{{Mellon} \& {Li}}{{Mellon} \&
  {Li}}{2008}]{mellonli08}
{Mellon} R.~R.,  {Li} Z.-Y.,  2008, \apj, 681, 1356

\bibitem[\protect\citeauthoryear{{Mellon} \& {Li}}{{Mellon} \&
  {Li}}{2009}]{MellonLi2009}
{Mellon} R.~R.,  {Li} Z.-Y.,  2009, \apj, 698, 922

\bibitem[\protect\citeauthoryear{{Mestel} \& {Spitzer} Jr.}{{Mestel} \&
  {Spitzer}}{1956}]{mestelspitzer56}
{Mestel} L.,  {Spitzer} Jr. L.,  1956, \mnras, 116, 503

\bibitem[\protect\citeauthoryear{{Monaghan}}{{Monaghan}}{2002}]{monaghan02}
{Monaghan} J.~J.,  2002, \mnras, 335, 843

\bibitem[\protect\citeauthoryear{{Morales Ortiz}, {Ceccarelli}, {Lis}, {Olmi},
  {Plume} \& {Schilke}}{{Morales Ortiz} et~al.}{2014}]{MoralesortizEtAl2014}
{Morales Ortiz} J.~L.,  {Ceccarelli} C.,  {Lis} D.~C.,  {Olmi} L.,  {Plume} R.,
     {Schilke} P.,  2014, \aap, 563, A127

\bibitem[\protect\citeauthoryear{{Morris}}{{Morris}}{1996}]{morris96}
{Morris} J.~P.,  1996, \pasa, 13, 97

\bibitem[\protect\citeauthoryear{{Mouschovias} \& {Spitzer} Jr.}{{Mouschovias}
  \& {Spitzer}}{1976}]{mouschoviasspitzer76}
{Mouschovias} T.~C.,  {Spitzer} Jr. L.,  1976, \apj, 210, 326

\bibitem[\protect\citeauthoryear{{Myers} \& {Goodman}}{{Myers} \&
  {Goodman}}{1988}]{MyersGoodman1988}
{Myers} P.~C.,  {Goodman} A.~A.,  1988, \apjl, 326, L27

\bibitem[\protect\citeauthoryear{{Nakano}, {Nishi} \& {Umebayashi}}{{Nakano}
  et~al.}{2002}]{nnu02}
{Nakano} T.,  {Nishi} R.,    {Umebayashi} T.,  2002, \apj, 573, 199

\bibitem[\protect\citeauthoryear{{Nakano} \& {Umebayashi}}{{Nakano} \&
  {Umebayashi}}{1986}]{nakanoumebayashi86}
{Nakano} T.,  {Umebayashi} T.,  1986, \mnras, 218, 663

\bibitem[\protect\citeauthoryear{{Osterbrock}}{{Osterbrock}}{1961}]{Osterbrock1961}
{Osterbrock} D.~E.,  1961, \apj, 134, 270

\bibitem[\protect\citeauthoryear{{O'Sullivan} \& {Downes}}{{O'Sullivan} \&
  {Downes}}{2006}]{OsullivanDownes06}
{O'Sullivan} S.,  {Downes} T.~P.,  2006, \mnras, 366, 1329

\bibitem[\protect\citeauthoryear{{Pandey} \& {Wardle}}{{Pandey} \&
  {Wardle}}{2008}]{pandeywardle08}
{Pandey} B.~P.,  {Wardle} M.,  2008, \mnras, 385, 2269

\bibitem[\protect\citeauthoryear{{Pinto} \& {Galli}}{{Pinto} \&
  {Galli}}{2008}]{pintogalli08}
{Pinto} C.,  {Galli} D.,  2008, \aap, 484, 17

\bibitem[\protect\citeauthoryear{{Pollack}, {Hollenbach}, {Beckwith},
  {Simonelli}, {Roush} \& {Fong}}{{Pollack} et~al.}{1994}]{PollackEtAl1994}
{Pollack} J.~B.,  {Hollenbach} D.,  {Beckwith} S.,  {Simonelli} D.~P.,  {Roush}
  T.,    {Fong} W.,  1994, \apj, 421, 615

\bibitem[\protect\citeauthoryear{{Price}}{{Price}}{2007}]{price07}
{Price} D.~J.,  2007, \pasa, 24, 159

\bibitem[\protect\citeauthoryear{{Price}}{{Price}}{2010}]{price10}
{Price} D.~J.,  2010, \mnras, 401, 1475

\bibitem[\protect\citeauthoryear{{Price}}{{Price}}{2012}]{price12}
{Price} D.~J.,  2012, J. Comp. Phys., 231, 759

\bibitem[\protect\citeauthoryear{{Price} \& {Bate}}{{Price} \&
  {Bate}}{2007}]{pricebate07}
{Price} D.~J.,  {Bate} M.~R.,  2007, \mnras, 377, 77

\bibitem[\protect\citeauthoryear{{Price} \& {Federrath}}{{Price} \&
  {Federrath}}{2010}]{pricefederrath10}
{Price} D.~J.,  {Federrath} C.,  2010, \mnras, 406, 1659

\bibitem[\protect\citeauthoryear{{Price} \& {Monaghan}}{{Price} \&
  {Monaghan}}{2004}]{pricemonaghan04}
{Price} D.~J.,  {Monaghan} J.~J.,  2004, \mnras, 348, 123

\bibitem[\protect\citeauthoryear{{Price} \& {Monaghan}}{{Price} \&
  {Monaghan}}{2005}]{pricemonaghan05}
{Price} D.~J.,  {Monaghan} J.~J.,  2005, \mnras, 364, 384

\bibitem[\protect\citeauthoryear{{Price} \& {Monaghan}}{{Price} \&
  {Monaghan}}{2007}]{pricemonaghan07}
{Price} D.~J.,  {Monaghan} J.~J.,  2007, \mnras, 374, 1347

\bibitem[\protect\citeauthoryear{{Price}, {Tricco} \& {Bate}}{{Price}
  et~al.}{2012}]{PriceTriccoBate2012}
{Price} D.~J.,  {Tricco} T.~S.,    {Bate} M.~R.,  2012, \mnras, 423, L45

\bibitem[\protect\citeauthoryear{{Sano} \& {Stone}}{{Sano} \&
  {Stone}}{2002}]{sanostone02}
{Sano} T.,  {Stone} J.~M.,  2002, \apj, 570, 314

\bibitem[\protect\citeauthoryear{{Santos-Lima}, {de Gouveia Dal Pino} \&
  {Lazarian}}{{Santos-Lima} et~al.}{2012}]{SantoslimaEtAl2012}
{Santos-Lima} R.,  {de Gouveia Dal Pino} E.~M.,    {Lazarian} A.,  2012, \apj,
  747, 21

\bibitem[\protect\citeauthoryear{{Santos-Lima}, {de Gouveia Dal Pino} \&
  {Lazarian}}{{Santos-Lima} et~al.}{2013}]{SantoslimaEtAl2013}
{Santos-Lima} R.,  {de Gouveia Dal Pino} E.~M.,    {Lazarian} A.,  2013,
  \mnras, 429, 3371

\bibitem[\protect\citeauthoryear{{Seifried}, {Banerjee}, {Klessen}, {Duffin} \&
  {Pudritz}}{{Seifried} et~al.}{2011}]{SeifriedEtAl2011}
{Seifried} D.,  {Banerjee} R.,  {Klessen} R.~S.,  {Duffin} D.,    {Pudritz}
  R.~E.,  2011, \mnras, 417, 1054

\bibitem[\protect\citeauthoryear{{Seifried}, {Banerjee}, {Pudritz} \&
  {Klessen}}{{Seifried} et~al.}{2012}]{SeifriedBanerjeePudritzKlessen2012}
{Seifried} D.,  {Banerjee} R.,  {Pudritz} R.~E.,    {Klessen} R.~S.,  2012,
  \mnras, 423, L40

\bibitem[\protect\citeauthoryear{{Seifried}, {Banerjee}, {Pudritz} \&
  {Klessen}}{{Seifried} et~al.}{2013}]{SeifriedBanerjeePudritzKlessen2013}
{Seifried} D.,  {Banerjee} R.,  {Pudritz} R.~E.,    {Klessen} R.~S.,  2013,
  \mnras, 432, 3320

\bibitem[\protect\citeauthoryear{{Shu}, {Galli}, {Lizano} \& {Cai}}{{Shu}
  et~al.}{2006}]{shuetal06}
{Shu} F.~H.,  {Galli} D.,  {Lizano} S.,    {Cai} M.,  2006, \apj, 647, 382

\bibitem[\protect\citeauthoryear{{Springel} \& {Hernquist}}{{Springel} \&
  {Hernquist}}{2002}]{springelhernquist02}
{Springel} V.,  {Hernquist} L.,  2002, \mnras, 333, 649

\bibitem[\protect\citeauthoryear{{Tassis} \& {Mouschovias}}{{Tassis} \&
  {Mouschovias}}{2007}]{tassismouschovias07a}
{Tassis} K.,  {Mouschovias} T.~C.,  2007, \apj, 660, 388

\bibitem[\protect\citeauthoryear{{Tobin}, {Looney}, {Wilner}, {Kwon},
  {Chandler}, {Bourke}, {Loinard}, {Chiang}, {Schnee} \& {Chen}}{{Tobin}
  et~al.}{2015}]{TobinEtAl2015}
{Tobin} J.~J.,  {Looney} L.~W.,  {Wilner} D.~J.,  {Kwon} W.,  {Chandler} C.~J.,
   {Bourke} T.~L.,  {Loinard} L.,  {Chiang} H.-F.,  {Schnee} S.,    {Chen} X.,
  2015, \apj, 805, 125

\bibitem[\protect\citeauthoryear{{Tomida}, {Okuzumi} \& {Machida}}{{Tomida}
  et~al.}{2015}]{TomidaOkuzumiMachida2015}
{Tomida} K.,  {Okuzumi} S.,    {Machida} M.~N.,  2015, \apj, 801, 117

\bibitem[\protect\citeauthoryear{{Tomida}, {Tomisaka}, {Matsumoto}, {Hori},
  {Okuzumi}, {Machida} \& {Saigo}}{{Tomida} et~al.}{2013}]{tomidaetal13}
{Tomida} K.,  {Tomisaka} K.,  {Matsumoto} T.,  {Hori} Y.,  {Okuzumi} S.,
  {Machida} M.~N.,    {Saigo} K.,  2013, \apj, 763, 6

\bibitem[\protect\citeauthoryear{{Tomisaka}}{{Tomisaka}}{1998}]{Tomisaka1998}
{Tomisaka} K.,  1998, \apjl, 502, L163

\bibitem[\protect\citeauthoryear{{Tomisaka}}{{Tomisaka}}{2002}]{Tomisaka2002}
{Tomisaka} K.,  2002, \apj, 575, 306

\bibitem[\protect\citeauthoryear{{Tomisaka}, {Machida} \&
  {Matsumoto}}{{Tomisaka} et~al.}{2004}]{TomisakaMachidaMatsumoto2004}
{Tomisaka} K.,  {Machida} M.~N.,    {Matsumoto} T.,  2004, \apss, 292, 309

\bibitem[\protect\citeauthoryear{{Tricco} \& {Price}}{{Tricco} \&
  {Price}}{2012}]{triccoprice12}
{Tricco} T.~S.,  {Price} D.~J.,  2012, J. Comp. Phys., 231, 7214

\bibitem[\protect\citeauthoryear{{Tricco} \& {Price}}{{Tricco} \&
  {Price}}{2013}]{TriccoPrice2013}
{Tricco} T.~S.,  {Price} D.~J.,  2013, \mnras, 436, 2810

\bibitem[\protect\citeauthoryear{{Tsukamoto}, {Iwasaki} \&
  {Inutsuka}}{{Tsukamoto} et~al.}{2013}]{tii13}
{Tsukamoto} Y.,  {Iwasaki} K.,    {Inutsuka} S.-i.,  2013, \mnras, 434, 2593

\bibitem[\protect\citeauthoryear{{Tsukamoto}, {Iwasaki}, {Okuzumi}, {Machida}
  \& {Inutsuka}}{{Tsukamoto} et~al.}{2015a}]{TsukamotoEtAl2015b}
{Tsukamoto} Y.,  {Iwasaki} K.,  {Okuzumi} S.,  {Machida} M.~N.,    {Inutsuka}
  S.,  2015a, \apjl, 810, L26

\bibitem[\protect\citeauthoryear{{Tsukamoto}, {Iwasaki}, {Okuzumi}, {Machida}
  \& {Inutsuka}}{{Tsukamoto} et~al.}{2015b}]{TsukamotoEtAl2015}
{Tsukamoto} Y.,  {Iwasaki} K.,  {Okuzumi} S.,  {Machida} M.~N.,    {Inutsuka}
  S.,  2015b, \mnras, 452, 278

\bibitem[\protect\citeauthoryear{{Umebayashi} \& {Nakano}}{{Umebayashi} \&
  {Nakano}}{1980}]{UmebayashiNakano1980}
{Umebayashi} T.,  {Nakano} T.,  1980, \pasj, 32, 405

\bibitem[\protect\citeauthoryear{{Umebayashi} \& {Nakano}}{{Umebayashi} \&
  {Nakano}}{1990}]{umebayashinakano90}
{Umebayashi} T.,  {Nakano} T.,  1990, \mnras, 243, 103

\bibitem[\protect\citeauthoryear{{Umebayashi} \& {Nakano}}{{Umebayashi} \&
  {Nakano}}{2009}]{UmebayashiNakano2009}
{Umebayashi} T.,  {Nakano} T.,  2009, \apj, 690, 69

\bibitem[\protect\citeauthoryear{{Wardle}}{{Wardle}}{2007}]{Wardle2007}
{Wardle} M.,  2007, \apss, 311, 35

\bibitem[\protect\citeauthoryear{{Wardle} \& {Ng}}{{Wardle} \&
  {Ng}}{1999}]{wardleng99}
{Wardle} M.,  {Ng} C.,  1999, \mnras, 303, 239

\bibitem[\protect\citeauthoryear{{Wurster}, {Price} \& {Ayliffe}}{{Wurster}
  et~al.}{2014}]{WPA2014}
{Wurster} J.,  {Price} D.,    {Ayliffe} B.,  2014, \mnras, 444, 1104

\end{thebibliography}
\appendix
\section{Equation of state}
\subsection{Discontinuous temperatures}
\label{app:eos:dis} 
\begin{figure}
\begin{center}
\includegraphics[width=\columnwidth]{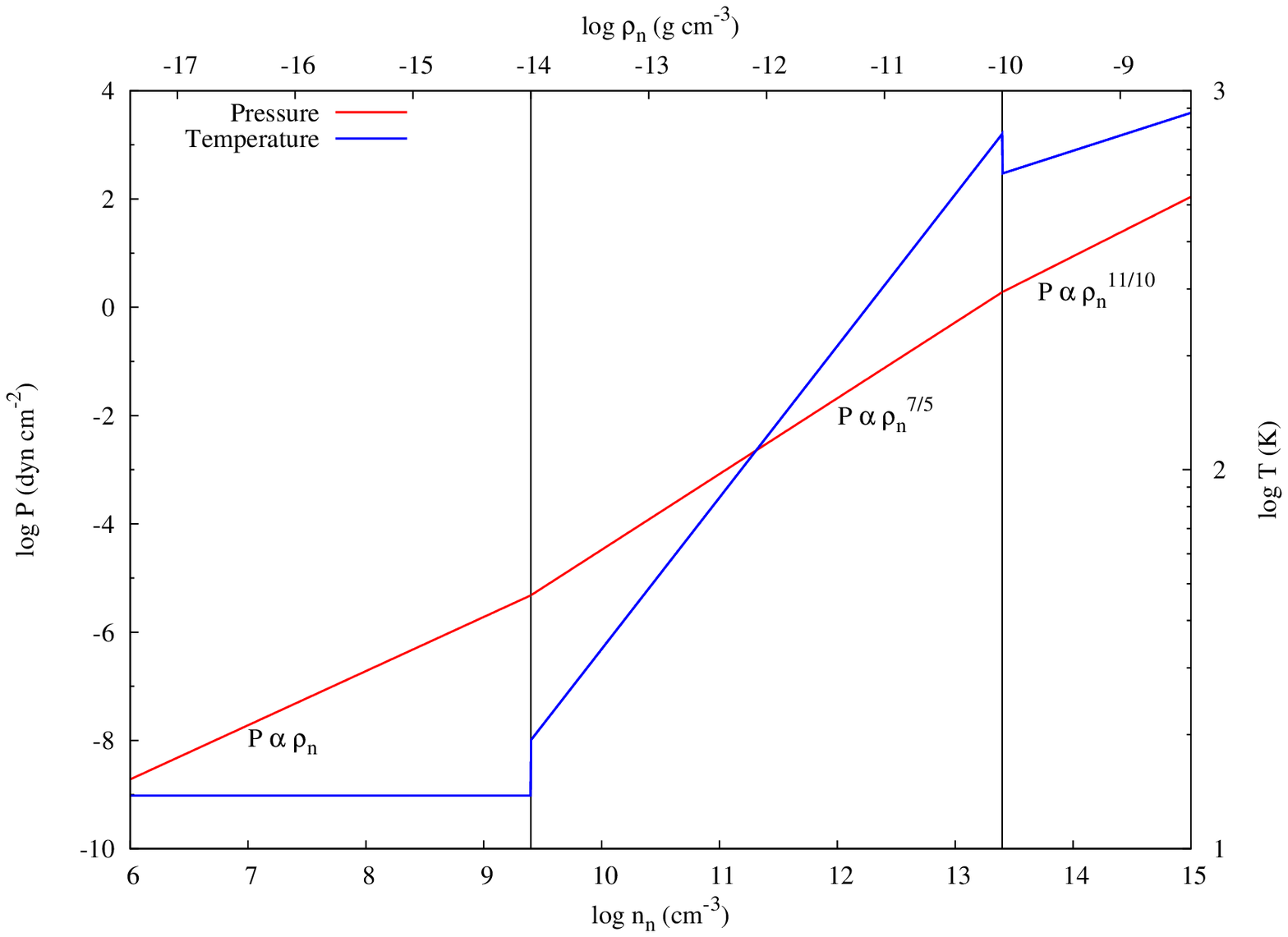}
\caption{Pressure (red) and temperature (blue) resulting from the barotropic equation of state (Eqn.~\ref{eq:eos}).  The vertical lines are the threshold densities, $\rho_\text{d}$ and $\rho_\text{c}$.  The pressure is a continuous function of density, while temperature is discontinuous at the threshold densities.}
\label{fig:PT}
\end{center}
\end{figure}

The barotropic equation of state given in \eqref{eq:eos} yields a continuous pressure, shown as the red curve in Fig.~\ref{fig:PT}.  
The local sound speed is given by
\begin{equation}
\label{app:cs}
c_s = \sqrt{\gamma\frac{P}{\rho}},
\end{equation}
where
\begin{equation}
\gamma = \left\{ \begin{array}{l l} 1;      &  \rho < \rho_\text{c}, \\
                                                  7/5;      &  \rho_\text{c} \leq \rho < \rho_\text{d}, \\
                                                  11/10;  &  \rho \geq \rho_\text{d}.
\end{array}\right.
\end{equation}
This local sound speed is converted into temperature using 
\begin{equation}
\label{eq:ap:T}
T = \frac{c_\text{s}^2 \mu m_\text{p}}{k_\text{B}},
\end{equation}
where $\mu$ is the mean molecular mass, $m_\text{p}$ is the proton mass, and $k_\text{B}$ is the Boltzmann constant.  It is this temperature that is used in the ionisation calculations, and is shown as the blue curve in Fig.~\ref{fig:PT}.  The discontinuities in temperature at the threshold densities are thus responsible for the discontinuities in grain charge, number densities, conductivities and resistivities in our test cases in Sections \ref{ssec:num:ionise} and \ref{ssec:num:conduct}.  Future work will test the effect of modifying \eqref{eq:ap:T} to be a continuous function of $n_\text{n}$.

\subsection{Isothermal equation of state}
\label{app:eos}
\begin{figure*}
\begin{center}
\includegraphics[width=0.75\textwidth]{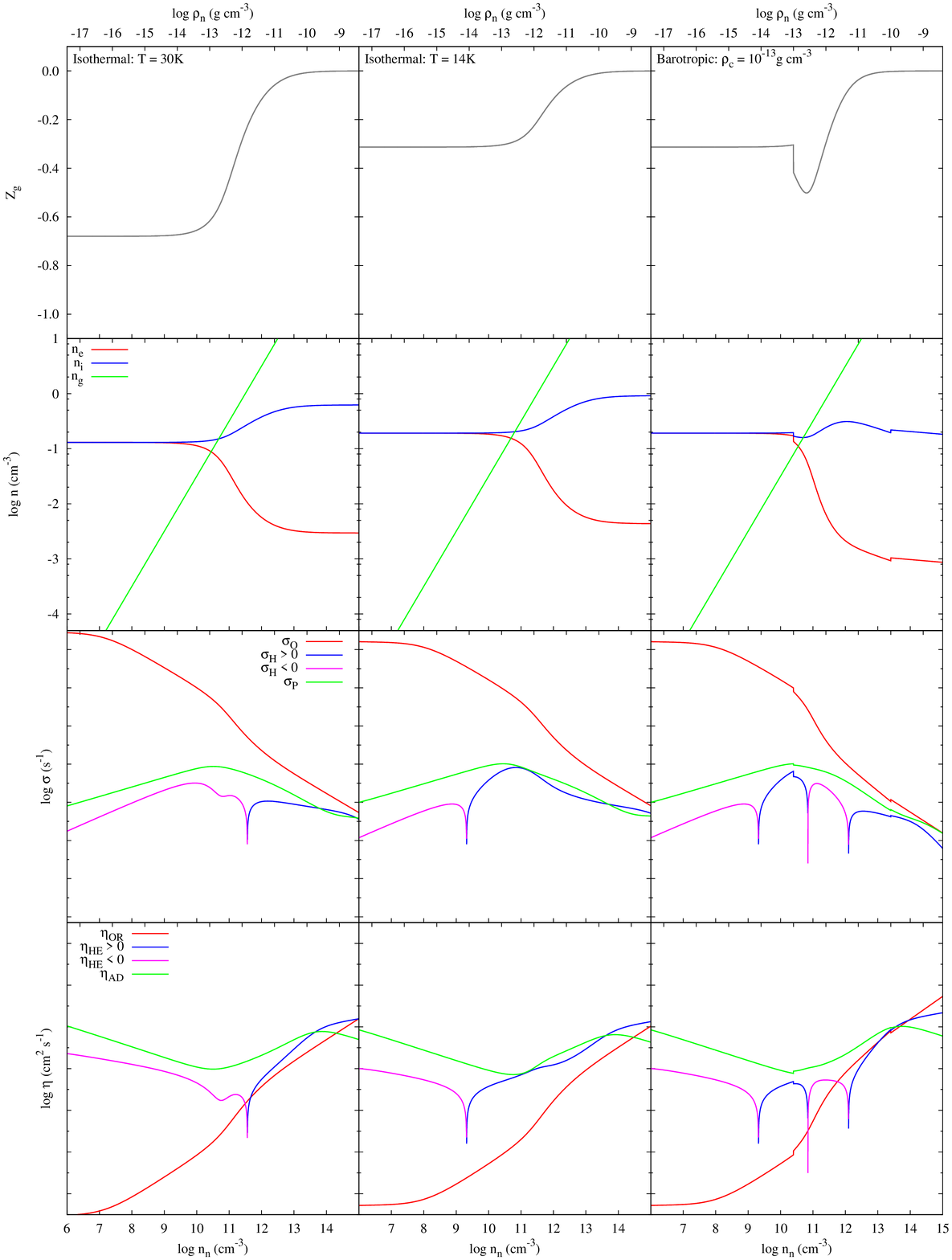}
\caption{\emph{Top to Bottom}: Grain charge, number densities, conductivities and resistivity coefficients for the isothermal equations of state.  In all cases, \zetaos.  The top ticks on each panel correspond to mass density (top scale), and the bottom ticks correspond to number density (bottom scale).  This is comparable to Fig.~\ref{fig:AllBaro}.  \emph{Left column}: Isothermal equation of state with $T = 30$ K to match \citet{wardleng99}.  The middle two panels are directly comparable to their fig. 1. \emph{Middle column}: Isothermal equation of state with $T = 14$ K to match our temperature for $\rho < \rho_\text{c} = 10^{-14}$ g cm$^{-3}$.  \emph{Right column}: Barotropic equation of state with $\rho_\text{c} = 10^{-13}$ g cm$^{-3}$.}
\label{fig:AllIso}
\end{center}
\end{figure*}

As shown in Sections \ref{ssec:num:ionise} and \ref{ssec:num:conduct}, the grain charge, species number densities, conductivities and resistivity coefficients are dependent on the value of $\zeta$.  Where possible, we have obtained the remaining parameters from experimental values, however the specific values will necessarily affect the results.  In this appendix we will briefly discuss the effect of the equation of state.  For our simulations, we have chosen a barotropic equation of state (c.f. Eqn.~\ref{eq:eos}) to mimic the expected equation of state in a star formation scenario without requiring a full radiative treatment.  In \citet{wardleng99} (herein WN99), an isothermal equation of state is used with $T=30$~K.  In the first column of Fig.~\ref{fig:AllIso}, we plot the grain charge, number densities of the charges species, conductivities, $\sigma$, and resistivity coefficients, $\eta$ for the isothermal equation of state.  For $\rho < \rho_\text{c} = 10^{-14}$~g~cm$^{-3}$, our equation of state assumes an isothermal temperature of $T=14$ K, which is plotted in the second column.  

At $T = 30$ K, our calculation of number densities, specifically the grain number density, varies from WN99, who obtain their values from \citet{umebayashinakano90}.  With the isothermal equation of state, the ion and electron number densities at $n_\text{n}~\sim10^{15}$~cm$^{-3}$  agree within a factor of $\sim$2 and $n_\text{i}$ and $n_\text{e}$ diverge at approximately the same neutral number density.  For low densities, our ion and electron number densities are constant whereas they slightly decline in WN99.  This discrepancy is likely a result of the grain charge:  Our grain charge is given in the top panel, while WN99's charge is $\pm1$.  Given the interdependency of number density, temperature and a grain charge in our algorithms, we can choose two values and then must self-consistently calculate the third.  

The Hall conductivity is the conductivity that is most modified by the equation of state.  By using an isothermal equation of state, the change of sign is at $n_\text{n} \approx 3.6\times 10^{11}$~cm$^{-3}$ compared to $\approx$$1.5\times 10^{12}$~cm$^{-3}$ for the barotropic equation of state.  As for \zetaos, the isothermal case does not have a second change of sign at lower densities.  The Pedersen conductivity is similar for both equations of state.  The Ohmic conductivity can vary by a factor of ten, and the conductivity calculated with the isothermal equation of state is in better agreement to the values presented in WN99.

\subsection{Barotropic equation of state with $\rho_\text{c} = 10^{-13}$ g cm$^{-3}$}
\label{app:Beos}
In the right-hand column of Fig.~\ref{fig:AllIso}, we plot the grain charge, number densities of the charges species, conductivities, $\sigma$, and resistivity coefficients, $\eta$ for the barotropic equation of state using $\rho_\text{c} = 10^{-13}$ g cm$^{-3}$.  The result is similar to Fig.~\ref{fig:AllBaro}, where  $\rho_\text{c} = 10^{-14}$ g cm$^{-3}$.  With the exception of grain charge, $\sigma_\text{H}$ and $\eta_\text{H}$, the values calculated by both values of $\rho_\text{c}$ are typically differ by less than a factor of two.  The grain charge does not become as negative with $\rho_\text{c} = 10^{-13}$ g cm$^{-3}$.  There is a third density at which  $\sigma_\text{H}, \  \eta_\text{H} \rightarrow 0$; moreover, these values remain negative below $n_\text{n} \approx 1.5\times 10^{12}$~cm$^{-3}$.
                
\section{The modified Hall parameter}
\label{app:beta}
\begin{figure}
\begin{center}
\includegraphics[width=\columnwidth]{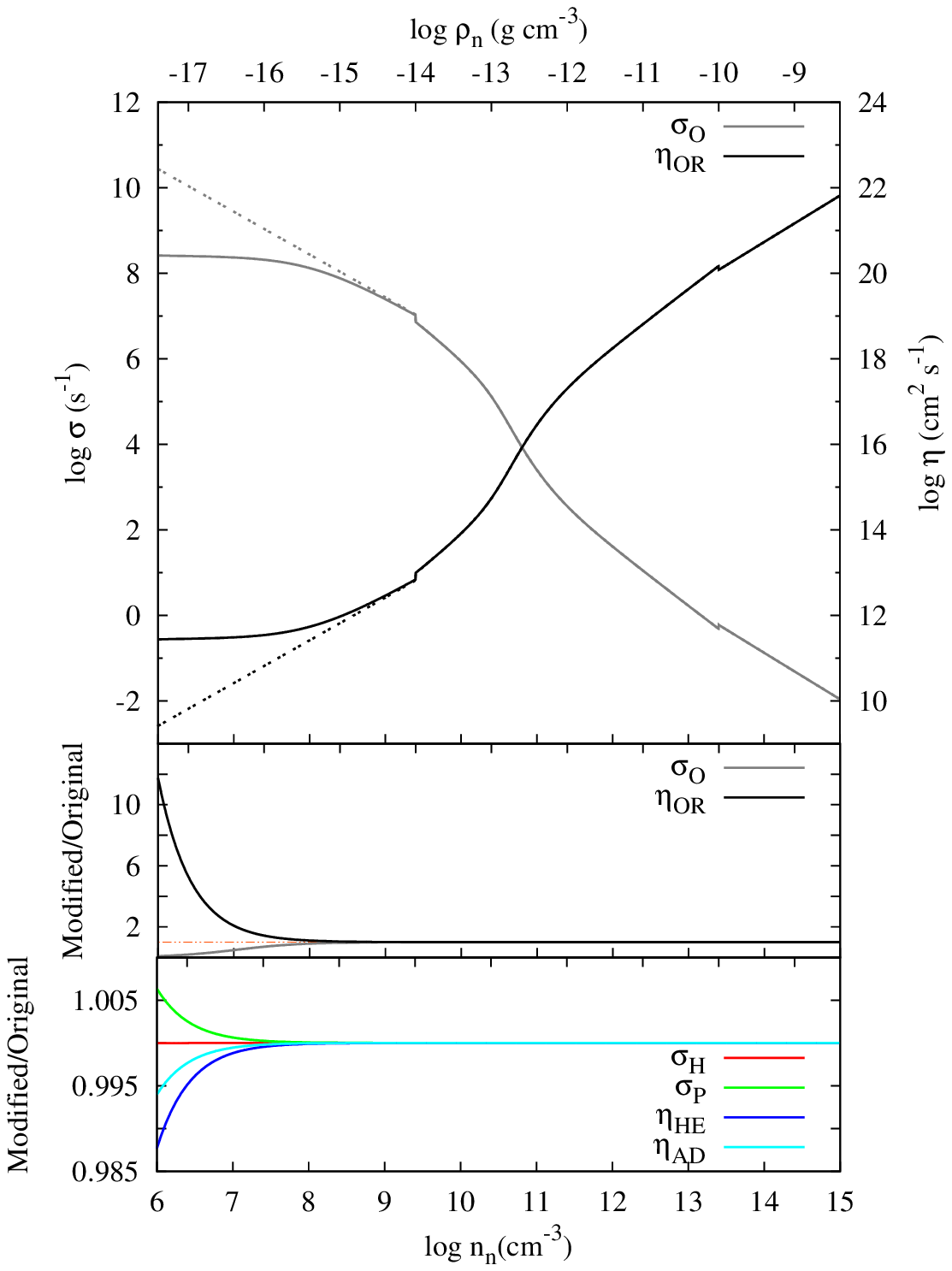}
\caption{\emph{Top}: Ohmic conductivity and resistivity, using the modified Hall parameters used in all calculations (solid) and the original Hall parameters (dashed).  In all cases, \zetaos.  The vertical range is different than in Figs.~\ref{fig:AllBaro} and \ref{fig:AllIso} to avoid truncating the curves.  \emph{Middle and bottom}: Ratios of the conductivities and resistivities calculated using  the modified Hall parameters to that using the original Hall parameters.  Note that the bottom two panels have different vertical scales from one another.  The top ticks on each panel correspond to mass density (top scale), and the bottom ticks correspond to number density (bottom scale).    At the lower end of the given number density range, Ohmic conductivity and resistivity differ significantly depending on the Hall parameters; the remaining conductivities and resistivities are relatively unaffected by the form of the Hall parameters.  At the densities where the choice of the Hall parameter affects the Ohmic conductivity and resistivity, Ohmic resistivity is the weakest non-ideal MHD effect by at least seven orders of magnitude.}
\label{fig:beta}
\end{center}
\end{figure}

The Hall parameter is defined in \eqref{eq:beta}, and our modified parameters are given in \eqref{eq:betamod}.  The conductivities, $\sigma$, and coefficients, $\eta$, are unaffected by the form of the Hall parameter for $n_\text{n} \gtrsim 10^9$~cm$^{-3}$.  Below this density, the values of the conductivities and coefficients calculated with the modified Hall parameters begin to diverge from their unmodified counterparts.  The maximum difference between the versions of $\sigma_\text{H}$, $\sigma_\text{P}$, $\eta_\text{HE}$ and $\eta_\text{AD}$ is 1.5 per cent, occurring at the minimum density tested, $n_\text{n} = 10^6$~cm$^{-3}$.  Both $\sigma_\text{O}$ and $\eta_\text{OR}$ yield larger divergences, as plotted in the top panel of Fig.~\ref{fig:beta}.  At $n_\text{n} = 10^6$~cm$^{-3}$, the two forms of $\sigma_\text{O}$ and $\eta_\text{OR}$ differ by a factor of $\sim$12.  Although the modified $\eta_\text{OR}$ has a larger value than its unmodified counterpart, both modified and unmodified values are more than seven orders of magnitude lower than the resistivities for the Hall effect or ambipolar diffusion.  Thus, the choice of the Hall parameter is not important for star formation.  For consistency with the $\eta_\text{OR}$ presented in \citet{pandeywardle08} and \citet{keithwardle14}, we use the modified versions.

\section{Verification of numerical methods}
\label{app:num}

\subsection{The Hall effect}
\label{app:num:hall}
The algorithm that governs ambipolar diffusion was thoroughly tested in \citet{WPA2014}.  The same general algorithm governs the Hall effect, but given its non-diffusive behaviour, we verify the algorithm here.  

\subsubsection{Wave Test}
\begin{figure}
\begin{center}
\includegraphics[width=\columnwidth]{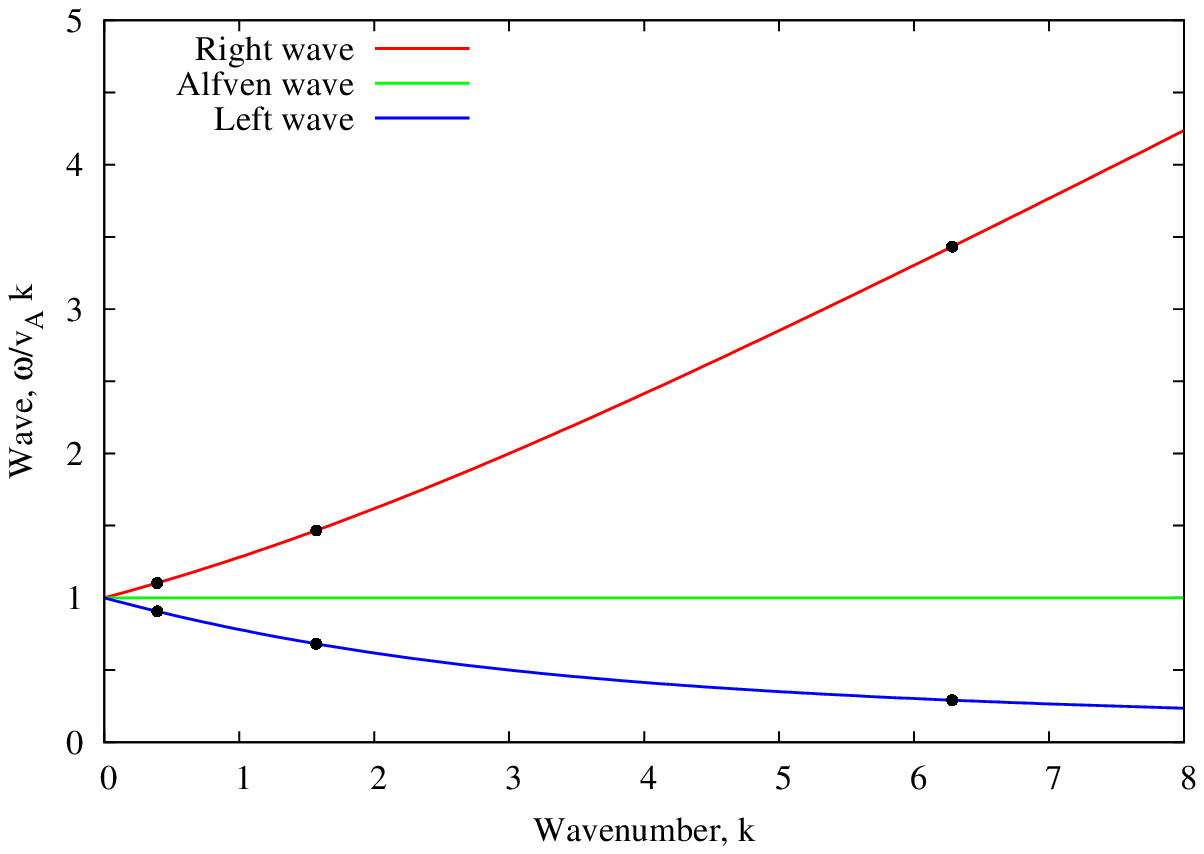}
\caption{Dispersion relation for the left- and right-circularly polarised wave, corresponding to  $\eta_\text{HE} < 0$ and  $> 0$, respectively.  The solid circles are the numerically calculated phase velocities.}  
\label{fig:HallWave}
\end{center}
\end{figure}

As in \citet{sanostone02}, we test our algorithm by comparing the numerically measured phase velocity with the linear dispersion relation, which is given by
\begin{equation}
\left(\omega^2 - v_\text{A}^2k^2\right)^2 = \eta_\text{HE}^2k^4\omega^2,
\end{equation}
where $\omega$ is the angular frequency of the wave and $k$ is the wavenumber.  In Fig.~\ref{fig:HallWave}, we plot the analytical dispersion relation for the left- and right-circularly polarised wave, which correspond to $\eta_\text{HE} < 0$ and  $\eta_\text{HE} > 0$, respectively.  We also plot our numerical results at fixed wavenumbers and constant Hall resistivities.  Our numerical results agree with the analytical prediction within one per cent.  

\subsubsection{Standing Shock Test}
\label{app:num:hallST}
As in \citet{Falle2003} and \citet{OsullivanDownes06}, we test our algorithm against the 1D isothermal steady-state equations for the strong Hall effect regime.  The analytical result is derived by setting $\frac{\partial}{\partial t} = 0$ and $\frac{\text{d}}{\text{d}y} = \frac{\text{d}}{\text{d}z} = 0$  in \eqref{eq:cty}--\eqref{eq:ind}.  Assuming $\frac{{\rm d} B_\text{x}}{\text{d} x} = 0$, the steady-state can be represented with the following two, coupled ordinary differential equations,
\begin{subequations}
\begin{flalign}
\frac{{\rm d} B_\text{y}}{\text{d} x} &= \frac{M_1 R_{22} -M_2 R_{12}}{R_{11}R_{22}-R_{21}R_{12}}, & \\
\frac{{\rm d} B_\text{z}}{\text{d} x} &= \frac{M_1 R_{21} -M_2 R_{11}}{R_{12}R_{21}-R_{22}R_{11}}, &
\end{flalign}
\end{subequations}
where each term on the right-hand side can be written in terms of only $\bm{B}$:
\begin{subequations}
\begin{flalign}
M_1 =& v_\text{x}B_\text{y} - v_\text{x,0}B_\text{y,0} + v_\text{y,0}B_\text{x,0} - v_\text{y}B_\text{x,0},  & \\
M_2 =& v_\text{x}B_\text{z} - v_\text{x,0}B_\text{z,0} + v_\text{z,0}B_\text{x,0} - v_\text{z}B_\text{x,0},  & \\
R_{11} =& \left(\eta^\text{c}_\text{OR} - \eta^\text{c}_\text{AD}\right)\frac{B_\text{z}^2}{B^2} + \eta^\text{c}_\text{AD},  & \\
R_{12} =& \left(\eta^\text{c}_\text{AD} - \eta^\text{c}_\text{OR}\right)\frac{B_\text{y}B_\text{z}}{B^2} + \eta^\text{c}_\text{HE}\frac{B_\text{x}}{B},  & \\
R_{21} =& \left(\eta^\text{c}_\text{AD} - \eta^\text{c}_\text{OR}\right)\frac{B_\text{y}B_\text{z}}{B^2} - \eta^\text{c}_\text{HE}\frac{B_\text{x}}{B},  & \\
R_{22} =& \left(\eta^\text{c}_\text{OR} - \eta^\text{c}_\text{AD}\right)\frac{B_\text{y}^2}{B^2} + \eta^\text{c}_\text{AD}.  &
\end{flalign}
\end{subequations}
Once the magnetic field is known, then the velocities are given by
\begin{subequations}
\begin{flalign}
v_\text{x} =& \frac{1}{2Q}\left( K_\text{x} - \frac{B^2}{2} - \sqrt{\left(K_\text{x} - \frac{B^2}{2}\right)^2 - 4c_\text{s}^2Q^2}\right),& \\
v_\text{y} =& \left( K_\text{y} + B_\text{x}B_\text{y}\right)/Q,& \\
v_\text{z} =& \left( K_\text{z} + B_\text{x}B_\text{z}\right)/Q,& 
\end{flalign}
\end{subequations}
where $c_\text{s}$ is the isothermal sound speed, and $K_\text{x}$, $K_\text{y}$, $K_\text{z}$ and $Q~=~\rho v_\text{x}$ are constants which can be calculated from the initial conditions.  The resistivities, $\eta^\text{c}$, are semi-constant, given by 
\begin{subequations}
\begin{flalign}
\eta^\text{c}_\text{OR} &= C_\text{OR}, &\\
\eta^\text{c}_\text{HE} &= C_\text{HE}B, &\\
\eta^\text{c}_\text{AD} &= C_\text{AD}\frac{B^2}{\rho}  \equiv \frac{v_\text{A}^2}{\gamma_\text{AD}\rho_\text{ion}}, \label{etacad}&
\end{flalign}
\end{subequations}
where $C_\text{OR}$, $C_\text{HE}$ and $C_\text{AD}$ are constants, $\gamma_\text{AD}$ is the collisional coupling constant between ions and neutrals and $\rho_\text{ion}$ is the ion density.  The final term of \eqref{etacad} matches the form presented in \citet{WPA2014}.

For our numerical test, we set up the shock where the values for the left and right sides are given by $\left(\rho_0, P_0, v_\text{x,0},v_\text{y},v_\text{z,0},B_\text{y,0},B_\text{z,0}\right) = 
\left(1.7942,0.017942,-0.9759,-0.6561,0.0,1.74885,0.0\right)$ and $
\left(1.0      ,0.01        ,-1.751, 0.0      ,0.0,0.6        ,0.0\right)$, respectively.  The $x$-magnetic field is constant at $B_\text{x} = 1$, and the isothermal sound speed is $c_\text{s} = 0.1$.  The coefficients are $C_\text{OR} = 1.12 \times 10^{-9}$, $C_\text{HE} =  -3.53  \times 10^{-2}$ and $C_\text{AD} =  7.83 \times 10^{-3}$, thus this evolution will be dominated by the Hall effect.  

The particles are set up on a closed-packed lattice with 512 particles in the $x$-direction on the left-hand side, and 12 and 13 particles in the $y$- and $z$-directions, respectively.   Initialising this idealised test on a three-dimensional lattice will yield instabilities as the system evolves \citep{morris96}; unlike the results presented in this paper, these particles are expected to evolve on the lattice, thus the regular shape will not be washed out.  To minimise the instabilities, we use the $\mathcal{C}^4$ Wendland kernel.

\begin{figure}
\begin{center}
\includegraphics[width=\columnwidth]{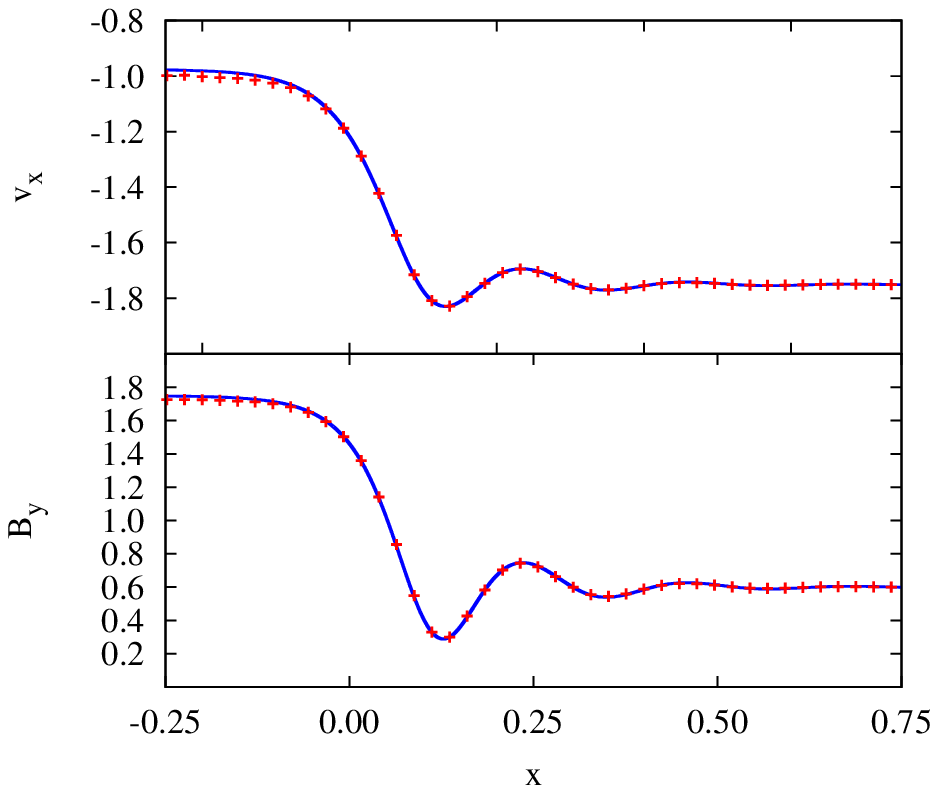}
\caption{The analytical (solid line) and numerical (crosses) results for the isothermal standing shock.  The initial conditions are given in the text.  At any given position, the analytical and numerical solutions agree within 3 per cent.}
\label{fig:HallShock}
\end{center}
\end{figure}

The analytical and numerical results are plotted in Fig.~\ref{fig:HallShock}.  At any given position, the analytical and numerical solutions agree within 3 per cent.  Similar results are obtained using different kernels and different initial lattice configurations.

\subsection{Super-timestepping}
\label{app:num:sts}
We have implemented super-timestepping into {\sc Phantom} for both global and individual particle timesteps.  In both cases, d$t'_\text{diff}$ is determined from the globally minimum $\min\left(\text{d}t_\text{OR}, \text{d}t_\text{AD}\right)$, while d$t$ is either the globally or locally minimum $\min\left(\text{d}t_\text{Courant}, \text{d}t_\text{HE}\right)$ for global and individual timesteps, respectively.  

We use the isothermal C-shock \citep{Draine80} with individual timesteps to test the effectiveness of our super-timestepping implementation.  We include ambipolar diffusion with the semi-constant resistivity given in \eqref{etacad}, setting $\gamma_\text{AD} =1$.  Given our implementation of super-timestepping, $k$ is the only free parameter, where a smaller $k$ yields a larger $N$.  Our tests are run using OpenMP on 12 nodes, and exclude Ohmic resistivity and the Hall effect.

Using $\rho_\text{ion} = 10^{-5}$, we run the C-shock using four values of $k$, as well as a fiducial run without super-timestepping.  In Table~\ref{table:sts5}, we summarise the results of these tests at $t_\text{final} = 14.5\tau_\text{AD}$, where $\tau_\text{AD} = \left(\gamma_\text{AD}\rho_\text{ion}\right)^{-1}$ is the characteristic timescale for ambipolar diffusion.  

\begin{table*}
\begin{center}
\begin{tabular}{c c c c c c c c c}
\hline
$\rho_\text{ion}$ & $k$   & $N_\text{total}$  & Runtime (hours) & $E_\text{total}$ ($\times 10^{19}$ code) & $\sum \left(\rho > \rho_0\right)$ ($\times 10^6$ code) & $N'_\text{max}$ & $N_\text{max}$ &$\nu\left(N_\text{max}\right) $\\
\hline
$10^{-5}$ & -       & 26 829                 &  6.06               &  2.273 & 2.708  & - & - &- \\
$10^{-5}$ &0.30 & 17 588                 &  6.05               &  2.274 & 2.705  & 4 & 4 & $1.73\times10^{-1}$\\
$10^{-5}$ &0.60 & 13 238                 &  4.88               &  2.276 & 2.677  & 3 & 3 & $6.47\times10^{-2}$\\
$10^{-5}$ &0.90 & 12 199                 &  4.83               &  2.277 & 2.696  & 3 & 3 & $9.72\times10^{-3}$\\
$10^{-5}$ &0.99 & 11 551                 &  4.55               &  2.277 & 2.696  & 3 & 3 & $8.67\times10^{-4}$\\ \\
$10^{-6}$ & -       & 142 203              &  24.50                &  2.4065 &  2.990 & -  & -  & - \\
$10^{-6}$ & 0.30 &   37 456              &  21.56                &  2.4061 &  2.992 &    0 & 11& $1.92\times10^{-2}$ \\
$10^{-6}$ & 0.60 &   41 936              &  23.94                &  2.4062 & 2.992 &   30 &  9 & $7.07\times10^{-3}$\\
$10^{-6}$ & 0.90 &   44 706              &  25.39                &  2.4061 & 2.992 &   62 &  9 & $1.06\times10^{-3}$\\
$10^{-6}$ & 0.99 &   61 445              &  34.44                &  2.4061 & 2.992 &  393& 23& $1.44\times10^{-5}$\\
\hline
\end{tabular}
\caption{The results from the isothermal C-shock test using ambipolar diffusion with constant resistivity using $\gamma_\text{AD} = 1$ and $\rho_\text{ion}$ as listed in the first column.  The results are presented for the final times of $t_\text{final} = 14.5$ and $1.25\tau_\text{AD}$ for $\rho_\text{ion} = 10^{-5}$ and $10^{-6}$, respectively.  The rows with a dash listed for $k$ are simulations without super-timestepping.  The third column is the total number of steps, where one step is defined as progressing time d$\tau$.  The fourth column is the runtime in hours, using individual timesteps and OpenMP on 12 nodes.  The fifth column is total energy at $t_\text{final}$, in code units.  The sixth column is the sum of the density of each SPH particle that satisfies $\rho > \rho_0$.  The seventh column is the maximum number of substeps required on any given d$t$ after subdividing d$t$ if required due to the signal velocity constraint.  The eighth column is the maximum number of sub-steps used in the super-timestepping algorithm, and the ninth column is its corresponding $\nu$.}
\label{table:sts5} 
\end{center}
\end{table*}

In each of the models with super-timestepping, the number of \emph{real} steps (where one real step is defined as progressing time d$t$) is 4576, which $\sim$5.9$\times$ lower than the number of real steps required without super-timestepping.  As expected, the \emph{total} number of steps (where one step is defined as progressing time d$\tau$) decreases for increasing $k$.  The required number of sub-steps per step varies as the simulation evolves, hence the non-linear relation between the total number of steps and $k$.  The maximum number of sub-steps is typically $N_\text{max}=3$; the corresponding $\nu$ is given in the final column of Table~\ref{table:sts5}.  At $t_\text{final}$, the total energy of each model differs by less than 0.18 percent.  For a second comparison, we sum the density of each SPH particle $i$ that satisfies $\rho_i > \rho_0 \equiv 1$; these sums differ by less than 0.45 per cent.  

We urge caution when comparing the runtimes to the model without super-timestepping.  In this test, all of the particles have timesteps that are constrained by ambipolar diffusion.  Thus, all particles are evolved with the super-timestepping algorithm using the globally minimum $\text{d}t_\text{AD}$ and $\text{d}t_\text{Courant}$.  Thus, in this model, the super-timestep algorithm essentially uses global timesteps.  The model without super-timestepping continues to optimise the individual timesteps; although more steps are required in total, not every particle is evolved on the shortest timestep, thus decreasing the wall time. 

We have also tested our algorithm in the more extreme condition of $\rho_\text{ion} = 10^{-6}$; the results are also presented in Table~\ref{table:sts5}.  Again, ambipolar diffusion was the limiting timestep for all particles, thus our super-timestepping algorithm essentially used global timesteps.  In this case, the large resistivity meant that large changes in the signal velocity occurred, hence our algorithm routinely decreased d$t$ such that d$\tau$ was never `too large.'  As expected, this occurred more frequently for larger $k$, thus smaller $k$ led to a greater number of steps.  Although super-timestepping is not efficient under these conditions, we can be confident that our algorithm can successfully handle steep shocks since the total energies and summed densities of each model differ by less than 0.017 and 0.33 per cent, respectively.  Further, given that the majority of the particles in the models presented in this paper are not constrained by Ohmic resistivity or ambipolar diffusion, our super-timestepping algorithm continues to optimise the individual timesteps to decrease the runtime.

In the models presented in the paper, the number of steps taken by the super-timestep algorithm is typically less than ten, but a few iterations use upwards of 30 sub-steps; these correspond to $\nu = 2.26\times 10^{-2}$ and $9.47\times10^{-5}$, respectively.  Given the tests presented here, we are confident that our results are not negatively impacted by super-timestepping.    

\label{lastpage}
\enddocument